\documentclass[12pt,oneside]{book}
\usepackage{amsmath,latexsym,amssymb,graphics,theorem}
\newtheorem{teo}{Theorem}[chapter]
\newtheorem{defi}[teo]{Definition}

{\theoremstyle{plain}\theoremheaderfont{\bfseries\itshape}
\newtheorem{prop}[teo]{Proposition}}

\newtheorem{lemma}[teo]{Lemma}
\newtheorem{cor}[teo]{Corollary}

{\theoremstyle{plain}\theorembodyfont{\rmfamily}
\theoremheaderfont{\itshape} \newtheorem{remark}[teo]{Remark}}

\numberwithin{equation}{section}

\long\def\symbolfootnote[#1]#2{\begingroup%
\def\thefootnote{\fnsymbol{footnote}}\footnote[#1]{#2}\endgroup}

\newcommand{\R}{\mathbb{R}}
\newcommand{\N}{\mathbb{N}}

\newcommand{\ii}{\'{\i}}
\newcommand{\scri}{{\cal J}}
\newcommand{\Par}{\par}

\begin{document}

\title{Global properties of asymptotically de Sitter and Anti de
Sitter spacetimes}
\author{Didier A. Solis}
\date{May,  17, 2006}
\maketitle

\pagestyle{plain}

\pagenumbering{roman}

\addtocounter{page}{2}

\begin{center}
\par\vspace{4cm}
\textit{Ad Majorem Dei Gloriam.}
\end{center}

\vspace{1cm}

\begin{center}
To all those who seek to fulfill God's will in their lives.
\end{center}

\vfill\eject

\begin{center}
ACKNOWLEDGEMENTS
\end{center}

I would like to express my deepest gratitude to:

\begin{itemize}

\item My advisor Dr. Gregory Galloway, for his constant support
and help. This dissertation would not had seen the light of day
without his continued encouragement and guidance.

\item Dr. N. Saveliev, Dr. M. Cai and Dr. O. Alvarez for their
valuable comments and thorough revision of this work.

\item  The Faculty and staff at the Math Department for all the
knowledge and affection they shared with me.

\item CONACYT (Consejo Nacional de Ciencia y Tecnolog{\ii}a) for
the financial support granted to the author.

\item My family: Douglas, Rosalinda, Douglas Jr, Rosil\'u and
Josu\'e for all their prayers and unconditional love. They are
indeed the greatest blessing in my life.

\item All my friends, especially Guille,  for always being there
for me to share the struggles and joys of life.

\item To the CSA gang, for being a living example of faith lived
in love and hope.

\item Last but by no means least, to God, without whom there is no
math, laughter or music.

\end{itemize}

\vspace{2cm}

\tableofcontents

\chapter{Introduction}

\pagestyle{myheadings}

\pagenumbering{arabic}

As a means to acquaint the reader to the subject, we begin this
work by providing a quick survey on definitions and standard
results in Lorentzian geometry. The proofs of most results listed
in this chapter can be found in any of the standard references
$\cite{BE,HE,ON,P}$. For consistency, we will follow as much as we
can the conventions given in $\cite{ON}$.

\section{Lorentz vector spaces}

In the special relativity, spacetime is modelled after Minkowski
space $\mathbb{M}^n$, that is, $\R^n$ endowed with the Lorentz
product $L:\mathbb{M}^n\times \mathbb{M}^n\to \R$,
\begin{equation}
L(v,w)=-v_0w_0+v_1w_1+\ldots +v_{n-1}w_{n-1}.
\end{equation}

In this context, the first coordinate $v_0$ of $v\in\mathbb{M}^n$
represents time while the remaining coordinates represent space.

\vfill\eject

From the algebraic point of view, $L$ is just a symmetric bilinear
form on the vector space $\R^n$. Moreover $L$ is non-degenerate,
in the sense that the linear functional $L_v\colon \mathbb{M}^n\to
\R$, $L_v(w)=L(v,w)$ is identically zero if and only if $v=0$.

In greater generality, a non-degenerate symmetric bilinear form
$g\colon V\times V\to\R$ on a real vector space $V$ is called an
\textit{inner product} in $V$. By a well known result in linear
algebra, every inner product $g$ admits an orthonormal basis, i.e.
a basis $\{e_i\}$, $i=1,2,\ldots ,n$ satisfying
\begin{equation}
g(e_i,e_j)=\begin{cases} -1 & \text{if $i=j$ and $1\le i\le r$}\\
1 &\text{if $i=j$ and $r+1\le i\le n$}\\ 0 &\text{if $i\neq j$}
\end{cases}
\end{equation}
Further, the number $r$ is an invariant of $g$, called the
\textit{signature} of $g$.

\begin{defi}
A Lorentz vector space $(V,g)$ is a vector space $V$ of dimension
$n\ge 2$ together with an inner product of signature $r=1$.
\end{defi}

Hence, it follows that any Lorentz vector space of  dimension $n$
is isometric to $\mathbb{M}^n$. Therefore $V$ splits in three
disjoint sets, just as Minkowski space.

\begin{figure}
\begin{center}
\includegraphics{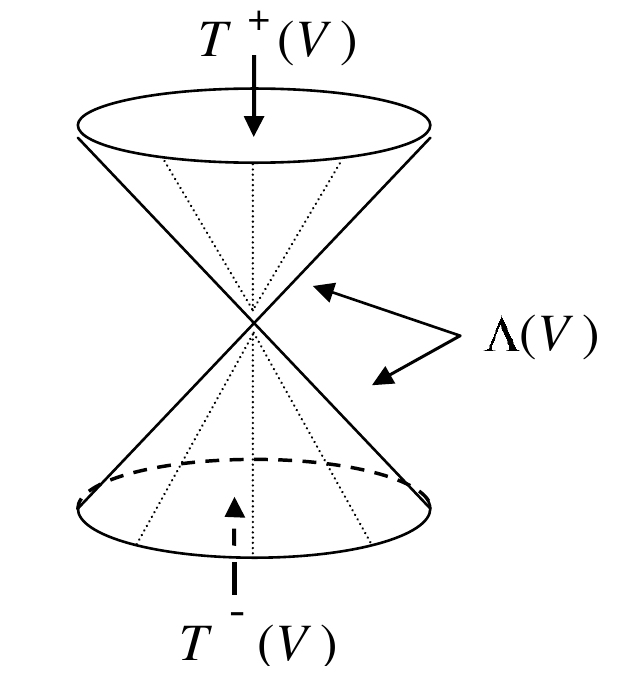}
\caption{\label{figmin00} {\footnotesize Minkowski space.}}
\end{center}
\end{figure}

\begin{defi}
Let $(V,g)$ be a Lorentz vector space, we say that $v\neq 0$ is
timelike, spacelike or null if $g(v,v)<0$, $g(v,v)>0$ or
$g(v,v)=0$, respectively. The zero vector will be considered
spacelike. A non-spacelike vector will be also called causal.
\end{defi}

Notice that the set of null vectors form a cone. This is true also
for timelike and causal vectors.

\vfill\eject

\begin{defi}
Let $(V,g)$ be a Lorentz vector space, the sets
\begin{equation}
\Lambda (V) = \{v\in V-\{0\}\mid g(v,v)=0\},\qquad  {\cal
T}(V)=\{v\in V\mid g(v,v)<0\},
\end{equation}
\begin{equation*}
C(V)=\{v\in V\mid g(v,v)\le 0\},
\end{equation*}
are called the  null, timelike and causal cone of $V$,
respectively.
\end{defi}

The existence of these cones is a distinctive feature of
Lorentzian vector spaces. At this point it is also important to
observe that ${\cal T}(V)$ has exactly two connected components.
This simple observation enables us to develop the notion of  past
and  future directions.

\begin{defi}
A time orientation on a Lorentz vector space $V$ of dimension
$n\ge 3$ is a choice of one connected component ${\cal T}^+(V)$ of
${\cal T}(V)$. This component will be called the future timelike
cone of $V$, while ${\cal T}^-(V)={\cal T}(V)-{\cal T}^+(V)$ is
called the past timelike cone of $V$.
\end{defi}

Since any null vector can be approximated by a sequence of
timelike vectors, a time orientation on $V$ induces in a natural
way a choice of a component $C^+(V)$ of the causal cone $C(V)$.

The following proposition enumerates some of the most important
properties relating timelike cones (see lemmas 5.29-5.31 in
$\cite{ON}$)

\begin{prop}\label{cones0}
Let $(V,g)$ be a Lorentz vector space and for $v\in C(V)$ define
$\vert v\vert :=\sqrt{-g(v,v)}$. Thus the following results hold:
\begin{enumerate}
\item $C^+(V)$ and $C^-(V)$ are convex sets.\item $v\in C^+(V)$ if
and only if $-v\in C^-(V)$.\item Let $v,w\in C^+(V)$, then $v$ and
$w$ are in the same causal cone if and only if $g(v,w)<0$ or $v$
and $w$ are two collinear null vectors.\item $\vert g(v,w)\vert\ge
\vert v\vert\cdot \vert w\vert$ for all $v,w\in C(V)$, with
equality if and only if $v$ and $w$ are collinear.\item $\vert
v+w\vert\ge \vert v\vert +\vert w\vert$ for all $v,w\in C^+(V)$.
\end{enumerate}
\end{prop}

To finish our discussion on Lorentz vector spaces, let us notice
that a Lorentz metric $g$ on $V$ does not always induce a Lorentz
metric on a vector subspace $W\subset V$. There are actually three
different possibilities: $(g\vert_W,W)$ can be either a Lorentz
vector space, a Riemann vector space (that is, $g\vert_W$ is a
positive definite bilinear form on $W$) or $g\vert_W$ may be a
degenerate form on $W$. In the latter case, we say that $W$ is a
\textit{null} subspace of $V$.

The next proposition gives a criterion to determine in which
category $W$ falls, depending on the causal vectors of $V$.

\begin{prop}\label{causalchar}
Let $(V,g)$ be a Lorentz vector space and $W\subset V$ a vector
subspace of $V$ then
\begin{enumerate}
\item $(W,g\vert_W)$ is a Riemann vector space if and only if $W$
contains no causal vector.\item $(W,g\vert_W)$ is a Lorentz vector
space if and only if $W$ contains a timelike vector.\item
$(W,g\vert_W)$ is null if $W$ contains a null vector but no
timelike vector.
\end{enumerate}
\end{prop}

\begin{figure}
\begin{center}
\includegraphics{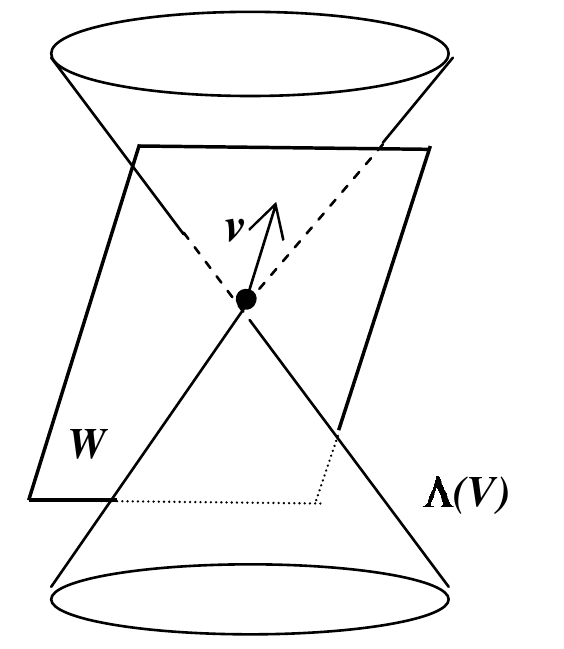}
\caption{\label{figmin01}{\footnotesize $ W\cap (\Lambda
(V)\cup\{0\})=\{tv\mid t\in\R\}$.}}
\end{center}
\end{figure}

\begin{remark}\label{degiso}
If $(W,g\vert_W)$ is a degenerate subspace of $V$ then $W$
intersects the null cone $\Lambda (V)$ in a distinctive null
direction. More precisely, $W\cap (\Lambda (V)\cup \{0\})$ is a
one dimensional (null) vector subspace of $V$. As a consequence,
if $v\in W\cap \Lambda (V)$ then the map $w\mapsto v+w$ is a
linear monomorphism from $W$ to $T_v\Lambda (V)$. (See figure
$\ref{figmin01}$).
\end{remark}

\section{Spacetimes}\label{spacetimes1}

We move now into studying spaces that locally look like Minkowski
space.

A \textit{metric} $g$ on a connected smooth manifold $M$ is a
non-degenerate, symmetric bilinear smooth tensor field of type
$(0,2)$ on $M$. In other words, $g$ smoothly assigns to each $p\in
M$ an inner product $g_p$ on the vector space $T_pM$. Notice that
by the connectedness of $M$ the signatures of $g_p$ are all equal,
and hence this common signature is referred to as the signature of
$g$. A smooth manifold endowed with a metric is called a
\textit{semi-Riemannian manifold.}

\begin{defi}
A Lorentzian manifold is a semi-Riemannian manifold of signature
$r=1$.
\end{defi}
Then the metric $g$ on a Lorentzian manifold $M$ can be thought as
a device that transplants smoothly the geometry of Minkowski space
to the tangent spaces of $M$.

For a Lorentz manifold to adequately describe our universe, it
must have a globally defined ``future".  As discussed above, we
can give a time orientation to each tangent space on a Lorentzian
manifold $(M,g)$.   Moreover, in light of proposition
$\ref{cones0}$, a time orientation on $V$ is equivalent to the
selection of a timelike vector $v\in{\cal T}(V)$. Thus the
existence of a global time orientation on $(M,g)$ is equivalent to
the existence of a smooth timelike vector field on $M$.

\begin{defi}
A Lorentzian manifold is said to be time orientable if there
exists a smooth timelike vector field $X$ on $M$.
\end{defi}

\vfill\eject

Notice that the existence of a time orientation is independent of
the existence of an orientation of $M$ as a smooth manifold.
Nevertheless, in analogy with non-orientable smooth manifolds, if
$(M,g)$ does not admit a time orientation, then it has a time
orientable double cover.

Because of its physical significance, on chapters 2 and 3 we will
be dealing exclusively with spacetimes.

\begin{defi}
A spacetime is an oriented and time oriented Lorentzian manifold.
\end{defi}

\begin{figure}
\begin{center}
\includegraphics{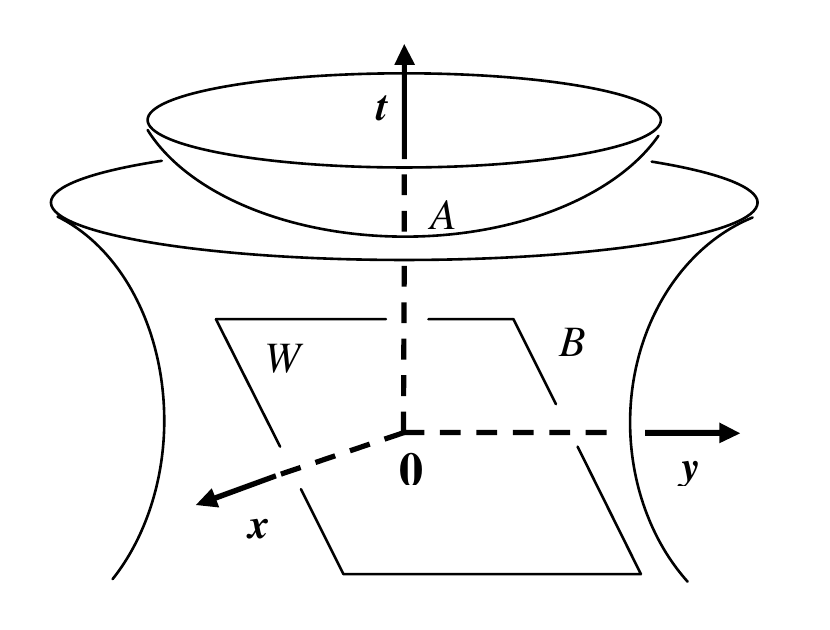}
\caption{\label{figchar}}
\begin{quote}
\begin{quote}
 {\footnotesize Three submanifolds of
$M=\mathbb{M}^3$ with different causal character. $A$ is the upper
half of the hyperboloid $L(v,v)=-1$, $B$ is the hyperboloid
$L(v,v)=1$, and $W$ is the plane $y+t=0$. $A$ is spacelike, $B$ is
timelike and $W$ is null (degenerate).}
\end{quote}
\end{quote}
\end{center}
\end{figure}

We finish this section by making the following observation:

\begin{remark}
Any embedded submanifold $\imath\colon N\hookrightarrow M$ of a
Riemannian manifold $(M,g)$ is a Riemannian manifold itself since
the pullback $\imath^*g$ is a positive definite symmetric form on
$N$. This does not hold in the Lorentzian setting, since the
pullback $\imath^*g$ may be degenerate. Further, it may not have a
constant signature; and even if it does, $\imath^*g$ could be
either Lorentzian or Riemannian. In the former case we say
$N\hookrightarrow M$ is \textit{timelike} while in the latter we
say $N$ is \textit{spacelike.} (See figure $\ref{figchar}$).
\end{remark}

\subsection{The Levi-Civita connection}

Just as in the more familiar Riemannian setting, a Lorentz metric
$(M,g)$ not only gives us a way to measure lengths of tangent
vectors, but it also can be used to build the machinery that
enables us to differentiate vector fields.

\begin{defi}
A connection on a smooth manifold $M$ is a function $\nabla\colon
{\cal X}(M)\times {\cal X}(M)\to {\cal X}(M)$, $(X,Y)\mapsto
\nabla_XY$ such that
\begin{enumerate}
\item $\nabla_{fX+gY}Z=f\nabla_XZ+g\nabla_YZ$ for all $f,g\in{\cal
C}^{\infty}(M)$.\item $\nabla_X(Y+Z)=\nabla_XY+\nabla_XZ$.\item
$\nabla_XfZ=f\nabla_XZ+X(f)Z$ for all $f\in{\cal C}^{\infty}(M)$.
\end{enumerate}
\end{defi}

Thus a connection is a device that takes ``directional
derivatives" of smooth vector fields. Notice that besides the
notion of differentiability introduced by a connection, we also
have a simpler notion of derivation given by the Lie bracket
$[X,Y]$. The difference between these two ways of taking
derivatives is given by a tensor, called the \textit{torsion}
\begin{equation}
\text{Tor}(X,Y)=\nabla_XY-\nabla_YX-[X,Y]\qquad X,Y\in{\cal X}(M).
\end{equation}
A connection is \textit{torsion free} if $\text{Tor}\equiv 0$. In
the case of a connection on a semi-Riemannian manifold $(M,g)$, it
is desirable that $\nabla$ relates to the metric $g$ via a
``product rule". We will say that $\nabla$ \textit{is adapted to}
$g$ if
\begin{equation}
Z(g(X,Y))=g(\nabla_ZX,Y)+g(X,\nabla_ZY)\qquad\text{for all }
X,Y,Z\in{\cal X}(M)
\end{equation}

A fundamental result due to T. Levi-Civita  establishes the
existence and uniqueness of a torsion free connection adapted to
the metric in any semi-Riemannian manifold.

\begin{teo}
Let $(M,g)$ be a semi-Riemannian manifold $(M,g)$. Then there
exists a unique torsion free connection $\nabla$ on $M$ adapted to
$g$. Such connection is called the Levi-Civita connection of $M$.
\end{teo}

Just as in Riemannian geometry, curvature is measured as the
difference between $[\nabla_X,\nabla_Y]$ and $\nabla_{[X,Y]}$.

\begin{defi}
Let $(M,g)$ be a semi-Riemannian manifold. The Riemann curvature
morphism is the map that assigns to each pair of smooth vector
fields $(X,Y)$ the endomorphism $R(X,Y)\colon {\cal X}(M)\to {\cal
X}(M)$ given by
\begin{equation}
R(X,Y)Z=\nabla_X\nabla_YZ-\nabla_Y\nabla_XZ-\nabla_{[X,Y]}Z.
\end{equation}
\end{defi}
Thus if ${\cal U}$ is a coordinate chart we have
$R(\partial_{\alpha},\partial_{\beta})\partial_{\gamma}={R^{\zeta}}_{\alpha\beta\gamma}\partial_{\zeta}$
for some ${R^{\zeta}}_{\alpha\beta\gamma}\in {\cal
C}^{\infty}({\cal U})$\symbolfootnote[2]{Recall  Einstein
summation convention:  summation is performed over an abstract
index when it appears twice in the same expression, once up and
once down.}

A number of different tensors can be constructed using the Riemann
morphism.

\begin{defi}$\ $\par

\begin{enumerate}
\item The Riemann curvature tensor is the $(0,4)$ tensor
metrically related to $R(X,Y)Z$, that is
$R(X,Y,Z,W):=g(R(X,Y)Z,W)$. \item The Ricci tensor $\text{Ric}$ is
found by contracting the Riemann endomorphism. In coordinates
${\text{Ric}}_{\alpha\beta}={R^{\zeta}}_{\alpha\beta\zeta}$. \item
The Weyl tensor $W$ is the trace free part of the Riemann
curvature tensor, hence it is characterized by the requirement
that any of its contractions vanishes. It is given by the
formula\begin{equation}
\displaystyle{W:=R-\frac{2}{n-2}g\circledast\text{Ric}+\frac{R}{(n-1)(n-2)}g\circledast
g},\end{equation}where $\circledast$ denotes the Kulkarni-Nomizu
product (see paragraph ${1.110 }$ in $\cite{besse}$).\item The
scalar curvature $R$ is the trace of the Ricci tensor. Thus
$R=g^{\alpha\beta}{\text{Ric}}_{\alpha\beta}$. \item Let
$\Pi\subset T_pM$ be a two dimensional plane of non-zero area. The
sectional curvature $K$ of $\Pi$ is defined by
\begin{equation}{\displaystyle K(\Pi
):=\frac{g(R(v,w)v,w)}{g(v,v)g(w,w)-g(v,w)^2}},\end{equation}
where $v,w\in T_pM$ are any two vectors spanning $\Pi$.
\end{enumerate}
\end{defi}

\begin{defi}
An {Einstein manifold} is a semi-Riemannian manifold $(M,g)$ for
which $\text{Ric}=kg$, for some constant $k$.
\end{defi}
It follows that Einstein manifolds  have constant scalar
curvature. As the name suggests, Einstein manifolds arise as
solution of the Einstein equations of general relativity.

\begin{remark}\label{weyl}
An Einstein manifold whose Weyl tensor vanishes everywhere has
constant (sectional) curvature (see paragraphs $1.114$ and $1.118$
in \cite{besse}).
\end{remark}

\subsection{Covariant derivative}

The Levi-Civita connection can also be used to take derivatives on
tangent directions along curves.
\begin{teo}
Let $\alpha\colon I\to M$  be a smooth curve and let us denote by
$\overline{{\cal X}}(M)$ the set of smooth vector fields along
$\alpha$. There is a unique operator $D/dt\colon \overline{{\cal
X}}(\alpha )\to\overline{{\cal X}}(\alpha )$ satisfying
\begin{enumerate}
\item ${\displaystyle
\frac{D}{dt}(X+Y)=\frac{DX}{dt}+\frac{DY}{dt}}$ for all $X,Y\in
\overline{{\cal X}}(\alpha )$.\item ${\displaystyle
\frac{D}{dt}(fX)=\frac{df}{dt}X+f\frac{DX}{dt}}$ for all
$f\in{\cal C}^{\infty}(\alpha )$, $X\in\overline{{\cal X}}(\alpha
)$.\item ${\displaystyle
\frac{D}{dt}X\vert_{\alpha}(t)=\nabla_{\alpha^{\prime}(t)}X}$ for
all $X\in{\cal X}(M)$.
\end{enumerate}
\noindent $D/dt$ is called the covariant derivative along
$\alpha$.
\end{teo}

A vector field $X\in\overline{{\cal X}}(\alpha )$ is said to be
\textit{parallel} if $X^{\prime}:=DX/dt$ vanishes along $\alpha$.
Given $v\in T_pM$ with $p=\alpha (0)$, basic ODE theory guarantees
the existence of a unique parallel vector field $V\in
\overline{{\cal X}}(\alpha )$ with $V(0)=v$. If $q=\alpha (a)$
then $V(a)\in T_qM$ is called \textit{the parallel translate of
$v$ to $q$ along $\alpha$}. Hence by considering the parallel
translates of all vectors at $p$ along $\alpha$, we construct a
map $P\colon T_pM\to T_qM$ called \textit{parallel translation
from $p$ to $q$ along $\alpha$}. It is easy to check that $P$ is a
linear isometry.

\subsection{Geodesics}

By taking the covariant derivative of the tangent vector field to
a curve $\alpha$ we get an analog to the acceleration of $\alpha$.
From our experience in Euclidean space, a straight line can be
characterized as a curve with zero acceleration. We can extend
this idea to semi-Riemannian geometry as follows:

\begin{defi}
A geodesic is a smooth curve $\alpha\colon I\to M$ such that
$\alpha^{\prime\prime}\equiv 0$
\end{defi}

\enlargethispage{.3cm}

Geodesics play a fundamental role in geometry. From existence and
uniqueness results in ODE theory, it follows that given $p\in M$
and $v\in T_pM$ there exists a unique geodesic $\gamma_v\colon
I\to M$ with $\gamma_v (0)=p$ and $\gamma_v^{\prime}(0)=v$ whose
domain of definition is maximal. In other words, if $\alpha\colon
J\to M$ is another geodesic with $\alpha (0)=p$ and
$\alpha^{\prime}(0)=v$, then $J\subset I$ and
$\alpha=\gamma_v\vert_J$. Such a geodesic is said to be
\textit{inextendible.} Furthermore, it turns out that every
continuous extension of a geodesic $\gamma$ to one of its
endpoints is a geodesic itself (refer to lemma 5.8 in
$\cite{ON}$).

\begin{prop}\label{geoext}
Let $\gamma\colon [0,a)\to M$ be a geodesic and let
$\tilde{\gamma}\colon [0,a]\to M$ be a continuous curve with
$\tilde{\gamma}\vert_{[0,a)}=\gamma$.  Then  $\tilde{\gamma}$ is a
geodesic.
\end{prop}

We can group all the geodesics emanating from $p\in M$ in a single
map $\exp_p\colon T_pM\to M$ called the \textit{exponential map at
$p$}.

\begin{defi}
The exponential map at $p$, $\exp_p\colon T_pM\to M$, is defined
by\break $\exp_p (v)=\gamma_v(1)$.
\end{defi}

The differential of the exponential map $(\exp_p)_*\colon
T_0(T_pM)\to T_pM$ is the canonical isomorphism $T_0(T_pM)\simeq
T_pM$. Thus, by the inverse function theorem, there is a
neighborhood $\tilde{{\cal U}}$ of $0_p\in T_pM$ such that
$\exp_p\vert_{\tilde{\cal U}}$ is a diffeomorphism onto ${\cal
U}:=\exp_p (\tilde{{\cal U}})$. Such a neighborhood ${\cal U}$ is
called a \textit{normal neighborhood of $p$}. A more detailed
analysis shows every point $p\in M$ has a neighborhood ${\cal U}$
that is normal with respect to \textit{any} of its points. These
latter normal neighborhoods are called \textit{convex.} The
terminology comes from the fact that there exists a unique
geodesic segment contained in ${\cal U}$ joining any pair of
points $x,y\in{\cal U}$.

At this point we notice that the exponential map based at $p\in M$
is a particular case of a more general construction. Given a
semi-Riemannian submanifold $P\subset M$ of dimension $k$ and
$p\in P$, we have the splitting $T_pM=T_pP\oplus T_pP^\perp$,
where $T_pP^\perp=\{v\in T_pM\mid g(v,w)=0\ ,\ \forall w\in
T_pP\}$. Hence $NP:=\bigcup_{p\in P}T_pP^\perp$ is a rank $n-k$
vector bundle over $P$. We then define the \textit{normal
exponential map to $P$}, $\exp^{\perp}\colon NP\to M$ by
$\exp^{\perp}(v_p):=\gamma_{v_p}(1)$. Thus, if we consider $\{p\}$
as a $0$-dimensional submanifold of $M$, the normal exponential
map to $\{p\}$ is just $\exp_p$.

\vfill\eject

\begin{defi}
A Jacobi field on a geodesic  $\alpha$ is a vector field
$Y\in\overline{\cal X}(\alpha)$ that satisfies the Jacobi equation
$Y^{\prime\prime}+R(Y,\alpha^{\prime})\alpha^{\prime}=0$.
\end{defi}

Jacobi fields arise in a natural way in various branches of
semi-Riemannian geometry, such as comparison theory. They also
play a critical role in the calculus of variations, since the
variation vector field of a geodesic $\alpha$ by geodesics is a
Jacobi field on $\alpha$. The next result shows that the existence
of certain Jacobi fields are related to the degeneracy of the
exponential map.

\begin{defi}
Two points $p$ and $q$ along a geodesic $\alpha$ are conjugate if
there exists a nontrivial Jacobi field $Y$ on $\alpha$ such that
$Y_p=0=Y_q$.
\end{defi}

\begin{prop}\label{conj}
Let $\alpha$ be a geodesic with $\sigma (0)=p$ and $\sigma (a)=q$.
Then the following are equivalent:
\begin{enumerate}
\item $q$ is conjugate to $p$ along $\alpha$.\item There exists a
nontrivial variation through geodesics from $p$ whose variation
vector field vanishes at $q$.\item The exponential map $\exp_p$ is
singular at $a\alpha^{\prime}(0)\in T_pM$.
\end{enumerate}
\end{prop}

In analogy to the concept of conjugate points, we encounter the
notion of \textit{focal points} when considering the points where
the normal exponential map degenerates. Appendix B is devoted to
the treatment of null focal points to spacelike submanifolds.

Note that geodesics in a Lorentzian manifold posses a
\textit{causal character.} In other words, $\alpha$ is either
timelike, spacelike or null, according to the character of its
tangent vector field.

Moreover, geodesics in a Lorentzian manifold are radically
different from its Riemannian counterparts. More notably, timelike
geodesics on a Lorentzian manifold locally \textit{maximize} arc
length. That is, if $\alpha$ is a timelike geodesic in a
Lorentzian manifold $(M,g)$,  ${\cal U}$ is a normal neighborhood
around $p=\alpha (0)$ and $I\subset \R$ is an interval such that
$0\in I$ and $\alpha (I)\subset {\cal U}$ then the geodesic
segment $\alpha\vert_0^a$ is the unique longest timelike curve in
${\cal U}$ joining $p$ to $\alpha (a)$ for all $a\in I$.

\section{Causal theory}

\subsection{Definitions}

A time orientation can be used to causally relate pair of points
on a spacetime $(M,g)$. We say that $p$ \textit{ chronologically
precedes} $q$, denoted by $p\ll q$, if there is a future timelike
path\symbolfootnote[2]{A future timelike path is a piecewise
smooth curve with finitely many segments, all of which are future
timelike.} from $p$ to $q$. Likewise, we say $p$ \textit{causally
precedes} $q$ if there is a future causal path from $p$ to $q$ and
denote this by $p\le q$. These relations define the \textit{causal
structure} of $M$.

Notice that we can also define the relations $\ll$ and $\le$
{relative} to a set $A\subset M$. In other words, $p\ll_Aq$
($p\le_Aq$) will mean that there is a future timelike (causal)
path contained in $A$ from $p$ to $q$.

Intuitively, if $p\ll q$ then $q$ is at the future of $p$. We
formalize this observation in the next definition.

\vfill\eject

\begin{defi}
Let $(M,g)$ be a spacetime and $p\in M$. The timelike future of
$p$, $I^+(p)$, and causal future of $p$, $J^+(p)$, are given by:
\begin{equation}
I^+(p) = \{q\in M\mid p\ll q\},\qquad\qquad J^+(p)=\{q\in M\mid
p\le q\}.
\end{equation}
The timelike and causal pasts of $p$ are defined time dually:
\begin{equation}
I^-(p) = \{q\in M\mid q\ll p\},\qquad\qquad J^-(p)=\{q\in M\mid
q\le p\}.
\end{equation}
\end{defi}
Notice the above definitions extend to arbitrary sets $A\subset M$
in a natural way. For instance $I^+(A)=\bigcup_{p\in A}I^+(p)$.

The causal structure of $M$ is closely related to its manifold
topology.

\begin{prop}
Let $(M,g)$ be a spacetime and $p\in M$. Then
\begin{enumerate}
\item $I^+(p)$ is an open set in $M$. \item
$\mbox{int}(J^+(p))=I^+(p)$.\item
$J^+(p)\subset\overline{I^+(p)}$.
\end{enumerate}
\end{prop}

Notice that in general $J^+(p)$ is not a closed subset of $M$. On
the other hand, $J^+(p,{\cal U})$ is closed in ${\cal U}$, where
${\cal U}$ is a convex normal neighborhood of $p$.

\begin{figure}
\begin{center}
\includegraphics{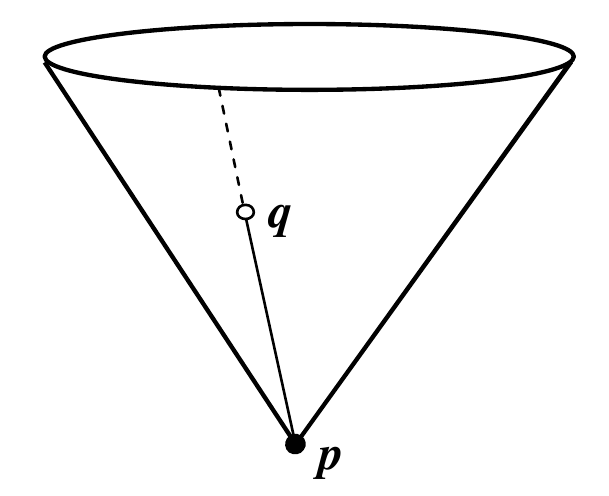}
\begin{quote}
\begin{quote}
\caption{\label{figj}} {\footnotesize Let $M=\mathbb{M}^3-\{q\}$
and $p$ be the origin, then $J^+(p)$ is not a closed subset of
$M$.}
\end{quote}
\end{quote}
\end{center}
\end{figure}

Note also that any timelike curve can be approximated by a
sequence of broken null geodesics. Conversely, using variational
arguments we can deform a causal path $\alpha$ into a timelike
curve while keeping its endpoints fixed, unless $\alpha$ is a
smooth null geodesic (see proposition 10.46 in $\cite{ON}$).

\begin{prop}
Let $(M,g)$ be a spacetime. If $\alpha$ is a future causal path
from $p$ to $q$ that is not a null geodesic, then there is a
timelike curve from $p$ to $q$ arbitrarily close to $\alpha$.
\end{prop}

The following results follow immediately:

\begin{prop}\label{ngeo1}
Let $q\in J^+(p)-I^+(p)$, then every future causal curve from $p$
to $q$ can be parameterized as a null geodesic.
\end{prop}

\begin{prop}
Let $p,q,r\in M$ with $p\le q$ and $q\ll r$, then $p\ll r$.
\end{prop}

\enlargethispage{.5cm}

\begin{prop}\label{2null}
Let $\alpha$ and $\gamma$ be two distinct future null geodesics
meeting at $q\in M$. If $p$ comes before $q$ along $\alpha$ and
$r$ comes after $q$ along $\gamma$, then $p\ll r$.
\end{prop}

Thus proposition $\ref{2null}$ implies that if two future directed
null geodesics emanating from $p$ meet at $q$, then every point to
the future of $q$ along any of them is on the timelike future of
$p$.

Moreover, since in light of proposition $\ref{conj}$ conjugate
points can be viewed as almost meeting points of geodesics, we
also have the following:

\begin{prop}\label{3null}
Let $\alpha$ be a null geodesic emanating from $p\in M$. If
$q\in\alpha$ comes after the first conjugate point to $p$ along
$\alpha$, then $q\in I^+(p)$.
\end{prop}

Notice also that by considering the normal exponential map to a
spacelike submanifold we get a similar result in terms of focal
points:
\begin{prop}\label{space2focal}
Let $P$ be a spacelike submanifold of $M$ and $\alpha\colon
[0,1]\to M$ a future causal curve from $p$ to $q\in S$. Then
either $I^+(p)\cap S\neq\emptyset$ or $\gamma$ is a null geodesic
normal to $S$ with no focal points to $S$ on $(0,1]$.
\end{prop}

We end up this section by defining the Lorentz distance function
and listing some of its properties.

\begin{defi}
Let $(M,g)$ be a spacetime, then  $\tau\colon M\times M\to
[0,\infty ]$ defined by
\begin{equation*}
\tau (p,q)=\left\{
\begin{array}{cc}
\sup \{L(\gamma )\mid \mbox{$\gamma$ is a causal curve from $p$ to
$q$\} }&\mbox{if $p\le q$}\\
0&\mbox{otherwise}
\end{array}\right.
\end{equation*}
is called the Lorentz distance function of $M$.
\end{defi}

\begin{prop} Let $(M,g)$ be a spacetime and $\tau $ its Lorentz
distance function, then
\begin{enumerate}
\item $\tau (p,q)>0$ if and only if $p\ll q$.\item If $p\le q$ and
$q\le r$ then $\tau (p,r)\ge \tau (p,q)+\tau (q,r)$.\item $\tau$
is lower semicontinuous, although it is not continuous in general.
\end{enumerate}
\end{prop}

\subsection{Global causality conditions}

For a spacetime $(M,g)$ to adequately depict the universe we live
in, it must satisfy certain causality conditions. For instance, it
is clear that $M$ should not contain any closed causal path.
Further, we would like $M$ not to have ``almost closed" causal
curves.

\begin{defi}
A set $A\subset M$ is causally convex if every causal curve
intersects it in a connected set. Strong causality is said to hold
at $p\in M$ if $p$ has arbitrarily small causally convex
neighborhoods. A spacetime is strongly causal if strong causality
holds in all of its points.
\end{defi}

Thus in strongly causal spacetimes, any future causal curve
beginning near $p$ must stay within a neighborhood of $p$ or else
leave it never to return. The following result characterizes
strong causality failure (see lemma 4.16 in $\cite{P}$).

\begin{prop}\label{penlemma}
Strong causality fails at $p\in M$ if and only if there exists
$q\le p$, with $p\neq q$ such that $x\ll y$ for all $x\in I^-(p)$,
$y\in I^+(q)$.
\end{prop}

Notice also that no future inextendible causal curve
\symbolfootnote[2]{A causal curve is said to be
\textit{inextendible} if it doesn't have any continuous
extension.} can be ``imprisoned" in a compact subset of a strongly
causal spacetime (refer to lemma 14.13 in $\cite{ON}$).

\begin{prop}\label{impri}
Suppose strong causality holds on a compact set $K\subset M$. If
$\alpha$ is a future inextendible causal curve with $\alpha (0)\in
K$ then $\alpha$ leaves $K$ never to return.
\end{prop}

Throughout this thesis a yet stronger form of causality is used.
For the most part, we will be studying globally hyperbolic
spacetimes.

\begin{defi}
A spacetime $(M,g)$ is said to be globally hyperbolic if it
strongly causal and the sets $J^+(p)\cap J^-(q)$ are compact for
all $p,q\in M$.
\end{defi}

Although global hyperbolicity seems to be  a very stringent
condition, as it turns out it is a reasonable assumption to be
imposed on a spacetime modelling our universe. From the results
presented in section $\ref{dd}$ below, it follows that a non
globally hyperbolic spacetime presents a breakdown in
predictability: complete knowledge of a snapshot of spacetime will
never be enough to fully determine the history of the universe.

From the mathematical point of view, globally hyperbolic
spacetimes have very nice properties.

\begin{prop}\label{globh} Let $(M,g)$ be a globally hyperbolic spacetime. Then
\begin{enumerate}
\item $(M,g)$ is causally simple. In other words, $J^+(p)$ and
$J^-(p)$ are closed for all $p\in M$.\item The sets $J^+(A)\cap
J^-(B)$ are compact for all compact $A,B\subset M$. \item The
Lorentzian distance function $\tau$ is continuous and finite on
$(M,g)$.\item Given $p,q\in M$ with $p\le q$, there is a future
causal geodesic $\gamma$ from $p$ to $q$ that realizes Lorentzian
distance, i.e. $L(\gamma )=\tau (p,q)$.
\end{enumerate}
\end{prop}

\subsection{Achronal sets}

\begin{defi}
A set $A\subset M$ is achronal if no two of its points are
chronologically related. In other words, there is no timelike
curve joining any two of its points. $A$ is said to be acausal if
no two of its points are causally related.
\end{defi}

We could be led to think that spacelike surfaces are achronal, but
this is not the case. Notice though that spacelike surfaces are
\textit{locally} achronal $\cite{budic}$. The following result
will prove useful in later chapters.

\begin{prop}\label{spacesc} $\ $\par
\begin{enumerate}
\item An achronal spacelike hypersurface is acausal.\item
 If $M$ is simply connected then every closed spacelike
hypersurface is achronal (hence acausal).
\end{enumerate}
\end{prop}

\begin{figure}
\begin{center}
\includegraphics{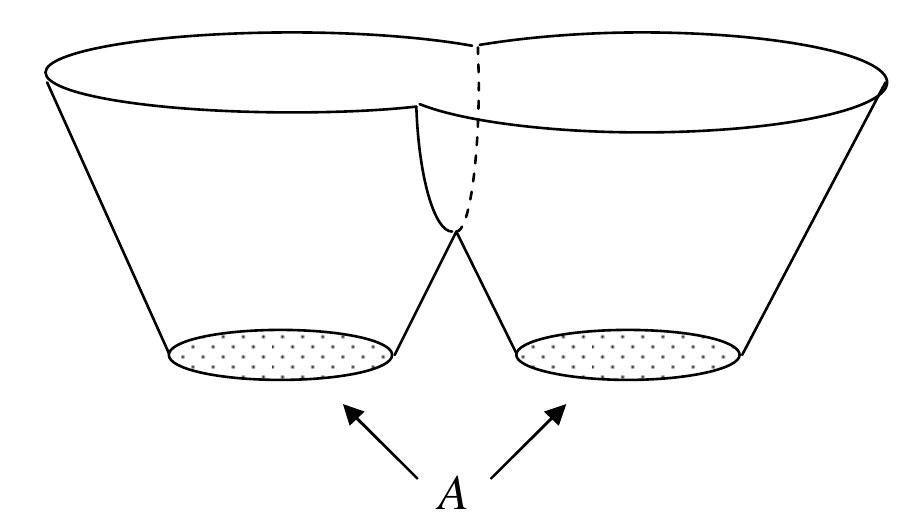}
\caption{\label{abdry}}
\begin{quote}
\begin{quote}
 {\footnotesize Let $A$ be the union of two
closed disks in $\mathbb{M}^3$. Notice $\partial I^+(A)$ is a
topological hypersurface but not a smooth manifold.  Note also
that all the null generators of $\partial I^+(A)$ have past
endpoints in $A$. Finally observe that $\text{edge}(\partial
I^+(A))=\emptyset$. }
\end{quote}
\end{quote}
\end{center}
\end{figure}

Achronal sets arise in many different settings, for instance, sets
of the form $\partial I^+(A)$ are achronal.  Notice also that
achronal subsets usually have ``edges", but if they don't then
they are topological hypersurfaces (see figure $\ref{abdry}$)

\begin{defi}
The edge $\text{edge}(A)$ of an achronal set $A$ consist on all
points $p\in\overline{A}$ such that every neighborhood ${\cal U}$
of $p$ contains a future timelike curve from $I^-(p,{\cal U})$ to
$I^+(p,{\cal U})$ that does not meet $A$.
\end{defi}

\begin{prop}
An achronal set $A\subset M$ is a topological hypersurface if and
only if $A\cap\text{edge}(A)=\emptyset$. Further, $A$ is a closed
topological hypersurface if and only if
$\text{edge}(A)=\emptyset$.
\end{prop}

The sets of the form $\partial I^+(A)$ are called \textit{achronal
boundaries} and play an important role in causal theory. We state
now two of the most important results pertaining achronal
boundaries (refer to results 3.15, 3.17 and 3.20 in $\cite{P}$).

\begin{prop}\label{pensep}
Let $M$ be a spacetime and let $B\neq\emptyset$ be an achronal
boundary. Then there exists a unique future set $F$ and a unique
past set $P$ such that $F$, $P$ and $B$ are pairwise disjoint and
$M=P\cup B\cup F$. Moreover, $B=\partial F=\partial P$.
\end{prop}

\begin{prop}
Let $A\subset M$, then $\partial I^+(A)$ is a topological
hypersurface. Further, $\partial I^+(A)-\overline{A}$ is ruled by
null geodesics. More precisely, if $x\in
\partial I^+(A)-\overline{A}$ then there exists a null geodesic
$\eta\subset \partial I^+(A)$ with future endpoint $x$ that is
either past inextendible or else has a past endpoint in
$\overline{A}$.
\end{prop}

\subsection{Domains of dependence and Cauchy surfaces}\label{dd}

Closely related to global hyperbolicity are the concepts of Cauchy
surface and domain of dependence. We begin by defining the domain
of dependence of an achronal set $A\subset M$.

\begin{defi}
Let $A\subset M$ be an achronal set. We define the future domain
of dependence $D^+(A)$ of $A$ as follows: $p\in D^+(A)$ if every
past inextendible causal curve starting at $p$ intersects $A$. The
past domain of dependence $D^-(A)$ is defined in a time dual
fashion. Finally, we define the domain of dependence $D(A)$ by
$D(A):=D^+(A)\cup D^-(A)$.
\end{defi}

On physical grounds, the future domain of dependence $D^+(A)$ is
the set of points of $M$ which are causally determined by $A$. In
other words, any signal sent to $p\in D^+(A)$ must have passed
through $A$ before reaching $p$, hence information stored on $A$
should suffice to predict what happens at $p$.

As it turns the topological interior of $D(A)$  is globally
hyperbolic.

\begin{teo}
Let $A\subset M$ be a closed achronal set. Then $\mbox{int}(D(A))$
is globally hyperbolic, if non-empty. Moreover, if $A$ is acausal
then $D(A)$ is open, hence globally hyperbolic.
\end{teo}

Similarly, we can also show (see proposition 6.6.6 in
$\cite{HE}$):

\begin{prop}\label{helemma}
Let $S$ be a closed achronal subset of $M$. If $p\in
\mbox{\text{int}}(D(S))$, then $J^+(p)\cap J^-(S)$ is compact.
\end{prop}

Now we turn our attention to Cauchy surfaces.

\begin{defi}
An achronal set $A\subset M$ is called a Cauchy surface if every
inextendible causal curve intersects it.
\end{defi}

As the name suggests, any Cauchy surface is necessarily a closed
topological hypersurface.

It is clear that not all spacetimes admit a Cauchy surface.
Actually, from the discussion above we have that $M$ admits a
Cauchy surface $S$ if and only if $M=D(S)$. Thus if a spacetime
$M$ admits a Cauchy surface, it is globally hyperbolic.

The converse is also true: every globally hyperbolic spacetime $M$
admits a Cauchy surface $S$. In this case, a result due to  R.
Geroch $\cite{geroch}$ establishes that $M$ is homeomorphic to
$\R\times S$. Recently, M. Sanchez and A. Bernal have settled the
long standing question as to if a globally hyperbolic spacetime
$M$ is \textit{diffeomorphic} to $\R\times S$ $\cite{BS}$. We
summarize our discussion in the next theorem:

\begin{teo}
A spacetime $M$ is globally hyperbolic if and only if it admits a
Cauchy surface $S$. Moreover, if $M$ is globally hyperbolic then
$M$ is \textit{diffeomorphic} to a product $\R\times S$, where
$\{t \}\times S$ is a spacelike Cauchy surface and $\nabla t$ is a
past timelike vector field on $M$.
\end{teo}

Notice that a function $f\in{\cal C}^{\infty}(M)$ for which
$\nabla f$ is past timelike is a \textit{time function}, i.e. $f$
is strictly increasing along any future causal curve. Thus, the
above result establishes the existence of a smooth time function
on every globally hyperbolic spacetime.

\section{Smooth null hypersurfaces}\label{snulls}

As discussed in section $\ref{spacetimes1}$, a \textit{null
hypersurface} $S$ in $(M,g)$ is a codimension one embedded
submanifold such that the pullback of the metric is degenerate. In
this section we briefly describe some aspects on the geometry of
null surfaces. For a detailed discussion consult
$\cite{kupeli01}$.

First notice that at any point $p$, there is a distinguished
future null direction in which $g$ degenerates. Thus any null
hypersurface admits a unique, up to a positive scaling factor $f$,
future directed null vector field $K$ such that $K_p^\perp=T_pS$
for all $p\in S$. It can be shown that the integral curves of $K$
are null geodesics, henceforth called the \textit{null generators}
of $S$.

Note also that by proposition $\ref{causalchar}$ above all vectors
in $T_pS$ not collinear to $K_p$ are spacelike. We define a
positive definite metric on a quotient of $T_pS$ as follows: given
$X,Y\in T_pS$ we say $X\sim Y$ if $X-Y=aK$ for some $a\in\R$.
$\sim$ is readily seen to be an equivalence relation on $T_pS$,
thus let us denote by $\overline{X}$ the equivalence class of $X$
and let us define $T_pS/K:= T_pS/\sim$ and
$TS/K:=\bigcup_pT_pS/K$. Observe $TS/K$ is actually independent of
the choice of the null vector field $K$. Finally, let $h$ be
defined by $h(\overline{X},\overline{Y})=g(X,Y)$. A
straightforward computation shows $h$ is a well defined symmetric
and positive definite bilinear form.

On the other hand, we can check the map $b_K\colon  T_pS/K\to
T_pS/K$, $b_K(\overline{X})=\overline{\nabla_XK}$ is well defined
and depends only on the value of $K$ at $p$. Moreover, in
similarity to the Riemannian setting, $b$ is self adjoint with
respect to $h$. Thus we call $b$ the \textit{Weingarten map of $S$
relative to $K$}; its associated bilinear form $B_K\colon
T_pS/K\times T_pS/K\to \R$,
$B_K(\overline{X},\overline{Y})=h(b_K(\overline{X}),\overline{Y})$
is called \textit{the null second fundamental form}.  Since
$b_{fK}=fb_K$, then the condition $B_K\equiv 0$ does not depend on
the choice of $K$, and thus it is an intrinsic property of $S$.

\begin{defi}
We say that $S$ is totally geodesic if $B_K\equiv 0$ for some null
future vector field $K$.
\end{defi}

As expected, if $S$ is totally geodesic then any geodesic in $M$
starting tangent to $S$ remains in $S$.

\begin{defi}
The trace of the Weingarten map $\theta_K =\text{Tr}(b_K)$ with
respect to $h$ is called {the null mean curvature}.
\end{defi}

Note $\theta_{fK}=f\theta_K$, hence null mean curvature
inequalities like $\theta_K \ge 0$ actually do not depend on the
choice of $K$.

Let $\Sigma$ be a codimension one submanifold of $S$ that is
transverse to $K$ near $p\in S$, thus $\Sigma$ is a spacelike
submanifold of dimension $n-2$. Let then $\{e_1,\ldots ,e_{n-2}\}$
be an orthonormal basis of $T_p\Sigma$. It follows that
$\{\overline{e}_1,\ldots ,\overline{e}_{n-2}\}$ is an orthonormal
basis of $T_pS/K$ and
\begin{equation}\label{nullmeanc}
\theta =
\sum_{i=1}^{n-2}h(b(\overline{e}_i,\overline{e}_i))=\sum_{i=1}^{n-2}g(\nabla_{e_i}K,e_i)=\text{div}_\Sigma
K.
\end{equation}
Therefore, we can interpret $\theta$ as the divergence towards the
future of the null generators of $S$.

Now let $\eta$ be  an affinely parameterized null generator of $S$
and let $b(s):=b_{\eta^{\prime}(s)}$. Then the one parameter
family of Weingarten maps $s\mapsto b(s)$ satisfy a Ricatti type
equation
\begin{equation}\label{rica1}
b^{\prime}+b^2+R=0
\end{equation}
where ${\overline{X}}^{\, \prime} :=\overline{X^{\prime}}$ denotes
covariant differentiation along $\eta$ and ${b}^{\prime}$, $R$ are
defined by
\begin{equation}
b^{\prime}(\overline{X}):=b(\overline{X})^{\prime}-b({\overline{X}}^{\,
\prime}),\qquad
R(\overline{X}):=\overline{R(X,\eta^{\prime})\eta^{\prime}}.
\end{equation}

Finally, by taking the trace of equation $\ref{rica1}$ we obtain
the Raychaudhuri equation
\begin{equation}
\theta^{\prime}=-\text{Ric}(\eta^\prime ,\eta^\prime
)-\sigma^2-\frac{1}{n-2}\theta^2
\end{equation}
where $\theta (s)=\theta_{\eta^{\prime}(s)}$, and the
\textit{shear}
\begin{equation}
{\displaystyle \sigma :=B-\frac{\theta}{(n-1)}h} \end{equation} is
the trace free part of $B$.

On the other hand, the trace free part of equation $\ref{rica1}$
gives rise to a propagation equation, where a  null tetrad is
chosen so that $e_0$  agrees with $\eta^{\prime}$. (Compare to
equation 4.36 in $\cite{HE}$).
\begin{equation}\label{propagation}
\sigma_{\alpha\beta}^{\prime}=-W_{\alpha 0\beta
0}-\theta\sigma_{\alpha\beta}-\sigma_{\alpha\gamma}\sigma_{\gamma\beta}+\delta_{\alpha\beta}\sigma^2,
\end{equation}

\section{The Einstein equations}\label{matterintro}

A cornerstone in the theory of general relativity is the idea that
gravity and matter shape the universe we live in. According to
general relativity, our universe is modelled by a spacetime
$(M,g)$. The effects of gravity are accounted by the Lorentz
metric $g$, while the matter content of spacetime is described by
the \textit{energy-momentum tensor} $T$. The interaction between
gravity and matter is given by the Einstein equations:
\begin{equation}
\text{Ric}-\frac{1}{2}Rg+\Lambda g=T.
\end{equation}
where $\Lambda$ is a constant called \textit{the cosmological
constant.}

Notice that the left hand side is purely geometrical, whereas the
right hand side is physical in nature. Thus the Einstein equations
establish a dictionary between the physics and geometry of
spacetime.

To illustrate this correspondence, let us notice that the
\textit{Einstein tensor}
\begin{equation}
G:=\text{Ric}-\frac{1}{2}Rg
\end{equation}
is divergence free; hence by the Einstein equations we have the
relation $\text{div}T\equiv 0$, which in physical terms is
interpreted as the Law of Conservation of Energy.

Conversely, note that some restrictions have to be imposed to $T$
in order for it to describe physically reasonable matter. Among
such restrictions  we find the so called \textit{energy
conditions}. For instance, \textit{the null energy condition}
asserts that the local contribution of energy must be
non-negative, i.e. $T(X,X)_p\ge 0$ for all null vectors $X\in
T_pM$. Thus, via the Einstein equations, the null energy condition
translates to the curvature inequality $\text{Ric}(X,X)\ge 0$ for
all null $X\in T_pM$.

In section $\ref{mattersec}$ we will be dealing primarily with a
particular form of matter model, called \textit{perfect fluid.}
The idea behind this model is that galaxies move like particles of
a homogeneous fluid with no viscosity. In this case the
energy-momentum tensor takes the form
\begin{equation}
T=(\rho +p)U^*\otimes U^*+pg
\end{equation}
where $U$ is a timelike vector field representing the fluid's
flow, and $\rho ,p\in{\cal C}^{\infty}(M)$ represent the density
and pressure of the fluid, respectively.

As pointed out in $\cite{SW}$, a large class of models (called
quasi-gases) that include perfect fluids have non positive trace.
In the latter case we have $\text{Tr}(T)=(n-1)p-\rho$, hence the
pressure is bounded by the density: $p\in [0,\rho /(n-1)]$. The
case $p=0$ is called \textit{dust} and it represents incoherent
matter. On the other hand, the case $p=\rho /(n-1)$ models the
effects of \textit{pure radiation} and is particulary useful in
the study of the early universe arising from a big-bang.

A universe devoid of matter is called \textit{vacuum}. Such a
spacetime satisfies the Einstein equations with $T\equiv 0$.
Notice that in this case by contracting both sides of the Einstein
equations we find $R=2n\Lambda /(n-2)$. Thus $\text{Ric}=\lambda
g$, with $\lambda =2\Lambda /(n-2)$. In other words, if $(M,g)$
satisfies the Einstein vacuum equations then it is an Einstein
manifold.


\chapter{Asymptotically de Sitter Spacetimes}

A Lorentzian space form is a geodesically complete and connected
Lorentzian manifold of constant sectional curvature $C$. A
classical result in semi-Riemannian  geometry states that two
simply connected Lorentzian space forms of the same dimension
having the same constant curvature $C$ must be isometric
$\cite{ON,Wolf}$.

De Sitter space $dS^n$ is the simply connected $n$-dimensional
space form of constant curvature $C\equiv 1$. In this chapter we
will analyze some global properties of spacetimes that look like
de Sitter space ``close to infinity". To make this notion precise,
we first introduce the concept of conformal infinity $\cite{Pen}$.


\section{Conformal boundaries}\label{confbdry}


As a means to motivate the definition of conformal infinity, we
start by looking at a Riemannian example: the hyperbolic half
space $H_n$. Recall $H_n$ is just the manifold
 \begin{equation} H_n=\{(x_1,\ldots
x_n)\in{\R}^n\mid x_n>0\}
\end{equation}
endowed with the metric
\begin{equation}
ds^2=\frac{dx_1^2+dx_2^2+\ldots+dx_n^2}{x_n^2},
\end{equation}
hence $ds^2$ is conformal to the standard Euclidean metric
\begin{equation}
\widetilde{ds^2}=dx_1^2+dx_2^2+\ldots+dx_n^2. \end{equation}
Moreover, notice that $H_n$ has a topological boundary
$\scri:=\partial H_n$ and that it is defined in terms of the
conformal factor $x_n$, namely
\begin{equation}
\scri =\{ (x_1,\ldots x_n)\in{\R}^n\mid x_n=0\}.
\end{equation}

As it can be checked, the inextendible geodesics in $H_n$ are
either circles orthogonal to $\scri$ or straight vertical lines.
Further, all these geodesics are complete. Hence $\scri$ is
actually ``at infinity" since it is made of the endpoints of
complete rays.

Now we introduce the concept of conformal infinity as developed by
Penrose $\cite{Pen}$:


\begin{defi}\label{conformalbdry}
We say that a spacetime $(\tilde{M},\tilde{g})$ admits a conformal
boundary $\scri$ if there exists a spacetime with non-empty
boundary $(M,g)$ such that
\begin{enumerate}
\item $\tilde{M}$ is the interior of $M$ and $\scri =\partial M$,
thus $M=\tilde{M}\cup\scri$. \item There exists $\Omega\in{\cal
C}^\infty({M})$ such that
\begin{enumerate}
\item ${g}=\Omega^2\tilde{g}$ on $\tilde{M}$, \item $\Omega >0$ on
$\tilde{M}$, \item $\Omega =0$ and $d\Omega \neq 0$ on $\scri$.
\end{enumerate}
\end{enumerate}
In this setting $g$ is referred to as the unphysical metric,
$\scri$ is called the conformal boundary of $\tilde{M}$ in $M$ and
$\Omega$ its defining function.
\end{defi}


Assume that the future inextendible null geodesic
${\gamma}_1\colon [0,a_1)\to \tilde{M}$ has a future endpoint
$p\in\scri$. Let $\gamma_2\colon [0,a_2)\to M$ the future
inextendible null geodesic in $(M,g)$ with the same initial
conditions as $\gamma_1$. By the conformal invariance of null
geodesics $\gamma_1$ and $\gamma_2$ coincide on $\tilde{M}$, hence
there exists $b\in (0,a_2)$ such that $\gamma_2(b)=p$. Since the
affine parameters $\lambda_i$ of $\gamma_i$ are related by
\begin{equation}
\frac{d\lambda_1}{d\lambda_2}=\frac{C}{\Omega^2}
\end{equation}
for some constant $C$ $\cite{Wald}$, we have $\lambda_1\to\infty$
as $\lambda_2\to b$. Thus $a_1=\infty$, so then $p$ is ``at
infinity" and hence the use of the term \textit{conformal
infinity} when referring to $\scri$ is justified.

Now we turn back our attention to de Sitter space and show it
admits a \textit{spacelike} conformal boundary. First recall that
$dS^n$ can be realized as the hyperboloid
\begin{equation}\label{hyperboloid}
-x_0^2+x_1^2+\ldots+x_n^2=1
\end{equation}
embedded in $\mathbb{M}^{n+1}$. By letting $\sinh t=x_0$ the
metric of $dS^n$ takes the form
\begin{equation} ds^2=-dt^2+\cosh^2t\,
d\omega^2
\end{equation}
where $d\omega^2$ is the standard metric in $S^{n-1}$. By further
setting $\tan (u/2+\pi /4) = e^t$ we get
\begin{equation}
ds^2 =\frac{1}{{\cos}^2u}(-du^2+d\omega^2).
\end{equation}

From here it is clear that $dS^n$ admits a \textit{spacelike}
conformal boundary
\begin{equation}
\scri=\{u=\pm \pi/2\}
\end{equation}
 in the Einstein static universe
$(E^n,g)=(\R\times S^{n-1},-du^2+d\omega^2)$

\begin{figure}
\begin{center}
\includegraphics{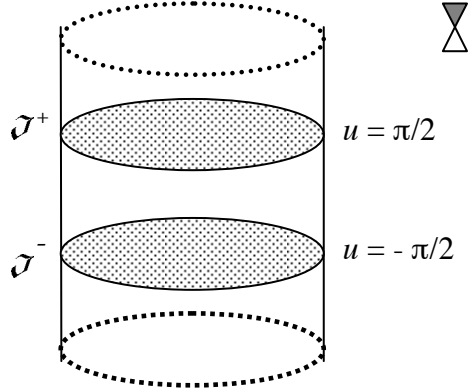}
\caption{\label{fig1} {\footnotesize De Sitter space embedded in
$E^n$.}}
\end{center}
\end{figure}

Hence the following definition formalizes the intuitive idea of a
spacetime ``having a structure at infinity similar to de Sitter
space".
\begin{defi}
A spacetime $(\tilde{M},\tilde{g})$ with conformal boundary
$\scri$ is said to be asymptotically de Sitter if $\scri$ is
spacelike.
\end{defi}

We present now the Schwarzchild de Sitter spacetime $SdS^n$ as our
first nontrivial example of an asymptotically de Sitter spacetime.
\begin{figure}
\begin{center}
\includegraphics{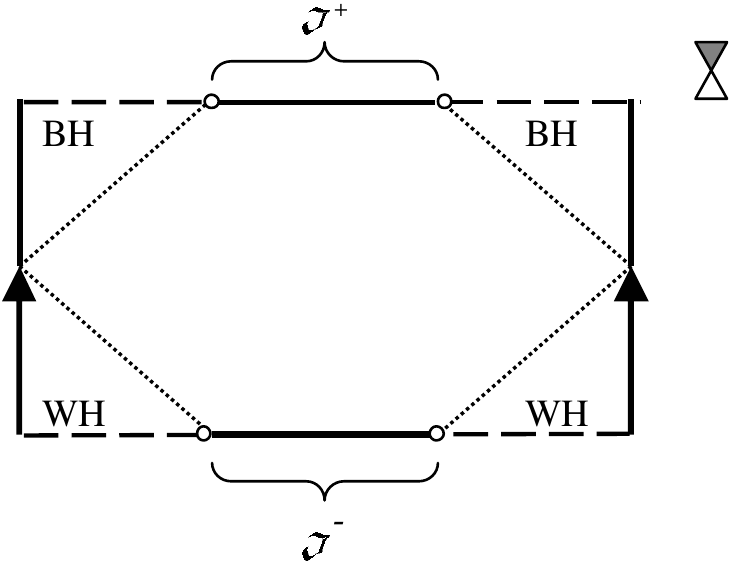}
\caption{\label{fig2} {\footnotesize Penrose diagram of
Schwarzschild de Sitter space.}}
\end{center}
\end{figure}

Physically, this spacetime represents a black hole sitting in a de
Sitter background. Mathematically, it is a particular case of the
so called Kottler metrics $\cite{kot}$. The line element is given
by
\begin{equation}
ds^2=-\left( 1-\frac{2m}{r^{n-3}}-\frac{\Lambda}{3}r^2\right)
dt^2+\left(1-\frac{2m}{r^{n-3}}-\frac{\Lambda}{3}r^2\right)^{-1}dr^2+r^2d\omega^2
\end{equation}
where $m>0$ is a constant and $d\omega^2$ is the standard metric
on $S^{n-1}$.

Another source of examples are the Robertson Walker spacetimes.
They are used in physics to model an isotropic and spatially
homogeneous universe, hence are of key importance in the study of
cosmology.


\begin{defi}\label{RWstime}
The Robertson Walker spacetime $RW(a,k)$ is the warped product
$\R\times_aN_k$ where $a(t)$ is a smooth non-vanishing function on
$\R$ and $N_k$ is a Riemannian space form of constant curvature
$k$.
\end{defi}

Assume $(\tilde{M},\tilde{g})=RW(a,0)$ satisfies the Einstein
equations with $\Lambda >0$. By imposing conditions on the
energy-momentum tensor (for example, when $T$ represents vacuum,
dust or radiation) we get an asymptotic behavior of $a(t)$ that
forces $\tilde{M}$ to be asymptotically de Sitter. This can be
done explicitly in particular examples by making the coordinate
change
\begin{equation}
u=\int\frac{dt}{a(t)}.
\end{equation}

To illustrate this process consider
$(\tilde{M},\tilde{g})=RW(e^t,0)$ and notice that under the
coordinate change $u=-e^{-t}$ the Robertson Walker metric
$ds^2=-dt^2+e^{2t}d\sigma^2$ is transformed to
\begin{equation}
ds^2=\frac{1}{u^2}(-du^2+d\sigma^2)
\end{equation}
which is clearly asymptotically de Sitter.

Because of the spacelike character of $\scri$ in an asymptotically
de Sitter spacetime, every timelike vector $v\in T_pM$ with
$p\in\scri$ is transversal to $\scri$. Thus, since
$\nabla\Omega\vert_{\scri}$ is outward pointing and timelike,
$\scri$ can be decomposed as the union of the disjoint sets
\begin{eqnarray}
\scri^+&=& \{p\in\scri\mid\nabla\Omega_p\ \textrm{is future
pointing}\}\\
\scri^-&=& \{p\in\scri\mid\nabla\Omega_p\ \textrm{is past
pointing}\}\nonumber
\end{eqnarray}
and as a consequence, $\scri^+\subset I^+(\tilde{M},M)$,
$\scri^-\subset I^-(\tilde{M},M)$. It follows as well that both
sets $\scri^+$, $\scri^-$ are acausal in $M$.

Notice though that one of the sets $\scri^\pm$ might be empty. We
say that $(\tilde{M},\tilde{g})$ is a \textit{past asymptotically
de Sitter} spacetime  if $\scri^-\neq\emptyset$. \textit{Future
asymptotically de Sitter} spaces are defined similarly. For
instance $RW(e^t,0)$ is future but not past asymptotically de
Sitter.

A very special feature of de Sitter space is that all inextendible
null geodesics in $dS^n$ have endpoints in $\scri$ and hence are
complete. This observation in turn motivates the following
definition:

\begin{defi}
A spacetime $(\tilde{M},\tilde{g})$ admitting a conformal boundary
$\scri$ is asymptotically simple if every inextendible null
geodesic has endpoints on $\scri$.
\end{defi}


As a counterexample, we can easily see from the Penrose diagram
that $SdS^n$ is not asymptotically simple, since there are future
inextendible null geodesics entering the black hole region and
hence never intersecting $\scri^+$.

Asymptotic simplicity is a strong global condition. As it was
pointed out before, asymptotically simple spacetimes are null
geodesically complete. Moreover, it also follows that
asymptotically simple and de Sitter spacetimes satisfy
$\scri^-\neq\emptyset\neq\scri^+$.

It was shown in $\cite{larsgal}$ that every asymptotically simple
spacetime $(\tilde{M},\tilde{g})$ is globally hyperbolic as well.
We include the statement of this result for further reference.

\begin{teo}
Let $(\tilde{M},\tilde{g})$ be an asymptotically simple and de
Sitter spacetime. Then $(\tilde{M},\tilde{g})$ is globally
hyperbolic.
\end{teo}

\section{Rigidity of de Sitter space}

Some of the most important results in semi-Riemannian geometry
require a priori bounds on geometrical quantities. However, it may
happen that one of these results involving a strict bound fails to
be true \textit{only under very special circumstances} when one
relaxes the strict bound condition to a weak inequality. Such
results are known as \textit{rigidity theorems}. As an example of
the rigidity philosophy we have the Cheeger-Gromoll Splitting
Theorem in Riemannian geometry:

\eject

\begin{teo}
Let $(N,h)$ be a complete Riemannian manifold  with
$\textrm{Ric}(v,v)\ge 0$ for all $v\in TM$ and which contains a
line (that is, a complete geodesic that minimizes distance between
any of its points). Then $(M,g)$ can be written uniquely as a
metric product $N^{\prime}\times{\R}^k$ where $N^{\prime}$
contains no lines and ${\R}^k$ is given the standard flat metric.
\end{teo}

Here, rigidity comes in the following way: it has been established
$\cite{gro}$ that a complete Riemannian manifold $M$ satisfying
$\textrm{Ric}(v,v)>0$, $\forall v\in TM$ is connected at infinity;
but if one relaxes this condition to $\textrm{Ric}(v,v)\ge 0$ and
assumes that $M$ is disconnected at infinity, then one is able to
prove the existence of a line in $M$ and hence the Splitting
Theorem asserts that $M$ is isometric to a product. In a nutshell,
connectedness at infinity will fail when the curvature bound is
relaxed \textit{only} in the very special case of a product
$N^{\prime}\times{\R}^k$ having all the properties stated at the
conclusion of the Splitting Theorem.

In the Lorentzian setting, a number of rigidity results have been
proven in recent times. For instance, the Lorentzian Splitting
Theorem proven in the 1990's $\cite{esch,galls,newman}$ is the
Lorentzian analog of the Cheeger-Gromoll theorem. More recently,
G. Galloway proved the following Null Splitting Theorem
$\cite{gal01}$.

\begin{defi}
A null line  is an achronal inextendible null geodesic; or
equivalently, an inextendible null geodesic that maximizes
Lorentzian distance between any of its endpoints.
\end{defi}

\begin{teo}
Let $(M,g)$ be a null geodesically complete spacetime which obeys
the null energy condition. If $M$ admits a null line $\eta$, then
$\eta$ is contained in a smooth properly embedded, achronal and
totally geodesic null hypersurface $S$.
\end{teo}

\begin{remark}
Let $\partial_0I^{\pm}(\eta )$ be the connected components of
$\partial I^{\pm}(\eta)$ containing $\eta$. The proof of the Null
Splitting Theorem shows $\partial_0I^+(\eta )=S=\partial_0
I^-(\eta )$. Moreover, the proof also shows that future  null
completeness of $\partial_0I^-(\eta )$ and past null completeness
of $\partial_0I^+(\eta )$ are sufficient for the result to hold.
\end{remark}

This result is a consequence of the Maximum Principle for $C^0$
Null Hypersurfaces $\cite{gal01}$. We will be needing later on a
weaker version of this result, namely the Maximum Principle for
Smooth Null Hypersurfaces.


\begin{teo}\label{MP}
Let $S_1$ and $S_2$ be smooth null hypersurfaces in a spacetime
$(M,g)$. Suppose that $S_1$ and $S_2$ meet in $p$ and

 1. $S_2$ lies to the future of $S_1$ near $p$.

 2. The null mean curvature scalars $\theta_i$ satisfy $\theta_2\le
0\le \theta_1$.

\noindent Then $S_1$ and $S_2$ coincide near $p$ and this common
hypersurface has null mean curvature $\theta =0$.
\end{teo}

As an application of the Null Splitting Theorem, the following
rigidity result for asymptotically de Sitter spacetimes has been
established $\cite{gal02}$.

\begin{teo}\label{rigds}
Let $(\tilde{M},\tilde{g})$ be an asymptotically simple and de
Sitter spacetime of dimension $n=4$ that satisfies the vacuum
Einstein equations with positive cosmological constant. If
$\tilde{M}$ contains a null line, then $\tilde{M}$ is isometric to
de Sitter space $dS^4$.
\end{teo}

This theorem can be interpreted in terms of the initial value
problem in the following way: H. Friedrich has shown that the set
of asymptotically simple solutions to the Einstein equations with
positive cosmological constant is open in the set of all maximal
globally hyperbolic solutions with compact spatial sections
$\cite{F2}$. As a consequence, by slightly perturbing the initial
data on a fixed Cauchy surface of $dS^4$ we should get an
asymptotically simple solution of the Einstein equations different
from $dS^4$. Thus in virtue of theorem $\ref{rigds}$, such a
spacetime \textit{has no null lines}. In other words, a small
perturbation of the initial data destroys \textit{all} null lines.

Alternatively, we could say that no other asymptotically simple
solution of the Einstein equations besides $dS^4$ develops
\textit{eternal observer horizons}. By definition, an observer
horizon ${\cal A}$ is the past achronal boundary $\partial
I^-(\gamma )$ of a future inextendible timelike curve, thus ${\cal
A}$ is ruled by future inextendible null geodesics. In the case of
de Sitter space, observer horizons are eternal, that is, all null
generators of ${\cal A}$ extend from $\scri^+$ all the way back to
$\scri^-$ and hence are null lines.

Since the observer horizon is the boundary of the region of
spacetime that can be observed by $\gamma$, the question arises as
to whether at one point $\gamma$ would be able to observe the full
space. More precisely, we want to know if there exists $q\in
\tilde{M}$ such that  $I^-(q)$ would contain a Cauchy surface of
spacetime. S. Gao and R. Wald were able to answer this question
affirmatively for globally hyperbolic spacetimes with compact
Cauchy surfaces, assuming null geodesic completeness, the null
energy condition and the null generic
condition\symbolfootnote[2]{The null generic condition is the
statement that each null geodesic contains a point at which
$k_{[\alpha}R_{\beta]\gamma\delta [\eta}k_{\zeta]}k_\gamma
k_\delta\neq 0$ where $k_\alpha$ denotes the tangent to the
geodesic.)} $\cite{GW}$. Refer also to $\cite{bou}$ for a
discussion on the relationship between the existence of eternal
observer horizons and entropy bounds on asymptotically de Sitter
spacetimes.

Though no set of the form $I^-(q)\subset dS^n$ contains a Cauchy
surface, $I^-(q)$ gets arbitrarily close to do so as
$q\to\scri^+$. However, notice that de Sitter space is not a
counterexample to $\cite{GW}$, since $dS^n$ does not satisfy the
null generic condition. Actually, the latter remark enables us to
interpret theorem $\ref{rigds}$ as a rigidity result in the
asymptotically simple context: by dropping the null generic
hypothesis in $\cite{GW}$ the conclusion will only fail if
$(\tilde{M},\tilde{g})$ is isometric to $dS^4$.

Notice also that the stability of the null generic condition under
small perturbations of the initial data has not yet been
established. In this sense, the null generic condition is not
``generic".

Since asymptotically simple spacetimes are globally hyperbolic
$\cite{larsgal}$, the above discussion prompts us to study  the
structure of globally hyperbolic and asymptotic de Sitter
spacetimes to later investigate the consequences of relaxing the
hypotheses of theorem $\ref{rigds}$.


\section{Globally hyperbolic and asymptotically de Sitter
spacetimes}\label{ghdss}

\begin{figure}
\begin{center}
\includegraphics{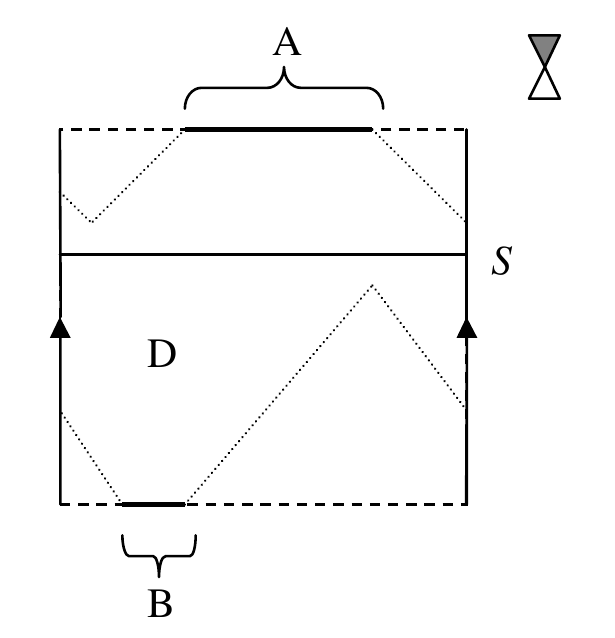}
\caption{\label{fig3}}
\begin{quote}
\begin{quote}
{\footnotesize $D=I^+(B)\cap I^-(A)$ is an asymptotically de
Sitter and globally hyperbolic open subset of $dS^n$ with Cauchy
surface $S$.}
\end{quote}
\end{quote}
\end{center}
\end{figure}

Notice that surjectivity doesn't hold  anymore when  the
hypothesis of asymptotic simplicity in theorem $\ref{rigds}$ is
weakened to global hyperbolicity. Actually, there are globally
hyperbolic and asymptotically de Sitter proper subsets of $dS^n$
admitting null lines with endpoints in $\scri$. (See fig.
$\ref{fig3}$ below).

The main goal of this section is to prove that surjectivity is the
\textit{only} conclusion of theorem $\ref{rigds}$ that does not
carry over to the more general globally hyperbolic setting. In
turn, by relaxing the hypothesis of asymptotical simplicity in
theorem $\ref{rigds}$ we broaden its scope of application, since
we will not be ruling out a priori the appearance of black holes.

To make proofs easier, it is convenient to embed the spacetime
with boundary $(M,g)$ in an open spacetime. This can always be
done in light of theorem \ref{openex}. In the case of globally
hyperbolic and asymptotically de Sitter spacetimes, this latter
result can be improved, as the next proposition shows.

\begin{prop}\label{ghex}
Let $(\tilde{M},\tilde{g})$ be a globally hyperbolic and
asymptotically de Sitter spacetime, then $(\tilde{M},\tilde{g})$
can be embedded in a globally hyperbolic spacetime $(N,h)$ such
that $\scri$ topologically separates $\tilde{M}$ and
$N-\tilde{M}$.
\end{prop}

\eject

\noindent\textit{Proof:} It suffices to show
$(\tilde{M},\tilde{g})$ can be extended past $\scri^-$, since a
similar procedure can be used to extend $(\tilde{M},\tilde{g})$
beyond $\scri^+$. Thus without loss of generality we can assume
$\scri=\scri^-$.

By proposition \ref{openex} there is an open spacetime
$(\overline{M},h)$ extending $(M,g)$. Since $\overline{M}$ is
obtained from $M$ by attaching collars, the separation part of the
proposition holds.

As a consequence, $\scri$ is acausal in $\tilde{M}$, then the
Cauchy development $D(\scri , \overline{M})$ is an open subset of
$\overline{M}$. Hence $N=M\cup D(\scri , \overline{M})$ is an open
spacetime containing $M$.

We proceed to show $(N,h)$ is globally hyperbolic. To this end,
let us note that $D:=D(\scri , N)$ is globally hyperbolic by
construction.

First we show $N$ is strongly causal.  Thus let $p\in N$. If $p\in
\tilde{M}$, we claim that any causally convex neighborhood ${\cal
V}$ of $p$ with respect to $\tilde{M}$ is causally convex with
respect to $N$ as well. In order to  prove this claim, let us
consider a future directed causal curve $\gamma$ with endpoints in
${\cal V}$. If $\gamma\subset \tilde{M}$ then $\gamma\subset{\cal
V}$ by the causal convexity of ${\cal V}$. On the other hand, if
$\gamma$ leaves $\tilde{M}$ then it has to intersect $\scri$  at
least twice to be able to come back to ${\cal V}$, thus violating
the acausality of $\scri$. A similar argument can be used in the
case $p\in N-M$ to show that any causally convex neighborhood of
$p$ with respect to $D^-(\scri, N)$ is causally convex with
respect to $N$.

Thus $p\in\scri$ is the only case that remains to be checked. We
proceed by contradiction. Assume strong causality fails at $p$,
then by proposition $\ref{penlemma}$ there exists $q\in J^-(p,N)$,
$q\neq p$ with the property that $y\in I^+(x,N)$ for all $x\in
I^-(p,N)$, $y\in I^+(q,N)$. Thus, $q\in D^-(\scri ,N)-\scri$ by
the acausality and separating properties of $\scri$. Now, since
strong causality holds on $D$, it follows that we can choose $q$
in such a way that $p\not\in J^+(q,D)$. Now let $\{p_n\}$ be a
sequence in $I^-(p,N)$ converging to $p$ and $\{q_n\}\subset
I^+(q,N)$ with $q_n\to q$. Notice $q_n\in D^-(\scri ,N)-\scri$ for
large $n$ by the separating properties of $\scri$. Since $p_n\in
D^-(\scri ,N)-\scri$ for all $n\in\N$, then we can assume without
loss of generality that both sequences are in $D^-(\scri
,N)-\scri$. Now let $\gamma_n$ be an inextendible future timelike
curve emanating from $p_n$ and passing through $q_n$. By
definition, $\gamma_n$ must intersect $\scri$ at a point $z_n$.
Moreover, notice that $z_n$ has to lie at the future of $q_n$
along $\gamma_n$, since acausality of $\scri$ would be violated
otherwise. Further, note $J^+(q,D)\cap\scri$ is compact in virtue
of proposition $\ref{helemma}$, being a closed subset of the
compact set $J^+(q,D)\cap J^-(\scri ,D)$. Since $\{z_n\}\subset
J^+(q,D)\cap\scri$ then there is a subsequence $\{z_m\}$
converging to $z\in J^+(q,D)\cap\scri$. Finally notice that
$\gamma_m\vert_{p_m}^{z_m}\subset D^-(\scri ,N)\subset D$. Then
since $D$ is globally hyperbolic a strong version of the Limit
Curve Lemma applies (see corollary $3.32$ in $\cite{BE}$) to
guarantee the existence of a future causal curve from $p$ to $z$,
a clear contradiction to the acausality of $\scri$. Thus strong
causality holds in all of $N$.

\enlargethispage{.5cm}

Let $p,q\in N$, then we want to show $A:=J^+(p,N)\cap J^-(q,N)$ is
compact. If $p\in\tilde{M}$ then by observations above
$J^+(p,N)=J^+(p)\subset \tilde{M}$,\symbolfootnote[2]{In the
remaining sections of this chapter all causal relations are taken
with respect to the physical spacetime $(\tilde{M},\tilde{g})$
unless otherwise specified. For instance, $J^+(A)=J^+(A,
\tilde{M})$, $\forall A\subset \tilde{M}$.
 } hence as an easy
consequence we have  $A=J^+(p)\cap J^-(q)$, thus $A$ is compact
due to the global hyperbolicity of $\tilde{M}$. Likewise, if $q\in
D$ then $J^-(q,N)\subset D$ and hence $A=J^+(p,D)\cap J^-(q,D)$.
So compactness of $A$ follows as well since $D$ is globally
hyperbolic.

Therefore, the only case left to consider is when $p\in
N-\tilde{M}$ and $q\in N-D$. Take now a Cauchy surface $S$ of $D$
slightly to the future of $\scri$ so that $S\subset\tilde{M}$. Let
$S^{\prime}:=J^+(p,N)\cap J^-(q,N)\cap S$ and let us prove that
\begin{equation}\label{001}
A=[J^+(p,D)\cap J^-(S^{\prime},D)]\cup [J^-(q)\cap
J^+(S^{\prime})].
\end{equation}

First notice that one inclusion is trivial. On the other hand, let
$x\in A$ and consider a future causal curve $\gamma$ from $p$ to
$q$ passing through $x$. Since $p\in N-\tilde{M}$, $q\in N-D$ and
$S$ is a Cauchy surface of $D$, $\gamma$ must intersect $S$ in a
point $y\in S^{\prime}$. If $y$ comes before $x$ along $\gamma$
then $\gamma\vert_y^q\subset\tilde{M}$, thus $x\in J^-(q)\cap
J^+(S^{\prime})$. Alternatively, if $y$ comes after $x$ along
$\gamma$ then $x\in D$, since $J^-(S,N)=J^-(S,D)$. But then
$J^+(p,N)\cap D=J^+(p,D)$ implies $x\in J^+(p,D)\cap
J^-(S^{\prime},D)$, thus proving equation $(\ref{001})$.

To finish the proof, notice that $J^+(p,N)\cap S=J^+(p,D)\cap S $,
hence $J^+(p,N)\cap S$ is compact. Also note that $J^-(q,N)\cap
S=J^-(q)\cap S$, so then $J^-(q,N)\cap S$ is a closed set as well.
Hence $S^{\prime}$ is compact. Thus by propositions $\ref{globh}$
and $\ref{spacesc}$ we have that the sets $J^+(p,D)\cap
J^-(S^{\prime},D)$ and $J^-(q)\cap J^+(S^{\prime})$ are compact,
and the result follows. \hfill$\Box\quad$\par

We analyze now the structure of a globally hyperbolic and
asymptotically de Sitter spacetime near the past endpoint of a
causal curve.

\begin{lemma}\label{structurelemma0}
Let $(\tilde{M},\tilde{g})$ be a globally hyperbolic and
asymptotically de Sitter spacetime and $\eta$ a future directed
causal curve in $M$. Further assume $p\in\scri^-$ is the past
endpoint of $\eta$. Then
\begin{enumerate}\item $ { \partial I^+(\eta )=J^+(p,N)-(I^+(p,N)\cup\{ p\})}$,
\item $ J^+(N_p,N)\cap\tilde{M}\subset D^+(N_p,N)\cap\tilde{M}$,
\end{enumerate}
\noindent where $N_p:=\partial_NI^+(p,N)$ and $N$  is a globally
hyperbolic extension of $(M,g)$.
\end{lemma}

\noindent\textit{Proof:} First, let us consider a globally
hyperbolic extension $(N,h)$ of $(M,g)$ as described in
proposition $\ref{ghex}$. Then by global hyperbolicity the set
$J^+(p,N)$ is closed in $N$, and as a consequence
\begin{equation}\label{eqq01}
\partial_NI^+(p,N)=J^+(p,N)-I^+(p,N).
\end{equation}
Thus by the acausality of $\scri^-$ we have
\begin{equation}\label{eqq02}
\tilde{M}\cap\partial_NI^+(p,N)=\partial_NI^+(p,N)-\{p\}
\end{equation}

On the other hand, let us show $I^+(\eta )=I^+(p,N)$. It is clear
that $I^+(\eta )\subset I^+(p,N)$. Conversely, let $x\in
I^+(p,N)$. Since $I^-(x,N)$ is open and $p\in I^-(x,N)$, there is
$y\in\eta\cap I^-(x,N)$. But any future timelike curve from $y$ to
$x$ has to be contained in $\tilde{M}$ due to the separating
properties of $\scri^-$, hence $x\in I^+(\eta )$ and the inclusion
$I^+(p,N)\subset I^+(\eta )$ is proven. As a consequence $\partial
I^+(\eta )=\partial_{\tilde{M}}I^+(p,N)$.

Finally, since $I^+(p,N)$ is an open set in $N$, by a standard
topological result $\cite{munkres}$ we get
\begin{equation}\label{eqq03}
\partial_{\tilde{M}}I^+(p,N)=\tilde{M}\cap\partial_NI^+(p,N).
\end{equation}
Then the first assertion follows by combining equations
$(\ref{eqq01})$, $(\ref{eqq02})$ and $(\ref{eqq03})$.

To prove the second part of the lemma we proceed by contradiction.
Thus let us assume $x\in
J^+(N_p,N)\cap\tilde{M}-D^+(N_p,N)\cap\tilde{M}$, hence $x\in
J^+(N_p,N)-N_p$. It follows $x\in I^+(p,N)$. On the other hand,
since $x\not\in D^+(N_p,N)\cap\tilde{M}$ there is a past
inextendible causal curve $\gamma \colon [0,a)\to N$ starting at
$x$ that does not intersect $N_p$.  Notice $\gamma$ never leaves
$I^+(p,N)$, since otherwise it had to intersect
$N_p=\partial_NI^+(p,N)$. Thus $\gamma$ is contained in the
compact set $J^+(p,N)\cap J^-(x,N)$, which by proposition
$\ref{impri}$ contradicts strong causality. $\Box$
\par

In a time dual manner, we get $\partial I^-(\eta
)=J^-(q,N)-(I^-(q,N)\cup\{ q\})$ where $q\in\scri^+$ is a future
endpoint of $\eta$ in a future asymptotically de Sitter spacetime.

Now notice that in $dS^n$ the set $N_p$ is just the local causal
cone at $p$. This generalizes to globally hyperbolic and
asymptotically de Sitter spacetimes satisfying the null energy
condition.

\begin{lemma}\label{structurelemma1}
Let $(\tilde{M},\tilde{g})$ be a globally hyperbolic and
asymptotically de Sitter spacetime and let $\eta$ be a future
directed null line in $\tilde{M}$ having endpoints $p\in\scri^-$
and $q\in\scri^+$. Further assume $(\tilde{M},\tilde{g})$
satisfies the null energy condition. Then $\partial I^+(\eta )$ is
the diffeomorphic image under the exponential map $\exp_p$ of the
set $\Lambda_p^+\cap {\cal O}$ where $\Lambda_p^+\subset T_pM$ is
the future null cone based at $0_p$ and ${\cal O}$ is the biggest
open set on which $\exp_p$ is defined.
\end{lemma}

\noindent\textit{Proof:} Let $(N,h)$ be as in the previous lemma.
Hence by proposition $\ref{ngeo1}$ and lemma
$\ref{structurelemma0}$, any point in
$\tilde{M}\cap\partial_NI^+(p,N)$ is the future endpoint of a
future null geodesic segment emanating from $p$. Thus
\begin{equation}
\partial I^+(\eta )\subset\exp_p(\Lambda_p^+\cap {\cal O})\cap\tilde{M}.
\end{equation}

Now let $\gamma$ be a null generator of $\partial I^+(\eta )$
passing through $x\in \partial I^+(\eta )$. Let $y\in\gamma$ a
point slightly to the past of $x$ and notice
$y\in\partial_NI^+(p,N)$ by equation $(\ref{eqq02})$. On the other
hand, let $\overline{\gamma}(t)$ be a null geodesic emanating from
$p$ and passing through $y$. Then $\gamma$ coincides with
$\overline{\gamma}\subset\tilde{M}$ since otherwise we would have
two null geodesics meeting at an angle in $y$ and hence $x\in
I^+(p,N)$ by proposition $\ref{2null}$, a clear contradiction.
Thus, $\gamma$ can be extended to $p\in\scri^-$ and thus it is
past complete. In a time dual fashion, the generators of $\partial
I^-(\eta )$ are future complete.

Let $S$ be the component of $\partial I^+(\eta )$ containing
$\eta$. By the proof of the Null Splitting Theorem, $S$ is a
closed smooth totally geodesic null hypersurface in $\tilde{M}$.
As a consequence, the null generators of $S$ do not have future
endpoints in $\tilde{M}$ and hence are future inextendible in $S$.
Furthermore, by the argument in the previous paragraph, each of
these generators is the image under $\exp_p$ of the set
$\mathbf{V}\cap{\cal O}$, where $\mathbf{V}$ is an inextendible
null ray in $\Lambda^+_p$.

Let $\gamma$ be a generator of $S$, then $\gamma\cap
I^+(p,N)=\emptyset$. Thus $\gamma$ is conjugate point free and
does not intersect with any other generator of $S$, since
otherwise $\gamma$ would enter $I^+(p,N)$ (see propositions
$\ref{2null}$ and $\ref{3null}$). As a result we have that $S$ is
the diffeomorphic image under $\exp_p$ of an open subset of
$\Lambda_p^+$.

To check that $S$ encompasses the whole local null cone at $p$
consider a causally convex normal neighborhood ${\cal V}$ of $p$
and a spacelike hypersurface $\Sigma$ slightly to the future of
$\scri^-$. Thus $\Sigma_0:=\Sigma\cap \exp_p(\Lambda_p^+\cap {\cal
V})$ is connected. Moreover, by the way ${\cal V}$ and $\Sigma$
were chosen we have $\Sigma_0\subset J^+(p)-(I^+(p)\cup
\{p\})=\partial I^+(\eta )$. Thus
\begin{equation}
\exp_p(\Lambda_p^+\cap {\cal O})\cap\tilde{M}\subset S
\end{equation}
since every future null geodesic emanating from $p$, including
$\eta$, must intersect $\Sigma_0$. It follows $S=\partial I^+(\eta
)$ and the proof is complete. $\Box$\par


\begin{cor}\label{structurecor}
Let $(\tilde{M},\tilde{g})$ be a globally hyperbolic and
asymptotically de Sitter spacetime satisfying the null energy
condition. Let $\eta$ be a future directed null line in
$\tilde{M}$ having endpoints $p\in\scri^-$ and $q\in\scri^+$ and
let $N_p=\partial_NI^+(p,N)$,  $N_q=\partial_NI^-(q,N)$. Then
$N_p-\{p\}$ and $N_q-\{q\}$ agree and form a smooth totally
geodesic null hypersurface $S$. Furthermore, $S\cup\{p,q\}$ is
homeomorphic to $S^{n-1}$.
\end{cor}

\noindent\textit{Proof:} Let $\partial_0I^+(\eta)$,
$\partial_0I^-(\eta  )$ be the components of $\partial I^+(\eta
)$, $\partial I^-(\eta )$ containing $\eta$ respectively. By the
proof of the previous lemma and the Null Splitting Theorem, we
have $\partial_0I^+(\eta)=\partial_0I^-(\eta  )$, and this common
null hypersurface is closed, smooth and totally geodesic.
Moreover, by the previous lemma we also conclude $S:=\partial
I^+(\eta )$ is connected, thus $S=\partial I^+(\eta)=\partial_0
I^+(\eta )$. Lastly, by lemma $\ref{structurelemma0}$ we have
$\partial I^+(\eta )=N_p-\{p\}$ and $\partial I^-(\eta
)=N_q-\{q\}$. Thus $N_p-\{p\}=S=N_q-\{q\}$ as desired.

On the other hand, notice that $N_p-\{p\}=N_q-\{q\}$ in
conjunction with lemma $\ref{structurelemma1}$ imply that every
point in $S$ is at the same time the future endpoint of a null
geodesic emanating from $p$ and the past endpoint of a null
geodesic from $q$. These geodesic segments must form a single
geodesic, otherwise achronality of $\eta$ would be violated.
Hence, all future null geodesics emanating from $p$ meet again at
$q$. Then $S\cup\{p,q\}$ is homeomorphic to a sphere. $\Box$

\begin{cor}\label{structurecor1}
Let $(\tilde{M},\tilde{g})$ be a globally hyperbolic and
asymptotically de Sitter spacetime satisfying the null energy
condition. If $\tilde{M}$ has a future directed null line $\eta$
with  endpoints $p\in\scri^-$ and $q\in\scri^+$, then $\tilde{M}$
has Cauchy surfaces homeomorphic to $S^{n-1}$. In particular,
$\tilde{M}$ is simply connected.
\end{cor}

\noindent\textit{Proof:} Let ${\cal N}_p$ be a normal neighborhood
of $p$, let $w\in T_pM$ be a unit future timelike vector and
consider $r>0$ small enough so that $S_r=\exp_p(C_r)\subset{\cal
N}_p$, where $C_r$ is the slice given by  $C_r:=\{v\in C_p^+\mid
g(v,w)=-r\}$. Further let $\Lambda_r=\{v\in
\Lambda_p^+\cup\{0_p\}\mid -r\le g(v,w)\le 0\}$, then since $N_p$
is the future local null cone at $p$, by projecting
$\exp_p(\Lambda_r)$ onto $S_r$ we obtain a homeomorphism between
$N_p$ and  an \textit{achronal} set $N_p^{\prime}\subset
\tilde{M}$. Let then $N_q^{\prime}$ be an achronal set
homeomorphic to $N_q$ obtained in a similar fashion.

Thus ${\cal S}^{\prime}:=N_p^{\prime}\cup N_q^{\prime}$ is
homeomorphic to $N_p\cup N_q=S\cup\{p,q\}$. It follows from
corollary $\ref{structurecor}$ that ${\cal S}^{\prime}\approx
S^{n-1}$.

Finally, notice ${\cal S}^{\prime}$ is an embedded compact
hypersurface in $\tilde{M}$. Since it is achronal in $\tilde{M}$,
it has to be a Cauchy surface for $\tilde{M}$ $\cite{budic}$. Thus
$\tilde{M}$ is homeomorphic to ${\cal S}\times \R$, hence simply
connected. $\Box$

We now proceed to prove the main result of this section.


\begin{teo}\label{teo1}
Let $(\tilde{M},\tilde{g})$ be a globally hyperbolic and
asymptotically de Sitter spacetime of dimension $n=4$ satisfying
the vacuum Einstein equations with positive cosmological constant.
If $\tilde{M}$ has a null line with endpoints $p\in\scri^-$,
$q\in\scri^+$ then $(\tilde{M},\tilde{g})$ is isometric to an open
subset of de Sitter space containing a Cauchy surface.
\end{teo}

\noindent\textit{Proof:} Let $(N,h)$ be a globally hyperbolic
extension of $(M,g)$ and let $S=\partial I^+(\eta )$. Further let
$N_p=S\cup\{p\}$. Then by lemma $\ref{structurelemma0}$ we have
${\displaystyle I^+(S)\subset D^+(N_p,N)\cap \tilde{M}}$. In a
time dual fashion ${\displaystyle I^-(S^{\prime})\subset
D^-(N_q,N)\cap \tilde{M}}$ where $S^{\prime}=\partial I^-(\eta )$
and $N_q={ S}^{\prime}\cup\{q\}$.


Since $(\tilde{M},\tilde{g})$ satisfies the null energy condition,
by corollary  $\ref{structurecor}$ we have $S=S^{\prime}$, and as
a consequence of proposition $\ref{pensep}$ we get
$\tilde{M}=I^+(S)\cup S\cup I^-(S)$. Hence
\begin{equation}\label{101}
\tilde{M}\subset D^+(N_p,N)\cup D^-(N_q,N)
\end{equation}

We show now $(\tilde{M},\tilde{g})$ has constant sectional
curvature. First notice that corollary $\ref{structurecor}$
establishes that $S$ is a totally geodesic null hypersurface. As a
consequence the shear tensor $\tilde{\sigma}_{\alpha\beta}$ of $S$
in the physical metric $\tilde{g}$ vanishes. Since the shear
scalar $\tilde{\sigma}=
\tilde{\sigma}_{\alpha\beta}\tilde{\sigma}^{\alpha\beta}$ is a
conformal invariant we have $\sigma_{\alpha\beta}\equiv 0$ as
well. Recall the shear satisfies the propagation equation
$(\ref{propagation})$
\begin{equation}
\sigma_{\alpha\beta}^{\prime}=-W_{\alpha 0\beta
0}-\theta\sigma_{\alpha\beta}-\sigma_{\alpha\gamma}\sigma_{\gamma\beta}+\delta_{\alpha\beta}\sigma^2,
\end{equation}
 Then it follows $W_{\alpha 0\beta 0}=0$ on $S$.

In $\cite{F}$ H. Friedrich used the conformal field equations
\begin{equation}
\nabla_{\alpha}{d^{\alpha}}_{\beta\gamma\zeta}=0,\qquad
{d^{\alpha}}_{\beta\gamma\zeta}=\Omega^{-1}{W^{\alpha}}_{\beta\gamma\zeta}
\end{equation}
along with a recursive ODE argument to guarantee the vanishing of
the rescaled conformal tensor $d$ on $D^+(S\cup\{p\},N)$ given
that $W_{0000}$ vanishes on $S$.

Hence, we have shown $d\equiv 0$ on $D^+(N_p,N)$. Since
$\widetilde{W}=W$ on $\tilde{M}$ we have $\widetilde{W}\equiv 0$
on $D^+(N_p,N)\cap\tilde{M}$. By a time dual argument we conclude
$\widetilde{W}\equiv 0$ on $D^-(N_q,N)\cap\tilde{M}$, thus
$\widetilde{W}\equiv 0$ on $\tilde{M}$ by $(\ref{101})$ above.
Since $(\tilde{M},\tilde{g})$ is an Einstein manifold, the
vanishing of the Weyl tensor implies that $(\tilde{M},\tilde{g})$
has constant curvature $C>0$ (see remark $\ref{weyl}$). Thus,
without loss of generality we can assume $C=1$.

By corollary $\ref{structurecor1}$ $(\tilde{M},\tilde{g})$ is
simply connected, hence there exists a local isometry
$\Phi\colon\tilde{M}\to dS^4$ by the Cartan-Ambrose-Hicks Theorem
$\cite{CE,Wolf}$. Notice that in particular, $\Phi$ is an open
map.

Let us denote by ${{\cal S}}$ a fixed Cauchy surface of
$\tilde{M}$. In virtue of $\cite{BS}$ we can assume ${\cal S}$ is
spacelike and $\tilde{M}={\cal S}\times\R$. We proceed to show
that $\Phi_{\cal S}:=\Phi\vert_{{\cal S}}$ is a topological
embedding.

To fix some notation, let us consider the conformal embedding of
$dS^4$ into the Einstein static universe, as described in section
$\ref{confbdry}$. Let $\mathfrak{ S}$ be the Cauchy surface of
$dS^4$ given by $\mathfrak{S}=u^{-1}(0)$, hence $dS^n\approx
\mathfrak{ S}\times \R$. Let $\pi\colon dS^4\to \mathfrak{ S}$ be
the projection on the first factor. Note that the fibers of $\pi$
are all timelike curves. Further, let $\hat{\cal S}:=\Phi ({\cal
S})$.

We first show $\pi\vert_{\hat{\cal S}}$ is a local homeomorphism.
Since manifolds are locally compact Hausdorff spaces, it suffices
to show $\pi$ is locally one to one.

\enlargethispage{.5cm}

Thus let $y\in\hat{\cal S}$. Take then $x\in {\cal S}$ with $\Phi
(x)=y$ and consider a neighborhood ${\cal V}$ of $x$ such that
$\Phi\colon{\cal V}\to{\cal U}_0$ is an isometry. Further, since
$dS^4$ is globally hyperbolic there is a causally convex
neighborhood ${\cal U}$ of $y$ contained in ${\cal U}_0$. Let then
$a,b\in {\cal U}$ such that $\pi (a)=z=\pi (b)$. If $a\neq b$ let
us denote by $\gamma$ the portion of $\pi^{-1}(z)$ from $a$ to
$b$, then $\gamma$ is a timelike curve connecting $a$ and $b$.
Thus by causal convexity, $\gamma$ must be contained in ${\cal
U}\subset{\cal U}_0$. Hence $\Phi^{-1}(\gamma )\cap {\cal V}$ is a
timelike curve joining two points of ${{\cal S}}$. But ${{\cal
S}}$ is achronal, being a Cauchy surface for $\tilde{M}$. Thus
$a=b$ so $\pi\vert_{\hat{\cal S}\cap{\cal U}}$ is injective.

Hence $F\colon {\cal S}\to\mathfrak{S}$ defined by
$F=\pi\circ\Phi_{{\cal S}}$ is a local homeomorphism, therefore
$F$ is an open map, i.e. $F({\cal S})$ is an open subset of
$\mathfrak{S}$. On the other hand, since ${\cal S}$ is compact
then $F ({\cal S})$ is compact, hence closed in $\mathfrak{S}$. It
then follows from connectedness of $\mathfrak{ S}$ that $F$ is
surjective.

 Thus by a standard topological result (refer for instance to
 proposition 2.19 in
 $\cite{LEES}$ and notice that the proof works as well in the continuous setting)
 we have that $F$ is a topological covering map.
Moreover, by corollary $\ref{structurecor}$ we have
$\mathfrak{S}\approx S^3$. Thus $\mathfrak{S}$ is simply connected
and as a consequence, $F$ is a homeomorphism. Hence $\Phi_{{\cal
S}}$ is injective as well, therefore a topological embedding since
${\cal S}$ is compact.

Then $\hat{{\cal S}}$ is a closed embedded \textit{spacelike}
hypersurface. Thus since $dS^4$ is simply connected, we have that
$\hat{{\cal S}}$ is achronal by proposition $\ref{spacesc}$.

Let ${\cal S}_a$, $a\in\R$, be the foliation of $\tilde{M}$
induced by ${\cal S}$. Since $\hat{{\cal S}}_a:=\Phi ({{\cal
S}_a})$ is achronal for all $a\in\R$ it follows that no two of
these surfaces can intersect. Thus $\Phi$ is injective.

The result now follows since every injective local isometry is an
isometry into an open subset of the codomain. $\Box$

\begin{remark}
G. Mess points out in $\cite{messy}$ the existence of  simply
connected and locally de Sitter spacetimes of constant curvature
$\equiv 1$ that can not be embedded in 3-dimensional de Sitter
space. In $\cite{beng}$, I. Bengtsson and S. Holst were able to
construct a similar example in dimension four. Moreover, this
latter spacetime occurs as a Cauchy development (see section
$\ref{ivp}$ below) of a Cauchy surface $S$ with topology
$\mathbb{H}^2\times\R$, which is clearly non compact. On the other
hand, theorem $\ref{teo1}$ above shows that no such example can be
found in our setting.
\end{remark}

Theorem $\ref{teo1}$ can be interpreted in terms of the Cosmic
Censor Conjecture. This notion is due to R. Penrose $\cite{CCP}$
who conjectured the existence of a ``cosmic censor" who would
forbid the appearance of naked singularities. In its strong form,
the cosmic censor conjecture just says that all reasonable
spacetimes must be globally hyperbolic.

The weak form of the Cosmic Censor Conjecture states that apart
from a possible initial singularity (a ``big bang" singularity),
the region of spacetime away from black or white holes must be
globally hyperbolic. Thus, if a singularity is present it should
be contained either inside a black hole or inside a white whole.
In that sense ``naked" singularities are not allowed, since all of
them are hidden by an event horizon.

Equivalently, we have the following mathematical formulation:

\noindent\textbf{ The Weak Cosmic Censor Conjecture} Let
$(\tilde{M},\tilde{g})$ be a spacetime admitting a conformal
boundary $\scri$. Then the Domain of Outer Communication ${\cal
D}(\tilde{M}):=I^+(\scri^-,M)\cap I^-(\scri^+,M)$ is globally
hyperbolic.

Note  that both de Sitter space and Schwarzchild-de Sitter satisfy
the weak cosmic censor conjecture. Thus we have this immediate
corollary to theorem $\ref{teo1}$:

\begin{cor}
Let $(\tilde{M},\tilde{g})$ be an asymptotically de Sitter
spacetime satisfying the Einstein equations. Further assume
$(\tilde{M},\tilde{g})$ admits a null line with endpoints
$p\in\scri^-$, $q\in\scri^+$. If the weak cosmic censor conjecture
holds in $(\tilde{M},\tilde{g})$ then the domain of outer
communication ${\cal D}(\tilde{M})$ is isometric to an open subset
of de Sitter space that contains a Cauchy surface.
\end{cor}


\section{The initial value problem}\label{ivp}

Notice that when put in coordinate form, the Einstein equation
$G=T$ gives rise to a second order PDE system
$G_{\alpha\beta}=T_{\alpha\beta}$. Thus, from the viewpoint of PDE
theory we can ask ourselves when this system admits an initial
value formulation; that is, we want to find out under what
circumstances initial data will determine a unique solution to the
Einstein equations.

In the  above setting, we are aiming at solving for the functions
$g_{\alpha\beta}$, i.e. for spacetime itself. Thus from a physical
perspective, it is natural to assign our initial conditions on a
Riemannian manifold $S$, which represents a frozen picture of
spacetime. We then use the Einstein equations to determine how $S$
evolves in time, and thus we generate a spacetime $(M,g)$ in which
$S$ sits as a spacelike hypersurface.

At this point we need to determine the nature of the initial
conditions. Notice first that a spacetime $(M,g)$ induces in a
natural way two geometrically meaningful objects on a spacelike
hypersurface: the first and second fundamental forms. Thus, a good
candidate for an initial data set would be a Riemannian manifold
$S$ endowed with two tensor fields $h$ and $K$ that will be the
induced metric and second fundamental form of $S$ in $(M,g)$.

\enlargethispage{.5cm}

However, these initial conditions can not be freely chosen.
Indeed, since the Gauss-Codazzi equations must hold, $h$ and $K$
must satisfy the \textit{constraint equations}:
\begin{eqnarray}
r+{(\text{Tr}_hK)}^2-\vert K\vert^2 &=&2T_{00}\\
D^jK_{ij}-D_i\text{Tr}_hK &=& T_{0i} \nonumber
\end{eqnarray}
where $i,j=1,2,\ldots ,n-1$ and $r$, $D$ are the scalar curvature
and Levi-Civita connection of $(S,h)$ respectively.

The fundamental result of Y. Choquet-Bruhat $\cite{YCB}$ shows
that the initial value problem in  $(3+1)$ general relativity is
well posed for initial conditions $(S, h, K)$ satisfying the
constraint equations and the vacuum Einstein equation $G=0$.

Moreover, R. Geroch and Y. Choquet-Bruhat $\cite{YCBG}$ have
proven the existence of a maximal Cauchy development ${\cal M}^*$
relative to a initial data set $(S,h,K)$ satisfying the vacuum
Einstein equation. As pointed out in $\cite{larsgal}$, the
argument used in $\cite{YCBG}$ is valid when considering the
Einstein equations with cosmological constant $G+\Lambda g=0$.

Following $\cite{Wald}$ we summarize our discussion in the
following theorem:

\begin{teo}
Let $(S,h)$ be a 3-dimensional Riemannian manifold and $K$ a
smooth symmetric tensor field on $S$. Suppose $(S,h,K)$ satisfy
the constraint equations, then there exists a spacetime $({\cal
M}^*,g^*)$, called the maximal domain of dependence of $(S,h,K)$,
satisfying all of the following:
\begin{enumerate}
\item $({\cal M}^*,g)$ is a solution of the vacuum Einstein
equation with cosmological constant. \item $({\cal M}^*,g^*)$ is
globally hyperbolic with Cauchy surface $S$. \item The induced
metric and second fundamental form of $S$ are $h$ and $K$.\item
Any spacetime satisfying the above three conditions is isometric
to a subset of $({\cal M}^*,g^*)$. \end{enumerate}
\end{teo}

Notice that the above definition doesn't give us any information
on how big ${\cal M}^*$ is, since in principle  ${\cal M}^*$ could
be isometrically embedded in a strictly larger spacetime $M_0$.
However, if this happens then $S\subset M_0$ can not be a Cauchy
surface of $M_0$. In other words, ${\cal M}^*$ satisfies a domain
of dependence condition (refer to theorem $10.2.2$ in
$\cite{Wald}$).

\begin{teo}\label{domdep}
Let $(S_i,h_i,K_i)$, $i=1,2$, be two initial data sets with
maximal Cauchy developments $({\cal M}^*_i,g_i^*)$. Let
$A_i\subset S_i$ and assume there is a diffeomorphism sending
$(A_1,h_1,K_1)$ to $(A_2,h_2,K_2)$. Then $D(A_1,{\cal M}^*_1)$ is
isometric to $D(A_2,{\cal M}^*_2)$.
\end{teo}

Now, we can state theorem $\ref{teo1}$ in terms of the initial
value problem.

\begin{teo}
Let $(S,h,K)$ be an initial data set and $({\cal M}^*,g^*)$ its
maximal Cauchy development. Suppose $({\cal M}^*,g^*)$ is
asymptotically de Sitter and satisfies the null energy condition.
If $({\cal M}^*,g^*)$ contains a null line from $\scri^-$ to
$\scri^+$, then it is isometric to $dS^4$.
\end{teo}

\noindent\textit{Proof:} By theorem $\ref{teo1}$ there is an
isometry $\Phi\colon({\cal M}^*,g^*)\to {\cal A}$, where ${\cal
A}$ is an open subset of $dS^4$. Furthermore, by the proof of
theorem $\ref{teo1}$ we also know $\Phi (S)$ is a Cauchy surface
of $dS^4$, hence $D(\Phi (S),dS^4)=dS^4$. Then the result  follows
from theorem $\ref{domdep}$. $\Box$\Par


\section{Matter fields in asymptotically simple
spacetimes}\label{mattersec}


In this section we will be considering matter fields on an
asymptotically de Sitter spacetime $(\tilde{M},\tilde{g})$
satisfying all four of the following hypotheses:


\noindent \textbf{A. The Dominant Energy Condition.}

\begin{defi} The energy-momentum tensor $T$ satisfies the Dominant Energy Condition if for all
timelike $X\in{\cal X}(M)$, $T(X,X)\ge 0$ and the vector field
metrically related to $T(X,-)$ is causal.
\end{defi}

The dominant energy condition is believed to hold for all known
forms of matter. It may be interpreted as saying that no observer
detects a negative local energy density. Actually the dominant
energy condition is equivalent to the statement that the speed of
the energy flow is less or equal than the speed of light
$\cite{Wald,HE}$.

Notice that the dominant energy condition is stated solely in
terms of the causal structure of $(\tilde{M},\tilde{g})$. It is
also easy to check that the following equivalence holds as well
$\cite{HE,DI}$:


\begin{prop}
A symmetric tensor $T$ satisfies the dominant energy condition if
and only if its components $T_{\alpha\beta}$ with respect to an
orthonormal frame $\{e_{\alpha}\}$ in which $e_0$ is timelike
satisfy $T_{00}\ge \vert T_{\alpha\beta}\vert$.
\end{prop}

From this proposition it is easy to see that a perfect fluid
satisfies the dominant energy condition if and only if $\rho
\ge\vert p\vert $.



\noindent\textbf{B. $\widetilde{\text {Tr}}\, T\le 0$ on a
neighborhood of $\scri$.}

This hypothesis is satisfied for a wide verity of fields. It holds
for  photon gases,  electromagnetic fields $\cite{SW,kriele,Wald}$
as well as for quasi-gases $\cite{SW}$. In particular it holds for
dust, pure radiation and all perfect fluids satisfying $0\le p\le
\rho /(n-1)$.


\noindent\textbf{C. If $K$ is a null vector at $p\in \tilde{M}$
with $T(K,K)=0$, then $T\equiv 0$ at $p$.}

Recall that a Type I energy-momentum tensor is by definition
diagonalizable $\cite{HE}$. With the exception of a null fluid,
all energy-momentum tensors representing reasonable matter are
diagonalizible $\cite{Wald}$. Let $\{\rho ,p_1,\ldots ,p_{n-1}\}$
be the eigenvalues of such a tensor with respect to an orthonormal
basis $\{e_0,e_1,\ldots ,e_{n-1}\}$, where $e_0$ is timelike. Then
for a Type I tensor the existence of $\lambda\in (0,1)$ satisfying
$\lambda \rho\ge\vert p_i\vert $, $i=1,\ldots,n-1$ prevents the
vanishing of $T_x$ in null directions, unless $T_x\equiv 0$. In
particular, perfect fluids with $0\le p\le \rho /(n-1)$ satisfy
this condition.

Assumption \textbf{C} carries a very important consequence: it
guarantees the vanishing of the energy momentum tensor along
totally geodesic null hypersurfaces.

\begin{prop}\label{tes0}
Let $(\tilde{M},\tilde{g})$  be a spacetime satisfying assumption
$\textbf{C}$ in which the Einstein equations hold. Let $S$ be a
totally geodesic null hypersurface in $(\tilde{M},\tilde{g})$,
then $T\equiv 0$ on $S$.
\end{prop}

\noindent\textit{Proof:} Let $p\in S$. In virtue of assumption
$\textbf{C}$, it suffices to show that $T(K,K)=0$ for some null
vector $K\in T_p\tilde{M}$. Hence let us consider a future null
generator $\gamma$ of $S$ trough $p$ and recall from section
$\ref{snulls}$ that $\gamma^{\prime}$ satisfies the Raychaudhuri
equation
\begin{equation}
\frac{d\theta}{ ds}=-\textrm
{Ric}({\gamma}^{\prime},{\gamma}^{\prime}) -{\sigma}^2-\frac{1}
{n-2}{\theta}^2.
\end{equation}
Since $S$ is totally geodesic we must have $\theta \equiv 0$ and
$\sigma\equiv 0$, thus
$\textrm{Ric}(\gamma^{\prime},\gamma^{\prime})=0$. Finally, since
$\gamma^{\prime}$ is null the Einstein equations imply
$\textrm{Ric}(\gamma^{\prime},\gamma^{\prime})=T(\gamma^{\prime},\gamma^{\prime})$,
and thus $T(\gamma^{\prime},\gamma^{\prime})=0$, as desired.
$\Box$\par


\noindent\textbf{D. The  following falloff condition holds:}
\begin{equation}\label{falloff2}
\displaystyle{\lim_{p\to\scri}\Omega {T(\nabla \Omega
,\nabla\Omega )}_p=0}.
\end{equation}

As a way of motivation, let us check that perfect fluid models on
Robertson-Walker spacetimes (see definition $\ref{RWstime}$)
satisfy this condition.

Thus let $(\tilde{M},\tilde{g})=RW(a,k)$  and assume $\tilde{M}$
satisfies the Einstein equations with $\Lambda >0$ with matter
content given by a perfect fluid. In this setting, we have the
equations $\cite{DI,kriele}$
\begin{equation}\label{020}
\rho +\Lambda
=\frac{(n-1)(n-2)}{2}\left(\frac{{\dot{a}}^2+k}{a^2}\right)
\end{equation}
\begin{equation*}
(n-1)\frac{\dot{a}}{a}=-\frac{\dot{\rho}}{\rho +p}
\end{equation*}
so, if we specialize further and consider dust or radiation,
equations $\ref{020}$ integrates to give rise to  conservation
laws:
\begin{equation}\label{cons}
\rho a^{n-1}=C\qquad\text{(dust)}\qquad\qquad \rho
a^n=C\qquad\text{(radiation)}
\end{equation}
and hence equations $(\ref{020})$ can be written as

\begin{eqnarray}\label{022}
{\dot{a}}^2 &=& \frac{2a^{3-n}}{(n-1)(n-2)}+\frac{2\Lambda
a^2}{(n-1)(n-2)}-k\qquad
\text{(dust)}\\
{\dot{a}}^2 &=& \frac{2a^{2-n}}{(n-1)(n-2)}+\frac{2\Lambda
a^2}{(n-1)(n-2)}-k\qquad \text{(radiation)}\nonumber
\end{eqnarray}

As $t\to\infty$ the solution to either of these two ODE's grows
without bound. Hence the term involving $a^2$ in equations
$(\ref{022})$ dominates the remaining ones, thus in both cases we
have ${\dot{a}}^2\sim Pa^2$, with $P=\sqrt{2\Lambda /(n-1)(n-2)}$.
Thus $a\sim e^{Pt}$ and then $\tilde{g}$ approaches the metric
$-dt^2+e^{2Pt}d\sigma^2$. Hence
\begin{equation}
\tilde{g}\sim\frac{1}{(Pu)^2}(-du^2+d\sigma^2),\qquad\ u=-e^{-Pt}.
\end{equation}

Let $g=(Pu)^2\tilde{g}$, so $g\sim -du^2+d\sigma^2$. Notice
$e_0=u{\nabla} u $ is a unit timelike vector field with respect to
$\tilde{g}$. Thus
\begin{equation}
T(\nabla u,\nabla u)=\frac{1}{u^2}T(e_0,e_0)\sim\frac{\rho}{u^2},
\end{equation}
hence the conservation laws $(\ref{cons})$ imply $T(\nabla
u,\nabla u)\to 0$ as $u\to 0$ and thus the falloff condition
\textbf{D} holds as well.

Assumption \textbf{D} enables us to give a nice description of the
unphysical metric near $\scri$ as the following result shows.

\begin{lemma}\label{gauge}
Let $(\tilde{M},\tilde{g})$ be a past asymptotically de Sitter
spacetime satisfying the Einstein equations with $\Lambda >0$.
Assume that for a completion $(M,g)$ with defining function
$\Omega$ the past conformal boundary $\scri^-$ is compact. Further
assume the decay condition
\begin{equation}
{\lim_{p\to\scri^-}{\Omega T(\nabla \Omega ,\nabla\Omega )}_p=0}
\end{equation}
holds, then there exist a defining function
$\overline{\Omega}\in{\cal C}^{\infty}(M)$ satisfying
$\ref{falloff2}$ and a neighborhood ${\cal U}$ of $\scri^-$ such
that
\begin{equation}\label{013}
\tilde{g}=\frac{1}{{\overline{\Omega}}^2}[-d{\overline{\Omega}}^2+h(u)]\qquad\text{on
${\cal U}$}
\end{equation}
where $h(u)$ is a Riemannian metric on the slice
$S_u=\overline{\Omega }^{-1}(u)$.
\end{lemma}

\noindent\textit{Proof:} First notice that by a constant rescaling
of the physical metric we can assume
\begin{equation}\Lambda =
\frac{(n-1)(n-2)}{2}.
\end{equation}

Since $\scri ={\Omega}^{-1}(0)$ and $d\Omega \neq 0$ on $\scri$
then $\nabla\Omega$ is normal to ${\scri}$. Let $(x_0,x_1,\ldots
,x_{n-1})$ be the slice coordinates in a neighborhood ${\cal W}$
of $\scri^-$ adapted in such a way that
${\partial}_0{\vert}_{\scri^-}=\nabla\Omega{\vert}_{\scri^-}$. Let
$X=\Omega{\partial}_0$ and notice that the component functions
$X^{\alpha}$ satisfy
\begin{equation}
 X^{\alpha}=\Omega
{\nabla}^{\alpha}\Omega +O({\Omega}^2)\qquad\text{on ${\cal W}$},
\end{equation}
 thus the Einstein
equations in conjunction with proposition \ref{confcurva} yield
the following estimate
\begin{equation}\label{010}
T(X,X)=\left[ \frac{(n-1)(n-2)}{2}g(\nabla\Omega ,\nabla\Omega
)+\Lambda \right]\tilde{g}(X,X) +O(\Omega )\qquad \text{on ${\cal
W}$}.
\end{equation}

As we approach $\scri^-$ the falloff condition $(\ref{falloff2})$
implies
\begin{equation}
\lim_{p\to\scri^-}{T(X ,X )}_p=\lim_{p\to\scri^-}\Omega^2{T(\nabla
\Omega ,\nabla\Omega )}_p=0.
\end{equation}
On the other hand, by $(\ref{010})$ we have
\begin{equation}
T(X,X)\to\left[ \frac{(n-1)(n-2)}{2}g(\nabla\Omega ,\nabla\Omega
)+\Lambda \right]{g}(\nabla\Omega ,\nabla\Omega )\qquad \text{as
$\Omega\to 0$.}
\end{equation}
hence we must have
\begin{equation}
g(\nabla \Omega ,\nabla\Omega ){\vert}_{\scri^-}=-1.
\end{equation}

Consider now the conformally rescaled quantities
\begin{equation}
\overline{\Omega}=\frac{\Omega}{\theta},\qquad\qquad
\overline{g}=\frac{g}{{\theta}^2}
\end{equation}
then we want to find $\theta$ smooth in a neighborhood ${\cal U}$
of $\scri$ such that $\overline{\Omega}$ agrees with $\Omega$ on
$\scri^-$ and
\begin{equation}\label{090}
\overline{g}(\overline{\nabla}\,\overline{\Omega},\overline{\nabla}\,\overline{\Omega})=-1\qquad\text{on
${\cal U}$}.
\end{equation}
To do so, first notice
$\overline{\nabla}\,\overline{\Omega}=\theta\nabla\Omega
-\Omega\nabla\theta$ so then equation $(\ref{090})$ gives rise to
the first order PDE
\begin{equation}\label{011}
2\theta g(\nabla\Omega ,\nabla\theta )-\Omega g(\nabla\theta
,\nabla\theta )-\frac{{\theta}^2}{{\Omega}}(1+g(\nabla\Omega
,\nabla\Omega ))=0
\end{equation}
and since
\begin{equation}
\frac{1+g(\nabla\Omega ,\nabla \Omega )}{\Omega}\in {\cal
C}^{\infty}({\cal W})
\end{equation}
then a standard PDE result (refer to the generalization of theorem
10.3 on page 36 in $\cite{Spivak5}$) guarantees that equation
$(\ref{011})$ subject to the initial condition
$\theta{\vert}_{\scri^-}=1$ has a unique solution in a
neighborhood ${\cal U}$ of $\scri^-$. Notice that, by shrinking
${\cal U}$ if necessary, we can extend $\theta$ smoothly to a
positive function in all of $M$.

Observe that the integral curves of the gradient
$\overline{\nabla}\, \overline{\Omega}$ are unit speed timelike
curves in ${\cal U}$, hence geodesic segments with respect to the
unphysical metric $\overline{g}$. Moreover, all this geodesics
emanate from and are normal to $\scri^-$. By further restricting
${\cal U}$ to a normal neighborhood of $\scri^-$, we can take the
slices $S_u$ to be the normal gaussian foliation of ${\cal U}$
with respect to $\scri^-$. Hence $\ref{013}$ follows.

Finally, notice that
\begin{equation}
T(\overline{\nabla}\, \overline{\Omega}, \overline{\nabla}\,
\overline{\Omega})=\theta^2T(\nabla\Omega ,\nabla\Omega ) +
O(\Omega )\qquad\text{on ${\cal U}$}
\end{equation}
hence the fall-off condition $(\ref{falloff2})$ holds for
$\overline{\nabla}\, \overline{\Omega}$ as well. $\Box$\par

\enlargethispage{.5cm}

To finish our discussion, notice that  a strengthened version of
assumption \textbf{D} causes $\scri$ to be totally geodesic.

\begin{prop}
Let $(\tilde{M},\tilde{g})$ be a past asymptotically de Sitter
spacetime satisfying the Einstein equations with $\Lambda >0$.
Assume $T$ satisfies the dominant energy condition and that
$T(\nabla \Omega ,\nabla \Omega )$ is bounded in a neighborhood of
$\scri^-$. Then $\scri^-$ is totally geodesic.
\end{prop}

\noindent\textit{Proof:} As before, we can assume $\Lambda
=(n-1)(n-2)/2$. We proceed to show the Hessian of $\Omega$
vanishes at every point of $\scri$. Thus, without loss of
generality, let $p\in\scri^-$. Let ${ V}$ be a paracompact
neighborhood of $p$ in $\scri^-$ and let ${\cal V}=D(V,N)$. Hence
$({\cal V}\cap \tilde{M},\tilde{g}\vert_{{\cal V}\cap \tilde{M}})$
is a past asymptotically de Sitter spacetime in its own right with
defining function $\Omega$ and past conformal boundary $V$.

In virtue of lemma $\ref{gauge}$ we can choose the defining
function $\Omega$ in a way that $g(\nabla\Omega ,\nabla\Omega
)=-1$ in a neighborhood of $V$. Further, by contracting the
Einstein equations we get
\begin{equation}
\widetilde{\text{Tr}}\, T=\frac{n-2}{2}(n(n-1)-\tilde{R}).
\end{equation}
Since $e_0=\Omega\nabla\Omega$ is a unit timelike vector with
respect to $\tilde{g}$, the dominant energy condition implies
\begin{equation}
\vert\widetilde{\text{Tr}}\, T\vert\le n\Omega^2T(\nabla\Omega
,\nabla\Omega ),
\end{equation}
but $T(\nabla\Omega ,\nabla\Omega )$ is bounded as $\Omega\to 0$,
hence $\widetilde{\text{Tr}}=O(\Omega^2)$, which in turn leads us
to
\begin{equation}\label{040}
\tilde{R}-n(n-1)=O(\Omega^2 )
\end{equation}

On the other hand, since $g(\nabla\Omega ,\nabla\Omega )=-1$ near
$V$, proposition \ref{confcurva} implies
\begin{equation}
R+2(n-1)\frac{\triangle \Omega}{\Omega} =
\frac{1}{\Omega^2}(\tilde{R}-n(n-1)),
\end{equation}
thus by $(\ref{040})$ we have $\triangle\Omega = O(\Omega )$.

Finally by combining the Einstein equations and the formulas from
propositions \ref{confcurva} and  \ref{confdiv}  we get
\begin{equation}
T=G+\frac{n-2}{\Omega}{\text
Hess}_{\Omega}-\frac{n-1}{\Omega}\triangle\Omega ,
\end{equation}
then by considerations above ${\text Hess}_{\Omega}/\Omega$ is
bounded near $V$, hence ${\text Hess}_{\ \Omega}\equiv 0$ on $p$.
Thus the second fundamental form of $\scri$ in $M$ vanishes as
well. $\Box$\par


\begin{figure}
\begin{center}
\includegraphics{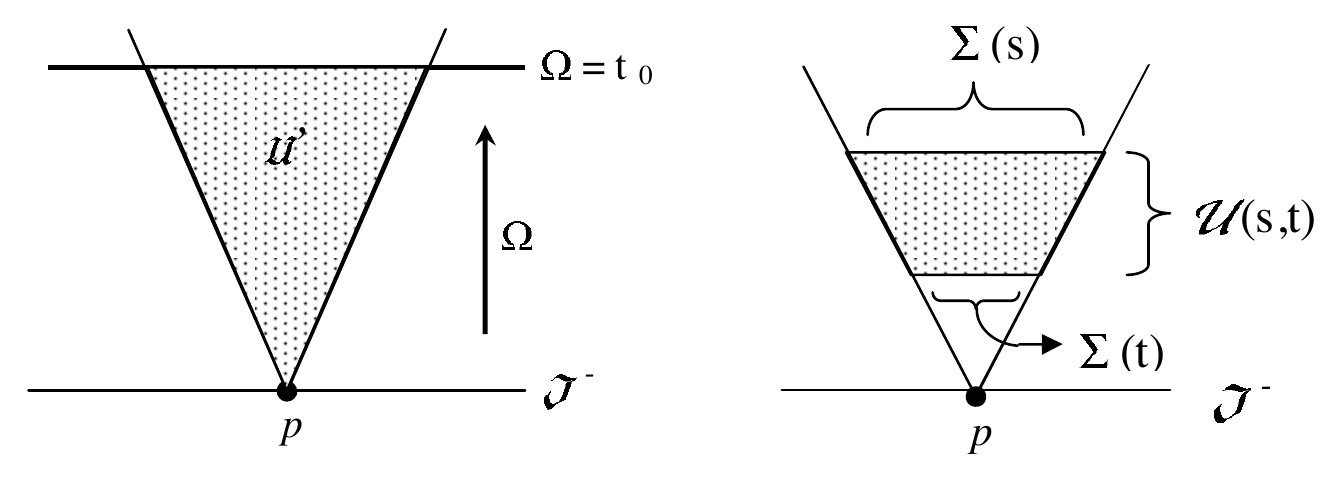} \caption{\label{fig4}{\footnotesize
Definition \ref{030}.}}
\end{center}
\end{figure}

Now let us consider a spacetime $(\tilde{M},\tilde{g})$ satisfying
all the assumptions of lemmas $\ref{structurelemma1}$ and
$\ref{gauge}$ above. That is,

\noindent $\bullet$ $(\tilde{M},\tilde{g})$ is globally hyperbolic
and asymptotically de Sitter.

\noindent $\bullet$ $(\tilde{M},\tilde{g})$ contains a null line
$\eta$ with past endpoint $p\in\scri^-$ and future endpoint
$q\in\scri^+$.

\noindent $\bullet$ $(\tilde{M},\tilde{g})$ satisfies the Einstein
equations with $\Lambda >0$.

\noindent $\bullet$ The energy momentum tensor $T$ has the decay
rate $(\ref{falloff2})$.

In this set up ${\cal S}:=\partial I^+(\eta )$ is just the future
null cone at $p$, i.e. ${\cal S}=\exp_p(\Lambda^+_p\cap{\cal
O})\cap\tilde{M}$ where ${\cal O}$ is the maximal set in which
$\exp_p$ is defined. Let us denote now the local causal cone at
$p$ by $\mathfrak{U}:=\exp_p(C^+_p\cap{\cal O})\cap\tilde{M}$,
hence $\mathfrak{U}-\{p\}$ is a manifold-with-boundary and
$\partial(\mathfrak{U}-\{p\})={\cal S}$. Let us choose now a
neighborhood ${\cal U}$ of $p$ and $\Omega$ so that ${\cal U}$ is
foliated by $\Omega$ and let $t_0>0$ such that
$\mathfrak{U}^{\prime}:=\mathfrak{U}\cap\Omega^{-1}([0,t_0])\subset
{\cal U}$.


\begin{defi}\label{030}
 For  $s,t\in (0,t_0)$ with $s<t$ we define ${\cal
U}(s,t):={\mathfrak{U}}^{\prime}\cap\Omega^{-1}([s,t])$, ${\cal
S}(s,t):={\cal S}\cap\Omega^{-1}([s,t])$ and $\Sigma
(t)={\mathfrak{U}}^{\prime}\cap\Omega^{-1}(t)$. (See figure
$\ref{fig4}$.)
\end{defi}


Thus ${\cal U}(s,t)$ is a compact manifold with corners
$\cite{LEES}$ and
\begin{equation}
\partial {\cal U}(s,t)={\cal S}(s,t)\cup\Sigma (s)\cup\Sigma (t)
\end{equation}

\begin{lemma}\label{structurelemma2}
Let $(\tilde{M},\tilde{g})$ be a spacetime satisfying the three
conditions listed above. Let $N_p=\partial I^+(p,N)$ and for
$0<t_1<t_0$ define ${\cal
S}^{\prime}:=[N_p-\Omega^{-1}([0,t_1))]\cup\Sigma (t_1)$,
$\mathfrak{
U}^{\prime\prime}:=\mathfrak{U}\cap\Omega^{-1}([0,t_1])$. Then
$J^+(p,N)\cap\tilde{M}-\mathfrak{U}^{\prime\prime}\subset
D^+({\cal S}^{\prime},N)\cap\tilde{M}$.
\end{lemma}

\enlargethispage{.5cm}

\noindent\textit{Proof:} Let $x\in
J^+(p,N)\cap\tilde{M}-\mathfrak{U}^{\prime\prime}$ and let
$\gamma$ be a past inextendible timelike curve with future
endpoint $x$. Since $J^+(p,N)\cap\tilde{M}\subset
D^+(N_p,N)\tilde{M}$ by lemma $\ref{structurelemma0}$, we have
that $\gamma$ must intersect $N_p$, say at $y$. If $\Omega (y)\ge
t_1$ then $y\in {\cal S}^{\prime}$. If $\Omega (y)<t_1$ then
notice that $\Omega (x)>t_1$ since
$x\not\in\mathfrak{U}^{\prime\prime}$. Now, since the function
$t\mapsto \Omega (\gamma (t))$ is continuous there exist a point
$z\in\gamma$ between $x$ and $y$ such that $\Omega (z)=t_1$. Hence
$z\in\Sigma (t_1)\subset  {\cal S}^{\prime}$. $\Box$

\begin{figure}
\begin{center}
\includegraphics{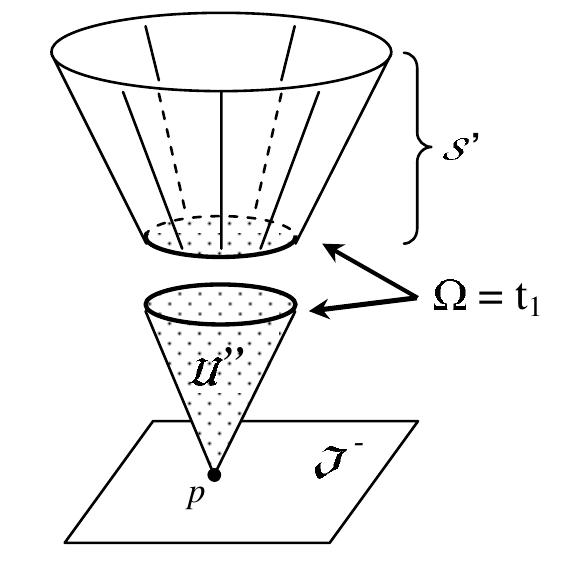}
\caption{\label{fig10}{\footnotesize Lemma
\ref{structurelemma2}.}}
\end{center}
\end{figure}


Now we can prove the main result of this section.

\begin{teo}\label{matterfield} Let $(\tilde{M}, \tilde{g})$ be an
asymptotically simple spacetime of dimension $n=4$ which is a
solution of the Einstein equations with positive cosmological
constant
\begin{equation}\label{einstein}
R_{\alpha\beta}-\frac{1}{2}Rg_{\alpha\beta}+\Lambda
g_{\alpha\beta}=T_{\alpha\beta},
\end{equation}
 where the
energy-momentum tensor $T$ satisfies the following:

\textbf{A.} The Dominant Energy Condition.

\textbf{B.} $\widetilde{\text {Tr}}\, T\le 0$ on a neighborhood of
$\scri$.

\textbf{C.} If $K$ is a null vector at $p\in \tilde{M}$ with
$T(K,K)=0$, then $T\equiv 0$ at $p$.

\textbf{D.} The falloff condition
$\displaystyle{\lim_{p\to\scri}\Omega{T(\nabla \Omega
,\nabla\Omega )}_p=0}$ holds.

\noindent If $(\tilde{M}, \tilde{g})$ contains a null line $\eta$
then $(\tilde{M}, \tilde{g})$ is isometric to de Sitter space.
\end{teo}


\noindent\textit{Proof:} We first show that the hypotheses of the
theorem imply $T\equiv 0$ on  asymptotically simple and de Sitter
spacetimes of any dimension. Then we can invoke theorem
$\ref{rigds}$ to conclude $\tilde{M}$ is de Sitter space when
$n=4$.

By asymptotic simplicity $\eta$ has a past endpoint $p\in \scri^-$
and a future endpoint $q\in\scri^+$.  We first show that the
assumptions on $T$ causes the energy momentum tensor to vanish in
a neighborhood of $p$.

Consider the notation of definition $\ref{030}$. For the time
being, let $s\in (0,t_0)$ be fixed and let  ${\cal U}(t):={\cal
U}(s,t)$, ${\cal S}(t):={\cal S}(s,t)$ for all $t\in (s,t_0)$. Let
$A$ be the vector field related to $T(\nabla \Omega ,-)$ via the
unphysical metric, in other words,
\begin{equation}
g(A, X)=T(\nabla \Omega ,X) \quad \text{for all $X\in{\cal
X}(\tilde{M})$.}
\end{equation}

Let $\textrm{dv}$ denote the volume element of $(M,g)$. Then
$d(i_{{A}}\text{dv})={\rm div}A\,\text{dv}\,$, hence by Stokes
theorem
\begin{equation}
 \int\limits_{{\cal U}(t)}{\rm div}A\, \text
{dv}=\int\limits_{{\cal U}(t)}d(i_{A}\text{
dv})=\int\limits_{\partial{\cal U}(t)}i_{A}\text{dv}.
\end{equation}
Furthermore, the integral over $\partial {\cal U}(t)$ can be
computed as
\begin{equation}\label{integral1}
 \int\limits_{\partial{\cal U}(t)}i_{A}\text{dv}
=\int\limits_{\Sigma (s )}i_{A}\text{ dv}+\int\limits_{\Sigma
(t)}i_{A}\text{dv}+\int\limits_{{\cal S}(t)}i_{A}\text{dv}.
\end{equation}
Notice $\nabla \Omega$ is a unit vector field normal to the
surfaces $\Sigma (a)$. Since $\nabla\Omega$ is outward pointing at
$\Sigma (s)$ and inward pointing at $\Sigma (t )$ we have
\begin{eqnarray}
i_A\text{dv}&=&-g(A,-\nabla \Omega ){\rm d}\, \sigma =T(\nabla
\Omega ,\nabla \Omega)\text{d}\sigma \qquad \text{at }\Sigma (t)\\
i_A\text{dv}&=&-g(A,\nabla \Omega ){\rm d}\, \sigma =-T(\nabla
\Omega ,\nabla \Omega)\text{d}\sigma \qquad \text{at }\Sigma (s
)\nonumber
\end{eqnarray}
where $\textrm{d}\sigma$ is the volume form on $\Sigma (a)$.

On the other hand, let $B =\{e_0,e_1,\ldots ,e_{n-1}\}$ be a
positively oriented frame at $p\in M$. A direct computation shows
\begin{equation}
i_A\text{dv}=\sum_{i=0}^{n-1}(-1)^{i}T(\nabla\Omega
,e_i)e^*_0\wedge\cdots \wedge e^*_{i-1}\wedge
e^*_{i+1}\wedge\cdots \wedge e^*_{n-1},
\end{equation}
but in virtue of proposition $\ref{tes0}$ we have $T\equiv 0$ on
$S$, hence the above equation implies $i_A\text{dv}\vert_S\equiv
0$. Thus
\begin{equation}\label{integrales}
\int\limits_{{\cal U}(t)}\text{ div}A  \,
\text{dv}=\int\limits_{\Sigma (t )}T(\nabla\Omega ,\nabla\Omega
)\text{d}\sigma -\int\limits_{\Sigma (s )}T(\nabla\Omega
,\nabla\Omega )\text{d}\sigma .
\end{equation}

Now let $\hat{T}$ be the $(1,1)$ tensor $g$-equivalent to $T$ and
let $C$ denote tensor contraction with respect to $g$. Since
$A=C(\hat{T}\otimes\nabla \Omega)$ we have
\begin{eqnarray}\label{cuentas05} \text{div}A &=& C\nabla A=C\nabla C (\hat{T}\otimes \nabla
\Omega )=C^2\nabla (\hat{T}\otimes \nabla \Omega )\nonumber\\ &=&
C^2(\nabla \hat{T}\otimes \nabla \Omega +\hat{T}\otimes \nabla
(\nabla \Omega )) \nonumber\\ &=& C^2(\nabla \hat{T}\otimes \nabla
\Omega )+C^2(\hat{T}\otimes \nabla (\nabla \Omega ) )\\ &=&
\text{div} T\, (\nabla \Omega ) +C^2(\hat{T}\otimes \nabla (\nabla
\Omega ))\nonumber .
\end{eqnarray}
Hence
\begin{eqnarray}\label{integr02} \int\limits_{{\cal U}(t)} \text{div} T\,  (\nabla \Omega )
\text{dv}\ \ &+&\int\limits_{{\cal U}(t)}C^2 (\hat{T}\otimes
\nabla (\nabla \Omega ) ) \text{dv}\nonumber \\
&=& \int\limits_{\Sigma (t)}\hskip -.2cm T(\nabla \Omega , \nabla
\Omega ) {\rm d}\sigma \, \text{dv} \ -\int\limits_{\Sigma
(s)}\hskip -.2cm T(\nabla \Omega , \nabla
\Omega ) {\rm d}\sigma\nonumber \\
\end{eqnarray}

Since $\mathfrak{ U}^{\prime}$ is compact,  the components
${\Omega_{;\alpha}}^{;\beta}$ of $\nabla (\nabla \Omega )$ in any
$g$-orthonormal frame field are bounded from above, say by $Q$.
Similarly, $T(\nabla\Omega ,\nabla\Omega )\ge \vert
{T^{\alpha}}_{\beta}\vert$ on $\tilde{M}$ by the dominant energy
condition, hence by continuity ${\displaystyle \lim_{z\to
p}T(\nabla\Omega ,\nabla\Omega )_z\ge \lim_{z\to p}\vert
{T^{\alpha}}_{\beta}(z)\vert}$ as well. Then
\begin{equation}
C^2(\hat{T}\otimes \nabla (\nabla \Omega )
)={T^{\alpha}}_{\beta}{\Omega_{;\alpha}}^{;\beta}\le  P\, T(\nabla
\Omega ,\nabla \Omega )
\end{equation}
on $\mathfrak{U}^{\prime}$, where $P:=n^2Q$. Thus
\begin{equation}\label{integr03}
\int\limits_{{\cal U}(t)}C^2(\hat{T}\otimes \nabla (\nabla \Omega
) )\text{dv}\ \ \le \int\limits_{{\cal U}(t)}P\, T(\nabla \Omega
,\nabla \Omega ) \text{dv},
\end{equation}

On the other hand, recall the conformal relation \ref{confdiv}:
\begin{equation}\label{cuentas06}
\text{div}\, T(\nabla \Omega )=\frac{1}{{\Omega
}^2}\widetilde{\text{div}}\, T(\nabla \Omega ) +\frac{n-2}{\Omega
}T(\nabla \Omega ,\nabla \Omega )+\frac{1}{
\Omega^3}\widetilde{\text{Tr}}\, T.
\end{equation}
Since the physical metric satisfies the Einstein equations, the
energy-momentum tensor is divergence free. Thus
\begin{equation}
\widetilde{\text{div}}\, T(\nabla \Omega )\equiv 0\quad \text{in
$\tilde{M}$}
\end{equation}
Moreover, by assumption $\textbf{B}$ $\widetilde{\text{Tr}}\ T\le
0$, thus ($\ref{cuentas06}$) and ($\ref{enineq2}$) give rise to
the inequality
\begin{equation}\label{enineq3}
\int\limits_{{\cal U}(t)} \text{div} T\,  (\nabla \Omega )
\text{dv}\ \le \int\limits_{{\cal U}(t)}\frac{n-2}{\Omega}\,
T(\nabla \Omega ,\nabla \Omega ) \text{dv}
\end{equation}

Hence equation $(\ref{integr02})$ along with $(\ref{integr03})$
and $(\ref{enineq3})$ yield
\begin{equation}
\int\limits_{\Sigma (t)}\hskip -.2cm T(\nabla \Omega , \nabla
\Omega ) {\rm d}\sigma \, \text{dv} \ -\int\limits_{\Sigma (s
)}\hskip -.2cm T(\nabla \Omega , \nabla \Omega ) {\rm d}\sigma\
\le  \int\limits_{{\cal U}(t)}\left( \frac{n-2}{\Omega}+P\right)
\, T(\nabla \Omega ,\nabla \Omega ) \text{dv}
\end{equation}
and then Fubini's theorem implies
\begin{equation}\label{enineq2}
\int\limits_{\Sigma (t)}\hskip -.2cm T(\nabla \Omega ,\nabla
\Omega ) {\rm d}\sigma -\int\limits_{\Sigma (s )}\hskip -.2cm
T(\nabla \Omega ,\nabla \Omega ) {\rm d}\sigma \le
\int_{s}^t\int\limits_{\Sigma (\tau ) } \left(
\frac{n-2}{\Omega}+P\right) \, T(\nabla \Omega ,\nabla \Omega )
{\rm d}\sigma\,\, {\rm d}\tau.
\end{equation}

Now, we would like to analyze the limit of both sides of relation
($\ref{enineq2}$) as $s\to 0$. Let then $p(s)\in\Sigma (s)$ be
such that $T(\nabla\Omega_z ,\nabla\Omega_z )\le
T(\nabla\Omega_{p(s)},\nabla\Omega_{p(s)})$ for all $z\in\Sigma
(s)$. Such $p(s)$ always exists since $\Sigma (s)$ is compact.
Thus
\begin{eqnarray}\label{enineq4}
 \int\limits_{\Sigma (s )}
\frac{1}{\Omega }T(\nabla \Omega ,\nabla \Omega )  {\rm d}\sigma
&\le& \frac{1}{ s } T({\nabla \Omega}_{p(s )},{\nabla \Omega
}_{p(s )})\int\limits_{\Sigma (s ) }\text{d}\sigma\nonumber\\ &=&
\frac{1}{s}T({\nabla \Omega}_{p(s )},{\nabla \Omega }_{p(s
)})\text{Vol}(\Sigma (s))
\end{eqnarray}

Let us consider now a small normal neighborhood ${\cal N}$ around
$p$. It is known $\cite{volsr}$ that the metric volume of the
local causal cone truncated by a timelike vector $X_p$ is of the
same order as the volume of the corresponding truncated cone in
$T_pM$. As a consequence, if $g(X_p,X_p)=-r^2$ then the volumes of
the slices $S_r:=\exp_p(C_r)$ and $C_r:=\{v\in C_p^+\mid
g(v,X_p)=-r^2\}$ are also of the same order in $r$, and hence
$\text{Vol}(S_r)= O(r^{n-1})$. Hence by considering $s$ very small
and setting $X_p=s\nabla\Omega_p$ we get the estimate
\begin{equation}\label{volume}
\text{Vol}(\Sigma (s))= O(s^{n-1}),
\end{equation}
since $\text{Vol}(S_s)\sim\text{Vol}(\Sigma (s))$ for very small
$s>0$. Thus without loss of generality we can take $t_0>0$ such
that $\mathfrak{U}^{\prime}$ is contained in such a normal
neighborhood ${\cal N}$.

Therefore by letting $s\to 0$, equations ($\ref{enineq4}$) and
($\ref{volume}$) above imply
\begin{equation}\label{enineq7}
\frac{1}{s}T({\nabla \Omega}_{p(s )},{\nabla \Omega }_{p(s
)})\text{Vol}(\Sigma (s)) \le C\, T({\nabla \Omega}_{p(s
)},{\nabla \Omega }_{p(s )})s^{n-2}
\end{equation}
for some positive constant $C$. Hence
\begin{equation}
\lim_{s \to 0^{+}}\int\limits_{\Sigma (s ) } \frac{1}{\Omega }T(
\nabla\Omega ,\nabla \Omega )  {\rm d}\sigma =0
\end{equation}
in virtue of assumption \textbf{B}.

Thus
\begin{equation}
x(t): = \lim_{s\to 0} \int_{s}^t\int\limits_{\Sigma (\tau ) }
\left(\frac{n-2}{ \Omega }+P\right)T(\nabla \Omega ,\nabla \Omega
) {\rm d}\sigma\,\, {\rm d}\tau
\end{equation}
is a well defined function of $t$. Further, since inequality
($\ref{enineq2}$) holds for all $s>0$ it will hold on the limit as
well. Hence by taking limits to both sides of inequality
($\ref{enineq2}$) we get
\begin{equation}\label{integr05}
\int\limits_{\Sigma (t)}\hskip -.2cm T(\nabla \Omega ,\nabla
\Omega ) {\rm d}\sigma \le x(t)
\end{equation}
and further notice $\ref{integr05}$ can be expressed as
differential inequality:
\begin{equation}
\frac{dx}{ dt}\le \left(\frac{n-2}{ t}+P\right)x,\qquad t\in (0,
t_0)
\end{equation}
which in turn implies
\begin{equation}
\frac{d}{dt}\left( \frac{e^{-Pt}}{t^{n-2}}x
\right)=\frac{e^{-Pt}}{t^{n-2}}\left[ \frac{dx}{dt}-\left(
\frac{n-2}{ t}+P\right)x\right] \le 0
\end{equation}
i.e. the function
\begin{equation}
I(t)=\frac{x(t)e^{-Pt}}{t^{n-2}}
\end{equation}
is decreasing near $\scri^-$.

Thus, we analyze ${\displaystyle\lim_{t\to 0^+}I(t)}$. Notice
first that estimate ($\ref{enineq7}$) yields
\begin{equation}
\int\limits_{\Sigma (t ) } \left(\frac{n-2}{\Omega
}+P\right)T(\nabla \Omega ,\nabla \Omega ){\rm d}\sigma \le
C^{\prime}\, T(\nabla\Omega_{p(t)},\nabla\Omega_{p(t)})t^{n-2}
\end{equation}
for some constant $C^{\prime}\ge 0$. Thus by L'Hospital rule we
get
\begin{eqnarray}
\lim_{t \to 0^{+}}\frac{x(t )} {{t}^{n-2}} &=& \lim_{t \to
0^{+}}\frac{1}{ (n-2){t}^{n-3}}\int\limits_{\Sigma (t ) }
\left(\frac{n-2}{\Omega }+P\right)T(\nabla \Omega ,\nabla \Omega )
{\rm
d}\sigma\\
&\le&\frac{C^{\prime}}{n-2} \, t\,
T(\nabla\Omega_{p(t)},\nabla\Omega_{p(t)})\nonumber
\end{eqnarray}
then ${\displaystyle \lim_{t \to 0^{+}}\frac{x(t )}
{{t}^{n-2}}}=0$ and hence ${\displaystyle\lim_{t\to 0^+}I(t)}=0$.

It follows that $I(t)\equiv 0$ on $\mathfrak{U}^{\prime}$, and
consequently $T(\nabla \Omega ,\nabla\Omega )\equiv 0$ on
$\mathfrak{U}^{\prime}$. Therefore $T\equiv 0$ on
$\mathfrak{U}^{\prime}$ by the dominant energy condition.

Now let $0<t_1<t_0$ and consider ${\cal S}^{\prime}$ and
$\mathfrak{U}^{\prime\prime}$ as in lemma $\ref{structurelemma2}$.
Hence it is clear that $T\equiv 0$ on
$\mathfrak{U}^{\prime\prime}$. Further, let $x$ be in the
topological interior of $D^+({\cal S}^{\prime},N)$, hence
$W=J^-(x,N)\cap J^+({\cal S}^{\prime},N)$ is compact. Then
$T\equiv 0$ on $W$ by the conservation theorem of Hawking and
Ellis (cfr. page 93 in $\cite{HE}$), thus $T\equiv 0$ on
$\text{int}D^+({\cal S}^{\prime},N)$. Hence by continuity we have
$T\equiv 0$ on $D^+({\cal S}^{\prime},N)\cap\tilde{M}$.

On the other hand, by the lemmas $\ref{structurelemma0}$ and
$\ref{structurelemma2}$ we have the inclusions
\begin{equation}
I^+(S)\subset J^+(p,N)\cap\tilde{M}\subset
\mathfrak{U}^{\prime\prime}\cup (D^+({\cal
S}^{\prime},N)\cap\tilde{M})
\end{equation}
where $S=\partial I^+(\eta )$ as in lemma $\ref{structurelemma1}$.
Then we just showed $T\equiv 0$ on $I^+(S)$.

On a time dual fashion, we can show $T$ vanishes in a neighborhood
of $q$ and consequently on the whole set $I^-(S)$. To finish the
proof, recall that since $\partial I^+(\eta )=S=\partial
I^-(\eta)$ then $\tilde{M}=S\cup I^+(S)\cup I^-(S)$ in virtue of
proposition 3.15 in $\cite{P}$. Therefore $T\equiv 0$ on
$\tilde{M}$ and the result follows. $\Box$

Finally, notice that by theorem $\ref{teo1}$ and the proof of
theorem $\ref{matterfield}$ we have the following corollary:

\begin{cor} Let $(\tilde{M}, \tilde{g})$ be a
globally hyperbolic and asymptotically de Sitter spacetime of
dimension $n=4$, which is a solution of the Einstein equations
with positive cosmological constant
\begin{equation}
R_{\alpha\beta}-\frac{1}{2}Rg_{\alpha\beta}+\Lambda
g_{\alpha\beta}=T_{\alpha\beta},
\end{equation}
 where the
energy-momentum tensor $T$ satisfies conditions \textbf{A} to
\textbf{D} above. If $(\tilde{M}, \tilde{g})$ contains a null line
$\eta$ with endpoints $p\in\scri^-$ and $q\in\scri^+$ then
$(\tilde{M}, \tilde{g})$ is isometric to an open subset of de
Sitter space.
\end{cor}


\chapter{Asymptotically Anti de Sitter Spacetimes}

\begin{figure}
\begin{center}
\includegraphics{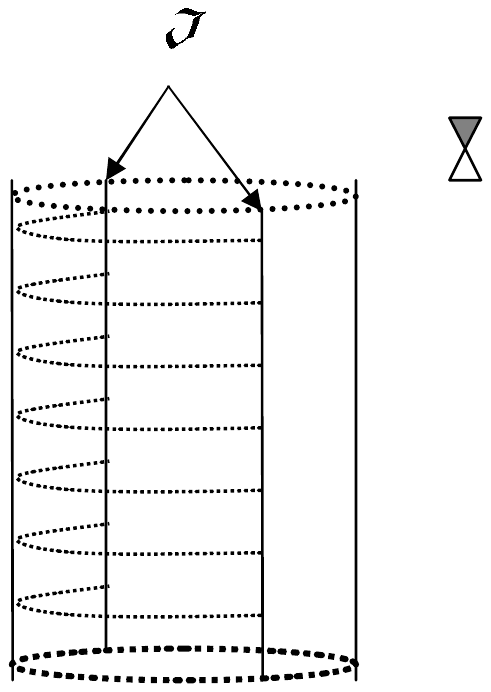}
\caption{\label{fig6}}
\begin{quote}
\begin{quote}
 {\footnotesize $AdS^n$ embedded in the
Einstein static universe. Notice $\scri\approx S^{n-2}\times R$ is
{timelike}. }
\end{quote}
\end{quote}
\end{center}
\end{figure}

Anti de Sitter space $AdS^n$ is the simply connected space form of
constant curvature $C\equiv -1$. It can be realized as the
hyperboloid
\begin{equation}
x_0^2-x_1^2-\ldots -x_{n-1}^2+x_n^2=1
\end{equation}
embedded in semi-Euclidean space $\R^{2,n-1}$. Then by considering
the parametrization
\begin{equation}
x_0=\cosh \rho\cos t\qquad  x_n=\cosh\rho\sin t
\end{equation}
the $AdS^n$ metric takes the form
\begin{equation}
ds^2=-{\cosh}^2\rho dt^2+d\rho^2+{\sinh}^2\rho d\omega^2.
\end{equation}
By further considering the change of variables $\sinh\rho
=\tan\sigma$ we get
\begin{equation}
ds^2=\frac{1}{{\cos}^2\sigma}(-dt^2+d\sigma^2+{\sin}^2\sigma
d\omega^2),
\end{equation}
thus $AdS^n$ is readily seen to admit a conformal boundary in the
Einstein static universe (see figure $\ref{fig6}$ above). Notice
that unlike de Sitter space, $AdS^n$ has a \textit{timelike}
conformal boundary $\scri$.

\section{Spacetimes with timelike boundary}

In order to be able to study in depth those spacetimes whose
structure at infinity is similar to $AdS^n$, we have to gain first
some understanding on the topology and causal structure of
spacetimes having a timelike boundary.  Thus the following
definitions arise naturally:

\begin{defi}
 A \textsl{manifold-with-timelike-boundary} $(M,g)$ is
a Lorentzian manifold-with-boundary in which the pullback
${\iota}^{*}g$, $\iota\colon\partial M\to M$ defines a Lorentzian
metric on $\partial M$.
\end{defi}

\begin{defi}\label{timeor}
A manifold with timelike boundary is \textsl{time-orientable} if
there exists a smooth timelike vector field $X$ defined in all
$M$. A \textsl{spacetime-with-timelike-boundary} is an orientable
and time-orientable manifold with timelike boundary.
\end{defi}

\begin{figure}
\begin{center}
\includegraphics{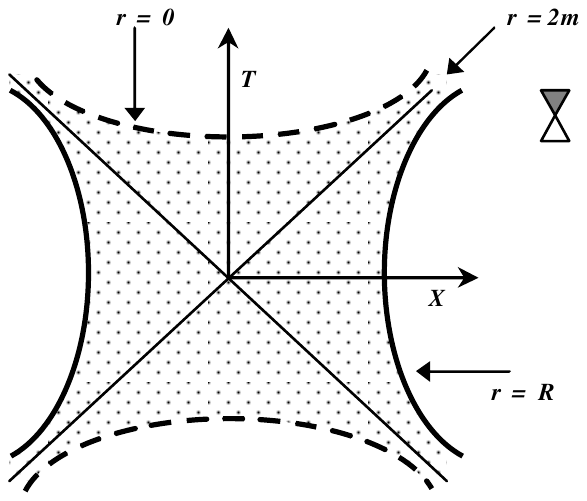}
\caption{\label{krus}}
\begin{quote}
\begin{quote}
 {\footnotesize Extended Scwarzchild spacetime showing the
  region $M=\{X^2-T^2\le C^2\}$, $C=(R/2m-1)\exp (R/2m)$, $R>2m$. Notice that $M$ is a
  spacetime-with-timelike-boundary
  and that $\partial M$ has two components.}
\end{quote}
\end{quote}
\end{center}
\end{figure}

As mentioned earlier, the conformal completion of $AdS^n$ in the
Einstein static universe is a spacetime-with-timelike-boundary.
Another example is provided by the region $X^2-T^2\le C^2$ in
Kruskal coordinates on the extended Schwarzchild spacetime. See
figure $\ref{krus}$.

Notice that since $\partial M$ is not necessarily given by a
defining function, there is no notion of ``infinity" in a
spacetime-with-timelike-boundary.

It is clear from definition $\ref{timeor}$ that $\partial M$ is
time-orientable as a Lorentzian manifold with the time orientation
given by $\tan{X}$, where $X$ is a time orientation for $M$ and
$\tan\colon T_pM\to T_p\partial M$ is the standard projection.
Recall also that an orientation on a smooth
manifold-with-timelike-boundary $M$ gives rise in  a natural way
to an orientation on $\partial M$, the so called Stokes
orientation. Hence we conclude that the boundary $\partial M$ of a
spacetime-with-timelike-boundary $M$ is a spacetime in its own
right.\par

In the remainder of this section we generalize important results
in Lorentzian geometry to the spacetime-with-timelike-boundary
scenario.


\subsection{Causal theory}

\begin{remark}
In order to make proofs simpler, from now on we will always assume
the spacetime-with-timelike-boundary $(M,g)$ is embedded in an
open spacetime $(\overline{M},\overline{g})$ (refer to theorem
\ref{openex}). Also, unless otherwise explicitly stated, all
causal relations and sets are taken with respect to $M$.
\end{remark}

We begin by proving that the relation $I^+$ is {open} in $M$, but
before we do so, let us prove a technical lemma:\Par

\enlargethispage{-.3cm}

\begin{prop}\label{open1}
Let $p\in M$, then $I^+(p)\cap {\partial}M$ is open in
${\partial}M$.
\end{prop}
\Par

\noindent \textit{Proof:} Let $x\in I^+(p)\cap {\partial}M$ and
consider a future pointing timelike curve $\gamma\colon
[0,1]\rightarrow M$ from $p$ to $x$. We first assume that
${\gamma\mid}_{1-\varepsilon}^1\subset\partial M$ for some
$\varepsilon >0$. Let $y=\gamma (1-\varepsilon )$, then $I^+(y,
{\partial}M)$ is an open neighborhood of $x$ in ${\partial}M$ and
$I^+(y, {\partial}M)\subset I^+(p)\cap {\partial}M$, which proves
the claim in this case.\Par

To deal with the general case, we first proceed to deform
${\gamma}$ into a timelike curve in $M$ intersecting ${\partial}M$
only once near $x$. This can be accomplished in the following way:
Let $w\in T_x\overline{M}$ be an inward pointing vector and let
$W$ be its parallel translate along ${\gamma}$. Then there is an
$\varepsilon
>0$ such that $x_0:=\gamma (1-\varepsilon )\in{\rm Int}(M)$ and $W$
is inward pointing whenever ${\gamma\mid}_{1-\varepsilon}^1$
intersects $\partial M$. Consider now the variation $(u,v)\mapsto
\mathbf{x}(u,v)$ of ${\gamma\mid}_{1-\varepsilon}^1$ with
variational vector field $fW$ where $f$ is a smooth bump function
vanishing at endpoints. By construction, for small values of $v$,
the longitudinal curves $u\mapsto\mathbf{x}(u,v)$ remain in $M$,
are timelike and  meet ${\partial}M$ only at the endpoint $x$.\Par

Now, let us consider a $\overline{M}$-neighborhood $\cal U$ of $x$
not containing $x_0$ and a past pointing timelike segment
$\alpha\colon [0,T]\to{\cal U}\cap\partial M$ starting at $x$.
First we extend ${\alpha}^{\prime}$ to a vector field $X\in{\cal
X}(\partial M)$ with $\textrm{supp}(X)\subset{\cal U}\cap\partial
M$, then we further extend $X$ to $\overline{X}\in{\cal
X}(\overline{M})$ with $\textrm{supp}(\overline{X})\subset{\cal
U}$.\Par

By the way $\overline{X}$ was constructed, all of its integral
curves starting at $\partial M$ remain on $\partial M$.  Thus by
uniqueness of ODE's, none of the integral curves with initial
points in ${\rm Int}\, (M)$ can intersect $\partial M$ in a
positive time, hence all these curves remain in ${\rm Int}\,
(M)$.\Par

For $v\in [1-\varepsilon,1]$  let $\varphi_v(t)$ be the image
under the flow $\varphi_{\overline X}$ at time $t$ of the point
$\gamma (v)\in M$, and notice that
${\varphi}_{1-\varepsilon}(t)=x_0$ for all $t$. By continuity of
the metric tensor, there is $t_0\in (0,T)$ such that the curves
$v\mapsto {\varphi}_v(t)$ are timelike for all $t\in [0,t_0]$. Let
$q={\varphi}_1(t_0)$, then $q\in I^+(x_0)$ hence $q\in I^+(p)$.
Notice also $\alpha\vert_x^q$ is a past pointing curve in
$\partial M$. Hence we are back to the first scenario and the
result follows. $\Box$

\begin{prop}\label{open}
The relation $I^+$ is open in $M$.
\end{prop}
\Par

\noindent\textit{Proof:} Let $x\in I^+(p)$. If $x\in {\rm Int}\,
(M)$ the result follows immediately. Thus let $x\in \partial M$,
so by lemma $\ref{open1}$  there is a ${\partial}M$-neighborhood
${\cal W}$ of $x$ such that ${\cal W}\subset I^+(p)$.\Par

Let $v\in T_x\overline{M}$ be a future timelike and inward
pointing vector, and extend it to a vector field $V\in {\cal
X}(\overline{M})$. By continuity we can find a $\partial
M$-neighborhood ${\cal V}$ of $x$ and $T>0$ so that the integral
curves ${\varphi}_y(t)$ of $V$ are future timelike for all
$t\in[0,T]$ and for all $y\in{\cal V}$ . Further, by considering
smaller ${\cal V}$ and $T$ if necessary, we assure $V$ is inward
pointing along ${\cal V}$ and hence ${\varphi}_y(t)\in
\textrm{Int}(M)$ for all $t\in (0,T)$.\Par

Thus, let ${\cal U}={\cal W}\cap{\cal V}$ and ${{\cal
U}}_0=\varphi_V({\cal U}\times [0,T))$. Clearly, ${{\cal U}}_0$ is
open in $M$  and $x\in {{\cal U}}_0$. Finally, the discussion
above shows ${{\cal U}}_0\subset I^+({\cal U})$. On the other
hand, ${\cal U}\subset {\cal W}\subset I^+(p)$, hence ${{\cal
U}}_0\subset I^+(p)$. $\Box$\Par

Notice that since $\partial M$ is a spacetime in its own right the
sets $I^{-}(q)$ are never empty, even when $q\in\partial M$. Thus,
as a consequence of proposition $\ref{open}$, we have that ${\{
I^+(p)\}}_{p\in M}$ is an \textit{open} cover of $M$.\Par

\begin{prop}\label{crono}
Let $p\ll q$ and $q\le r$. Then $p\ll r$.
\end{prop}

\noindent\textit{Proof:} First notice that the general case
reduces to the case in which $r\in\partial M$ and
$q\in\textrm{Int}(M)$. Thus, let $\alpha\colon [0,1]\to M$ be a
future causal curve from $q$ to $r$ and let us analyze first the
particular case in which $\alpha$ meets $\partial M$ only at $r$.
Thus let $w\in T_r\overline{M}$ be a past timelike and interior
pointing vector. Let $W$ be the parallel translate of $w$ along
the reverse of ${\alpha}$, thus by continuity
$g(W(1-t),{\alpha}^{\prime})>0$. Consider now $V(t)=(2-t)W(1-t)$
and observe  that $V(1)=W(0)=w$ and $V^{\prime}=-W(1-t)$, hence
$g(V^{\prime},{\alpha}^{\prime})=-
g(W(1-t),{\alpha}^{\prime})<0$.\Par

Let $(t,s)\mapsto \textbf{x}(t,s)$ be a variation of $\alpha$ with
variational vector field $V$. Then we can apply theorem 10.45 in
\cite{ON} to find $\varepsilon>0$, such that the longitudinal
curves ${\alpha}_s(t)=\textbf{x}(t,s)$ are timelike for all $s\in
(0,\varepsilon ]$, hence $\mathbf{x}(1,s)\in I^+(\mathbf{x}(0,s))$
for all $s\in (0,\varepsilon ]$. Notice as well that since the
final transversal curve $\beta (s)= \textbf{x}(1,s)$ is inward
pointing, we can assume $\textbf{x}(t,s)\in M$ for all $s\in
[0,\varepsilon ]$ and $t\in [0, 1]$. Further, since $w$ is past
pointing, we can also assume that $\beta$ is past timelike on the
interval $[0,\varepsilon]$, hence $r\in I^+(\beta (s))\subset
I^+(\mathbf{x}(0,s))$ for all $s\in (0,\varepsilon ]$.

Since $q\in I^+(p)$ and  $I^+(p)$ is open we know there exists a
$\varepsilon_0\in (0,\varepsilon )$ such that
$\textbf{x}(0,\varepsilon_0 )\in I^+(p)$. Hence by considerations
above we have $r\in I^+(p)$. This finishes the proof of the
special case.\Par

Now let us treat the more general case, thus let assume $A:=\alpha
\cap\partial M$ has at least two points. Further, let $B=A\cap
I^+(p)$ and by a slight abuse of notation, we identify $\alpha
(t)\in A$ with $t\in [0,1]$. Since $A\subset [0,1]$ is compact, it
has a minimum $t_0<1$, thus by the special case we just proved
above we have $t_0\in B$. Then $B\subset [0,1]$ is a non empty
bounded set, so let $T=\sup B$ and notice $T\in A$ since $A$ is
closed. We want to show $T=1$, but first we proceed to show $T\in
B$. To this end, let $v\in T_{\alpha (T)}\overline{M}$ be a past
timelike and inward pointing vector and let $V(s)$, $s\in
[-T,1-T]$ be its parallel translate along $\alpha$. By continuity,
there is a small interval $I$ around $T$ such that $(1+T-s)V(s)$
is past timelike and inward pointing for $s\in I\cap A$. On the
other hand, since $T=\sup B$ it follows $ I\cap B\neq\emptyset$.
If $T\in B$ there is nothing to prove, so let $T_0\in I\cap B$,
$T_0<T$. By the same variation argument used in the previous case
we can construct a future timelike curve with final point $T$ and
initial point arbitrary close to $T_0$. Therefore $T_0\in I^+(p)$
implies $T\in I^+(p)$, hence $T\in B$.

To finish the proof, assume $T<1$. Then since $I^+(p)$ is open and
$T\in I^+(p)$ there exists $\varepsilon >0$ such that
$T+\varepsilon <1$ and $[T,T+\varepsilon )\subset I^+(p)$.
Therefore $t\not\in A$ for $t\in (T,T+\varepsilon)$, that is,
$\alpha (t)\not \in\partial M$ for all $t\in (T,T+\varepsilon )$.
Thus if $t_1\le 1$ denotes the first point in $A$ after $T$, then
$\alpha\vert_{T+\varepsilon /2}^{t_1}$ intersects $\partial M$
only at $t_1$. Hence we are once again back to our special case
with  $q=\alpha (T+\varepsilon /2)$ and $r=\alpha (t_1)$, thus
$t_1\in I^+(T)$, hence $t_1\in I^+(p)$. and therefore $t_1\in B$,
a contradiction. $\Box$\Par

Notice that a similar argument can be used to prove $p\ll r$
whenever $p\le q$ and $q\ll r$. As the first application of
proposition $\ref{crono}$ we have the following result:\Par

\begin{prop}\label{closure}
Let $S\subset M$, then $J^+(S)\subset \overline{I^+(S)}$
\end{prop}

\noindent\textit{Proof:} Let $x\in J^+(S)$. We proceed to
construct a sequence $\{x_n\}$ in $I^+(S)$ converging to $x$.
Thus, let us consider $q\in I^+(x)$ and let $\gamma\colon [0,a]\to
M$ be a timelike segment from $x$ to $q$. Now let $t_n\searrow 0$
be a sequence in $(0,a)$, hence $x_n\to x$ and $x_n\in I^+(x)$,
where $x_n:=\gamma (t_n)$. Take now $p\in S$ with $x\in I^+(p)$,
therefore by proposition $\ref{crono}$ we have $x_n\in I^+(p)$.
$\Box$\Par



\begin{figure}
\begin{center}
\includegraphics{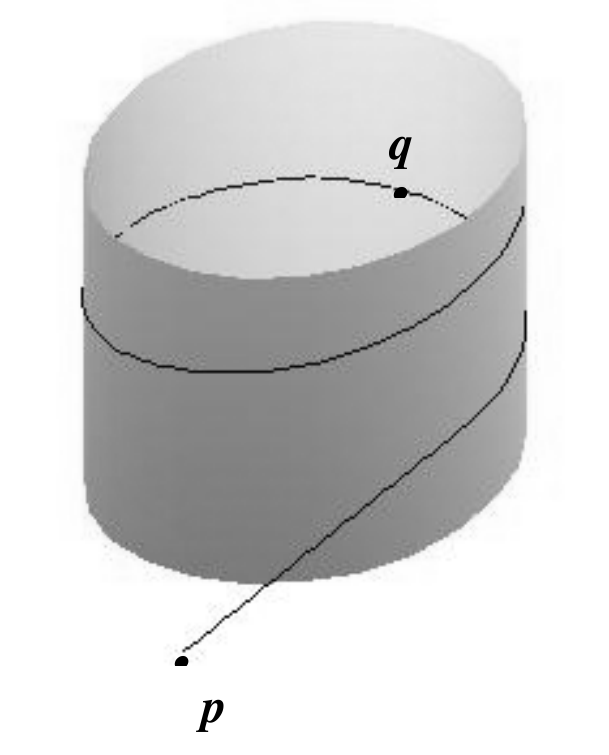}
\caption{\label{fig7}}
\begin{quote}
\begin{quote}
{\footnotesize Let $M$ be Minkowski space $\mathbb{M}^3$ with an
open cylinder removed. Observe that $q\in J^+(p)-I^+(p)$, but no
future null geodesic in $M$ connects $p$ to $q$.}
\end{quote}
\end{quote}
\end{center}
\end{figure}

At this point it is worthwhile noticing that the analog to
proposition $\ref{ngeo1}$ does not hold on the
spacetime-with-timelike-boundary case, as the example depicted in
figure $\ref{fig7}$ shows. However, we do get a statement in the
spirit of proposition $\ref{ngeo1}$ when our causal curve touches
$\partial M$ only at its endpoints.

\begin{prop}\label{geo1}
Let  $\gamma\colon [0,1]\to M$ be a future causal curve joining
$p$ to $q$ such that $\gamma (0,1)\subset\mbox{ Int}(M)$. Then
either $q\in I^+(p)$ or $\gamma$ is a smooth null geodesic.\Par
\end{prop}

\noindent\textit{Proof:} Let us assume $q\not\in I^+(p)$. Let
$t_n\searrow 0$ and $s_n\nearrow 1$ be two sequences on $[0,1]$
and let us define $p_n=\gamma (t_n)$, $q_n=\gamma (s_n)$. By
proposition $\ref{crono}$ and our assumption we have that $q_n\in
J^+(p_n)-I^+(p_n)$. Thus by proposition $\ref{ngeo1}$ we have
${\gamma\vert}_{p_n}^{q_n}\subset \textrm{Int}(M)$ is a null
geodesic segment. It follows that $\gamma\vert_{(0,1)}$ is a null
geodesic. Finally, consider a spacetime extension $\overline{M}$
and notice $\gamma$ continuously extends to its endpoints in
$\overline{M}$, then by proposition $\ref{geoext}$ $\gamma$ is a
geodesic in $\overline{M}$, hence it is a geodesic in $M$ as well.
$\Box$\Par

We can also generalize lemma $10.50$ in $\cite{ON}$ to the
spacetime-with-timelike-boundary context.

\begin{prop}\label{normal}
Let $S$ be a spacelike submanifold of $\partial M$ and
$\gamma\colon I\to M$ a causal curve joining $p$ to $q\in S$ that
intersects $\partial M$ only at $q$. Then either there is a
timelike curve from $p$ to $S$ or $\gamma$ is a null geodesic
meeting $S$ orthogonally.
\end{prop}

\noindent\textit{Proof:} By the previous proposition, $\gamma$
must be a null geodesic or otherwise $q\in I^+(p)$, hence we only
need to prove the statement about the orthogonality of $\gamma$.
Assume then that $\gamma$ is a null geodesic not normal to $S$. We
will show $S\cap I^+(p)\neq\emptyset$.\Par\vspace{.3cm}

Since $S\subset\partial M$ is spacelike, there is a neighborhood
$U^{\prime}$ of $q$ in $\partial M$ such that $S\cap U^{\prime}$
is acausal  and closed in $U^{\prime}$. Then $U:=D(S\cap
U^{\prime},U^{\prime})$ is a globally hyperbolic neighborhood of
$q$ in $\partial M$. Now, by means of the normal exponential map
we can extend $S\cap U$ to a spacelike hypersurface $S_0$ in
$\overline{M}$. Consider now a $\overline{M}$-neighborhood ${{\cal
V}_0}$ of $q$ such that ${{\cal V}_0}\cap\partial M\subset U$.
Since $S_0\cap{{\cal V}_0}\subset{{\cal V}_0}$ is spacelike, there
is a $\overline{M}$-neighborhood ${{\cal U}_0}\subset{{\cal V}}_0$
of $q$ such that $S_0\cap {{\cal U}_0}$ is acausal in ${{\cal
U}_0}$. As we did before, by replacing ${{\cal U}_0}$ by
$D(S_0\cap{{\cal U}_0},{{\cal U}_0})$ we can assume ${{\cal U}_0}$
is a globally hyperbolic $\overline{M}$-neighborhood of $q$.
\Par\vspace{.3cm}

Let us prove now $J^+(S_0\cap{{\cal U}_0},{{\cal
U}_0})\cap\partial M\subset I^+(S\cap U,\partial M)\cup S$. Let
$x\in J^+(S_0\cap{{\cal U}_0},{{\cal U}_0})\cap\partial M$, then
if $x\in S$ the claim follows immediately, so let us assume $x\not
\in S$ and let $\alpha\colon I\to U$ be a future directed and
inextendible timelike curve passing through $x$. Since $x\in U-S$
then by definition of domain of dependence, $\alpha$ must
intersect $S\cap U$ exactly once, say at $y$. Assume $y$ comes
strictly after $x$ along $\alpha$, then $\alpha$ must have left
${{\cal U}_0}$ at a point before $y$, otherwise the acausality of
$S_0\cap{{\cal U}_0}$ in ${{\cal U}_0}$ would be violated. Hence
$\alpha$ -considered as an inextendible curve in ${{\cal U}_0}$-
must intersect $S_0\cap {{\cal U}_0}$ to the past of $x$, say at
$z\in S\cap U$, thus $z\neq y$. But then $\alpha\vert_z^y$ would
be a future timelike curve in $U$ joining two points of $S\cap U$,
contradicting the acausality of $S\cap U$ in $U$. Therefore $y$
has to come before $x$ along $\alpha$, i.e $x\in J^+(y,U)$ and the
claim follows. Similarly we can show the time dual
$J^-(S_0\cap{{\cal U}_0},{{\cal U}_0})\cap\partial M\subset
I^-(S\cap U,\partial M)\cup S$.\Par

\eject

Thus let  $p^{\prime}\in\gamma\cap{{\cal U}_0}$, $p^{\prime}\neq
q$ be such that $\gamma\vert_{p^{\prime}}^q\subset {{\cal U}_0}$.
Hence, proposition $10.48$ of \cite{ON} applied to
$\gamma\vert_{p^{\prime}}^q$ and $S\cap{{\cal U}_0}$ assures the
existence of a future timelike curve $\overline{\sigma}$ in
${{\cal U}_0}$ joining $p^{\prime}$ to $q^{\prime}\in S\cap{{\cal
U}_0}$. For simplicity, assume $\overline{\sigma}$ intersects $S$
only at $q^{\prime}$.\Par

If $\overline{\sigma}$ stays in $M$ we are done. Otherwise
$\overline{\sigma}$ has to intersect $\partial M$ for the first
time at a point $r$ strictly before $q^{\prime}$. Notice $r\not\in
J^+(S_0\cap{{\cal U}_0},{{\cal U}_0})$ since $S_0\cap{{\cal U}_0}$
is acausal in ${{\cal U}_0}$, hence $r\in J^-(S_0\cap{{\cal
U}_0},{{\cal U}_0})$ and hence $r\in I^-(S\cap U,U)\cup S$ by the
claim above, hence the result follows in this scenario as well.
$\Box$\Par

With similar techniques we can prove  the following analogue  of
proposition $\ref{space2focal}$:\Par

\begin{prop}\label{focal}
Let $S$ be a spacelike submanifold of $\partial M$ and
$\gamma\colon I\to M$ a future causal curve joining  $p$ to $q\in
S$ that intersects $\partial M$ only at $q$. If $I^+(p)\cap
S=\emptyset$ then $\gamma$ is a null geodesic orthogonal to $S$ at
$q$ with no focal points of $S$ strictly after $p$.
\end{prop}\Par

Now we turn our attention to the study of the Lorentzian distance
function $\tau$. Recall that $\tau$ is known to be lower
semicontinuous for open spacetimes. The same result holds in the
spacetime-with-timelike-boundary case.

\begin{prop}\label{lower}
The Lorentzian distance $\tau$ is lower semicontinuous on $M$.
\end{prop}

\noindent\textit{Proof:} Let $p,q\in M$. If $\tau (p,q)=0$ the
result holds trivially, so let us assume $\tau (p,q)>0$. By
proposition $\ref{crono}$, this in turn implies $q\in I^+(p)$.\Par

Let $\varepsilon >0$ and suppose $\tau (p,q)<\infty$, then we want
to find neighborhoods ${\cal U}$ and ${\cal V}$ around $p$ and $q$
such that $\tau (x,y)\ge \tau (p,q)-\varepsilon$ for all
$x\in{\cal U}$ and $y\in{\cal U}$. Thus consider a future pointing
timelike curve $\gamma$ from $p$ to $q$ with $L(\gamma )>\tau
(p,q)-\varepsilon /3$.

Since the  the function $t\to L(\gamma\vert_0^t)$ is continuous
there is $a\in\gamma$, $a\neq p$ such that
$L({\gamma\vert}_p^a)<\varepsilon /3$. Similarly, we can find
$b\in\gamma$, $b\neq q$ with $L({\gamma\vert}_b^q)<\varepsilon
/3$.\Par

Let ${\cal U}=I^-(a)$ and ${\cal V}=I^+(b)$, then given $x\in
{\cal U}$ there is a future timelike curve $\alpha$ from $x$ to
$a$. Furthermore, such $\alpha$ trivially satisfies
$L({\alpha})>L({\gamma\vert}_p^a)-\varepsilon /3$. Likewise, given
$y\in{\cal V}$ there is a future timelike curve from $b$ to $y$
with $L({\beta})>L({\gamma\vert}_b^q)-\varepsilon /3$.

By concatenating $\alpha$, ${\gamma\vert}_a^b$ and $\beta$ we
obtain a future timelike curve ${\gamma}_0$ from $x$ to $y$ with
\begin{equation}
L({\gamma}_0)>L(\gamma )-2\varepsilon /3>\tau (p,q)-\varepsilon
\end{equation}
hence $\tau (x,y)>\tau (p,q)-\varepsilon$ for all $x\in {\cal U}$
and all $y\in{\cal V}$. \Par

If $\tau (p,q)=\infty$ then there are timelike curves of arbitrary
large length joining $p$ to $q$. Thus given $A>0$ there is a
causal curve from $p$ to $q$ with $L(\gamma )>A+2$. Following the
same procedure as above we find neighborhoods ${\cal U}$ and
${\cal V}$ around $p$ and $q$ such that $\tau (x,y)>A$, and the
result follows. $\Box$


\subsection{Global Hyperbolicity}

Continuing our program, now we extend to the
spacetime-with-boundary context the concept of global
hyperbolicity.

\begin{defi}
A set $B\subset M$ is causally convex if for any pair $x,y\in B$
we have that the condition $p\in J^+(x)\cap J^-(y)$ implies $p\in
B$. Thus, any causal curve with endpoints in $B$ stays in $B$.
\end{defi}

\begin{defi}
We say $M$ is strongly causal at $p\in M$ if $p$ has arbitrarily
small causally convex neighborhoods; that is, given a neighborhood
${\cal U}$ of $p$, there exists a causally convex neighborhood
${\cal V}$ of $p$ contained in ${\cal U}$. $M$ is said to be
strongly causal if it is strongly causal at any of its points.
\end{defi}

\eject

\begin{defi}
 A spacetime-with-timelike-boundary $(M,g)$ is said to be
globally hyperbolic if strong causality  holds on $M$ and the sets
$J^+(p)\cap J^-(q)$ are compact for all $p,q\in M$.
\end{defi}\Par

We first prove that the boundary of a globally hyperbolic
spacetime-with-timelike-boundary inherits this property.\Par

\begin{prop}\label{partialgh}
Let $(M,g)$ be a globally hyperbolic
spacetime-with-timelike-boundary. Then $\partial M$ is a globally
hyperbolic spacetime.
\end{prop}\Par

\noindent \textit{Proof:} It is clear that $\partial M$ is
strongly causal, so only the compactness property needs to be
shown. Thus, let $p,q\in\partial M$, then by lemma 4.29 in
\cite{BE} it suffices to show $A=J^+(p,\partial M)\cap J^-(q,
\partial M)$ has compact closure in $\partial M$. \Par

Notice that $J^+(p,\partial M)\subset J^+(p)\cap \partial M$. By
global hyperbolicity, $J^+(p)\cap J^-(q)$ is compact, hence closed
in $M$. Thus ${J^+(p)\cap J^-(q)}\cap\partial M$ is closed as
well. Then
\begin{equation}
\overline{J^+(p,\partial M)\cap J^-(q,\partial
M)}\subset{J^+(p)\cap J^-(q)}\cap\partial M
\end{equation}
where the upper bar indicates the closure with respect to $M$.
Therefore
\begin{equation}
B=\overline{J^+(p,\partial M)\cap J^-(q,\partial M)}
\end{equation} is compact, being a closed subset of the compact
set ${J^+(p)\cap J^-(q)}$.\Par

Further, the closure of $A$ in $\partial M$ is just the
intersection of the compact set $B$ and the closed set $\partial
M$, hence it is compact as well. The proof is complete. $\Box$\Par

\eject

\begin{prop}
Let $M$ be globally hyperbolic and $p\in M$, then  $J^+(p)$ is
closed.
\end{prop}
\Par

\noindent \textit{Proof:} Let $x\in \overline{J^+(p)}$ and
consider $q\in I^+(x)$. Let $\gamma$ be a future directed timelike
curve from $x$ to $q$, and let $\{ x_n\}$ be a sequence in
$\gamma$ such that $x_{n+1}\ll x_n$ and $x_n\to x$. Thus $x_n\in
J^+(p)\cap I^-(q)\subset J^+(p)\cap J^-(q)$ and hence $x\in
\overline{J^+(p)\cap J^-(q)}$. Since $M$ is globally hyperbolic,
the set $J^+(p)\cap J^-(q)$ is compact, hence closed, thus
$\overline{J^+(p)\cap J^-(q)}={J^+(p)\cap J^-(q)}$. Therefore
$x\in{J^+(p)\cap J^-(q)}\subset J^+(p)$. $\Box$\Par

\begin{prop}\label{causimple}
Let $M$ be globally hyperbolic, then $M$ is causally simple, i.e.
$J^+(A)$ is closed for all compact $A$.
\end{prop}
\Par

\enlargethispage{.5cm}

\noindent \textit{Proof:} Let $A\subset M$ be compact and let
$x\in J^+(A)$. Just as in the previous result, let $\{x_n\}$ be a
sequence in $I^+(x)$ converging to $x$ with $x_{n+1}\ll x_n$,
hence $J^-(x_{n+1})\subset J^-(x_n)$. Further, since $J^-(x_n) $
is closed and $A$ is compact, then $J^-(x_n)\cap A$ is compact.
Notice also that $x\in J^-(x_n)\cap A$ for all $n$, therefore $\{
J^-(x_n)\cap A\}$ is a nested sequence of nonempty compact sets.
Thus by a standard topological result $\bigcap_n (J^-(x_n)\cap
A)\neq\emptyset$, so let $p\in \bigcap_n (J^-(x_n)\cap A)$. It
follows that $x_n\in J^+(p)$ for all $n$. By the previous result
$J^+(p)$ is closed, so we have $x\in J^+(p)$ as desired.
$\Box$\Par

\begin{prop}\label{causimple1}
Let $M$ be globally hyperbolic, then $J^+(A)\cap J^-(B)$ is
compact for any pair of compact sets $A,B\subset M$.
\end{prop}

\noindent\textit{Proof:} First notice that by the previous result
$J^+(A)\cap J^-(B)$ is closed. Now, for each $a\in A$ let
$a^{\prime}\in M$ be a point in the chronological past of $a$.
Then $A\subset \bigcup I^+(a^{\prime})$. Since $A$ is compact and
each set $I^+(a^{\prime})$ is open, there are finitely many
$a^{\prime}_i$ such that $A\subset \bigcup_iI^+(a^{\prime}_i)$.
Likewise, there exist finitely many $b^{\prime}_j$ with $B\subset
\bigcup_jI^+(b^{\prime}_j)$. Thus $J^+(A)\cap J^-(B)\subset
\bigcup_{i,j}J^+(a^{\prime}_i)\cap J^-(b^{\prime}_j)$. Each set
$J^+(a^{\prime}_i)\cap J^-(b^{\prime}_j)$ is compact, thus
$J^+(A)\cap J^-(B)$ is a closed subset of a compact set, hence it
is compact. $\Box$\Par


\subsection{Limit of curves}

In this section we analyze how causal curves arise as limits of
causal curves in globally hyperbolic spacetimes. Our discussion
will follow the lines of chapter 3 in \cite{BE}. \Par

Since a limit of smooth curves need not be smooth, we need to
extend our definition of causality to continuous curves.\Par

\begin{defi}
A continuous curve $\gamma\colon I\to M$ on a strongly causal
spacetime-with-timelike-boundary $M$ is said to be future timelike
if for any $t_0\in I$ there exists a $\overline{M}$-convex
neighborhood ${{\cal U}}_0$ around $\gamma (t_0)$ and an interval
$[a,b]\subset I$ around $t_0$ such that for all $t,s\in [a,b]$
with $t\le t_0\le s$ we have $\gamma (s)\in I^+(\gamma (t),{{\cal
U}}_0\cap M)$.
\end{defi}

Obviously, replacing $I^+$ by $J^+$ in the previous definition
gives us the notion of future causal continuous curve. The time
dual concepts are defined in a similar way.\Par

\enlargethispage{.6cm}

\begin{remark}
Because of strong causality this concept is well defined, i.e.
this definition actually does not depend on the extension
$\overline{M}$.
\end{remark}
\begin{remark}
Causal relations are preserved. In other words, $q\in I^+(p)$ if
and only if there exists a \textit{continuous} future timelike
curve from $p$ to $q$. A similar statement holds for $J^+(p)$.
\end{remark}

\begin{defi}
Let $\{\gamma_n\}$ be a sequence of curves on $M$. A curve
$\gamma$ is a limit of $\{\gamma_n\}$ if there is a subsequence
$\{\gamma_m\}$ such that for all $p\in\gamma$, each neighborhood
of $p$ intersects all but finitely many of the $\gamma_m$'s. Such
subsequence is said to distinguish $\gamma$.
\end{defi}

As can be shown, a given sequence $\{\gamma_n\}$ may have many
different limit curves, or no limit curve at all. However, in the
open spacetime context the existence of such a limit curve is
guaranteed when $\{\gamma_n\}$ has an accumulation point (for a
proof, consult proposition $3.31$ in \cite{BE}). A similar result
holds in the spacetime-with-timelike-boundary case.\Par

\begin{lemma}\label{curvelemma0}
Let $\{\gamma_n\}$ be a sequence of future inextendible causal
curves in $M$ having an accumulation point $p$, then there exists
a limit curve $\gamma$ of $\{\gamma_n\}$. Further, $\gamma$ is
future inextendible causal and $p\in\gamma$.
\end{lemma}

\noindent \textit{Proof:} Consider an extension $\overline{M}$ and
apply the aforementioned Limit Curve Lemma to the sequence
$\{\gamma_n\}$ to obtain a future inextendible limit curve
$\gamma\colon \R\to \overline{M}$ passing through $p$. Finally,
since $M$ is a closed subset of $\overline{M}$ and each point on
$\gamma$ is a limit point of a sequence in $M$ we have
$\gamma\subset M$. $\Box$

The following is a straightforward modification of the analogous
results in \cite{BE} to fit in the
spacetime-with-timelike-boundary case.\Par

\begin{prop}\label{curvelemma}
Let $M$ a be globally hyperbolic spacetime-with-timelike-boundary.
Let $\{p_n\}$ and $\{q_n\}$ be two sequences converging to $p$ and
$q$, respectively with $p\neq q$. Further assume that $q_n\in
J^+(p_n)$ for all $n$. If $\{{\gamma}_n\}$ is a sequence of future
directed causal curves from $p_n$ to $q_n$, then $\{{\gamma}_n\}$
has a future directed causal limit curve $\gamma$ from $p$ to $q$.
\end{prop}

\Par

\noindent\textit{Proof:} First notice $q\in J^+(p)$. Suppose
otherwise, then since $J^+(p)$ is closed there is a neighborhood
${\cal U}$ of $q$ contained in $M-J^+(p)$. Let $a,b\in \cal{U}$
such that $a\in I^+(q)$ and $b\in I^+(a)$. Since $I^-(a)$ is open,
$q\in I^-(a)$ and $q_n\to q$ we have $q_n\in I^-(a)$ for all $n\ge
N$, hence $p_n\in J^-(a)$, $n\ge N$ as well. But, since $J^-(a)$
is closed and $p_n\to p$ we must have $p\in J^-(a)$ therefore
$b\in J^+(p)$, a contradiction.\Par

Consider now an extension $(\overline{M},\overline{g})$ of
$(M,g)$. For each $x\in M$ let ${{\cal U}}^{\prime}_x$ be a
precompact $\overline{M}$-convex neighborhood around it. Due to
strong causality there is a causally convex neighborhood ${\cal
U}_x$ around $x$ satisfying ${{\cal U}}_x\subset{{\cal
U}}^{\prime}_x\cap M$. Cover $J^+(p)\cap J^-(q)$ with a collection
$\{{\cal U}_{\alpha}\}$ of such neighborhoods. Then by compactness
$J^+(p)\cap J^-(q)$ can be covered by finitely many of them, say
$\{{\cal U}_i\}$ $1\le i\le n$.\Par

Let $h$ be a complete Riemannian metric on $\overline{M}$. Every
smooth curve $\sigma$ on a normal neighborhood ${\cal U}^{\prime}$
satisfies a Lipschitz condition, thus $h$-arc length is bounded in
any precompact convex neighborhood ${\cal U}^{\prime}$. Let
$B_i>0$ be such that $L_h(\sigma )\le B_i$ for all causal
$\sigma\colon I\to \overline{\cal U}_i$ and set $B=\sum B_i$,
$U=\bigcup_i {\cal U}_i$. Since each ${\cal U}_i$ is causally
convex we have $L_h(\sigma )\le B$ for all causal $\sigma\colon
I\to U$.\Par

Let ${\gamma}_n\colon [0,a_n]\to M$ be a future causal curve from
$p_n$ to $q_n$ parameterized with respect to $h$-arc length and
let $\overline{{\gamma}_n}\colon [0,\infty )\to \overline{M}$ be a
future inextendible extension of it. By applying the limit curve
lemma to $\overline{{\gamma}_n}$ with accumulation point $p$ we
get a (continuous) causal curve $\gamma\colon [0,\infty
)\to\overline{M}$ with $\gamma (0)=p$ and a subsequence
$\{\overline{{\gamma}_m}\}$ of $\{\overline{{\gamma}_n}\}$
converging uniformly to $\gamma$ in compact subsets of $[0,\infty
)$.\Par

First notice that $\gamma$ passes through $q$:  Since
$\gamma_m\subset U$ we have $0< a_m=L(\gamma_m)\le B$. Thus
$q_m\to q$ together with uniform convergence on $[0, B]$ yield
$q=\lim_m q_m=\lim_m{\gamma}_m (a_m)=\gamma (a)$ for some $a\in
(0,B]$. $\Box$\Par

In order to relate causality and the Lorentzian distance function,
we need to consider a  kind of convergence of curves that
guarantees favorable properties of the arc length functional.\Par

We topologize the space ${\cal C}(M)$ of all continuous causal
curves in $M$ in the following way: let ${\cal U,V,M}$ be open
sets in $M$, and let us denote by ${\cal C}_{\cal M}({\cal U},
{\cal V})$ the set of all continuous future causal curves
contained in ${\cal M}$ whose initial points lie ${\cal U}$ and
its final points lie in ${\cal V}$. The collection of all such
sets form a basis for a topology in ${\cal C}(M)$ $\cite{P}$.\Par

\begin{defi}
The topology on ${\cal C}(M)$ defined above is called the $C^0$
topology of curves in $M$.
\end{defi}

The two different types of convergence we have been discussing are
closely related as the following result shows. For a proof, refer
to proposition  in $\cite{BE}$ and apply the same type of
modifications we used in propositions $\ref{curvelemma0}$ and
$\ref{curvelemma}$. \Par

\begin{prop}\label{C0}
Let $M$ be a strongly causal and assume $\gamma\colon [a,b]\to M$
is a limit curve of the sequence of causal curves $\{\gamma_n\}$.
If $\gamma_n(a)\to \gamma (a)$ and $\gamma_n(b)\to \gamma (b)$
then there is a subsequence $\{\gamma_m\}$ that converges to
$\gamma$ on the $C^0$ topology of curves.
\end{prop}

To extend the notion of length to continuous causal curves we
first observe that any smooth causal curve can be approximated in
the $C^0$ topology of curves by a broken causal geodesic, thus its
length can be viewed as the limit of the lengths of such broken
geodesics.\Par

To be more precise, let $\gamma\colon I\to M$ be a future
continuous causal curve from $p$ to $q$ and consider a collection
of points $\Upsilon:=\{x_i\}$, $0\le i\le k$ on $\gamma$ such that
$x_0=p$, $x_k=q$ and there exist convex neighborhoods
$\{\overline{\cal U}i\}$ in an extension $\overline{M}$ with the
property that there is a geodesic segment
$\gamma_i\subset\overline{\cal U}_i$ joining $x_i$ to $x_{i+1}$.
If we denote by $\gamma_{\Upsilon}$ the concatenation of the
$\gamma_i$'s then we define the length of $\gamma$ by
\begin{equation}
L(\gamma )=\inf_{\Upsilon} L(\gamma_{\Upsilon}). \end{equation}
Notice as well that the infimum on the right hand side is
independent of the choice of the extending spacetime
$\overline{M}$.\Par

It is known that the arc length functional is upper semicontinuous
in the $C^0$ topology of curves in strongly causal spacetimes
(\cite{BE,Wald}). Only slight changes on the proof of theorem
$7.5$ in $\cite{P}$ are needed to prove the analogous result for
spacetimes-with-timelike-boundary.\Par

\begin{prop}\label{uppersemi}
The arc length functional $L$ is continuous on strongly causal
spacetimes-with-timelike-boundary. That is, for any sequence of
causal curves $\{\gamma_n\}$ converging to $\gamma$ in the $C^0$
topology of curves we have $L(\gamma )\ge\limsup L(\gamma_n)$.
\end{prop}

We finish this section with a couple of useful results regarding
the Lorentzian distance function and maximal curves in globally
hyperbolic spacetimes-with-timelike-boundary.\Par

\begin{prop}\label{maxldist}
Let $M$ be a globally hyperbolic spacetime-with-timelike-boundary,
$S\subset\partial M$ be a compact submanifold  and  $p\in M$ such
that $p\in J^+(S)$. Then there exists a causal curve $\gamma$ from
$S$ to $p$ such that $\tau(S,p)=L(\gamma )$.
\end{prop}

\noindent\textit{Proof:} For each $n\in\N$ there is $\gamma_n$
from $S$ to $p$ such that $\tau(S,p)-1/n<L(\gamma_n)$. Since $S$
is compact, the sequence $\{\gamma_n(0)\}$ has a subsequence
$\{\gamma_m(0)\}$ converging to $x\in S$.  By proposition
$\ref{curvelemma}$ there is a causal limit curve $\gamma$ of the
sequence $\{\gamma_m\}$ joining $x$ and $p$. Further, in light of
proposition $\ref{C0}$ we may assume $\{\gamma_m\}$ converges to
$\gamma$ in the $C^0$ topology of curves. Thus by the upper
semicontinuity of $L$ we have
\begin{equation}
{\tau}(S,p)\le\limsup L(\gamma_m)\le L(\gamma).
\end{equation}
It follows $\tau(S,p)=L(\gamma)$. $\Box$\Par

\begin{prop}
Let $M$ be a globally hyperbolic spacetime-with-timelike-boundary.
Then $\tau$ is continuous on $M$.
\end{prop}

\noindent\textit{Proof:} We know already $\tau$ is lower
semicontinuous (proposition $\ref{lower}$), thus we only need to
check $\tau$ is upper semicontinuous.

We proceed by contradiction. Let $p,q\in M$ and $\{p_n\}$,
$\{q_n\}$ be sequences in $M$ converging to $p$ and $q$
respectively, and assume there is  $\varepsilon
>0$  such that $\tau (p_n,q_n)\ge \tau (p,q)+\varepsilon$ for all
$n$. Thus $\tau (p_n,q_n)>0$ and hence $q_n\in J^+(p_n)$.
Therefore, by proposition $\ref{maxldist}$ above, there exist a
future causal curve $\gamma_n$ from $p_n$ to $q_n$ such that
$L(\gamma_n )=\tau (p_n,q_n)$. By propositions $\ref{curvelemma}$
and $\ref{uppersemi}$ there is a future causal curve $\gamma$ from
$p$ to $q$ satisfying $L(\gamma )\ge \limsup L(\gamma_n )$. Then
we find
\begin{equation}
L(\gamma )\ge \limsup L(\gamma_n ) =\limsup \tau (p_n,q_n)\ge \tau
(p,q)+\varepsilon
\end{equation}
which is a clear contradiction. $\Box$



\subsection{Covering Spacetimes}

Our main goal in this section will be to construct a globally
hyperbolic spacetime-with-timelike-boundary $\pi\colon\hat{M}\to
M$ having an isometric copy of a fixed component of $\partial M$.
In order to attain this goal, we need to prove a monodromy type
lemma first:\Par

\begin{lemma}\label{monodromy}
Let $\pi\colon \hat{M}\to M$ a topological covering map of a
globally hyperbolic spacetime-with-timelike-boundary $M$. Let
$\gamma_n\colon [0,t_n]\to \hat{M}$ be a sequence of curves
satisfying the following properties:

\noindent 1. The sequence $\{\gamma_n (0)\}$ converges to
$x\in\hat{M}$.

\noindent 2. The curves $\alpha_n:=\pi\circ\gamma_n$ are future
causal.

\noindent 3. The sequence $\{\alpha_n(t_n)\}$ converges to $q\in
M$, $q\neq \pi (x)$.

\noindent Then  a subsequence $\{\gamma_m(t_m)\}$ converges to
$y\in \pi^{-1}(q)$
\end{lemma}

\noindent \textit{Proof:} Let us fix some notation: let
$x_n=\gamma_{n}(0)$, $y_n=\gamma_n(t_n)$, $p_n=\alpha_n(0)$,
$q_n=\alpha_n(t_n)$ and $p=\pi (x)$.\Par

Let ${\cal U}$ be an evenly covered neighborhood around $p$ and
let $\hat{\cal U}$ be the unique component of $\pi^{-1}({\cal U})$
containing $x$. Since  $\pi\vert_{\hat{\cal U}}\colon\hat{\cal
U}\to {\cal U}$ is a homeomorphism and $x_n\to x$ we have $p_n\to
p$,  thus by proposition $\ref{curvelemma}$ there is a causal
limit curve $\alpha\colon [0,b]\to M$ of the sequence
$\{\alpha_n\}$ joining $p$ to $q$. Let $\gamma$ be the unique lift
of $\alpha$ to $\hat{M}$ with base point $x$. We shall show below
that  a subsequence of $\{y_n\}$ converges to $y=\gamma (b)$. \Par

Next let us consider an evenly covered neighborhood ${\cal U}_s$
around each $\alpha (s)$ and take ${\cal V}_s\subset{\cal U}_s$
causally convex with $\alpha (s)\in{\cal V}_s$. By compactness,
there will be finitely many such ${\cal V}_i$, $i=1,\ldots ,k$
covering $\alpha$. Further, we can choose ${\cal V}_i$ in such a
way that $p\in{\cal V}_1$, $q\in{\cal V}_k$ and ${\cal V}_i$ only
intersects ${\cal V}_{i-1}$ and ${\cal V}_{i+1}$. Now, let
$a_0=p$, $a_k=q$ and $a_i\in \alpha\cap{\cal V}_i\cap{\cal
V}_{i+1}$ for each $1\le i\le k-1$. Notice
${\pi}^{-1}(a_i)\cap\gamma$ consist of a single point, since
$\alpha$ would contain closed loops otherwise. Hence let us define
 $b_i:=\pi^{-1}(a_i)\cap\gamma$ and let $\hat{\cal U}_i$ be the
unique component of $\pi^{-1}({\cal U}_i)$ containing $b_{i-1}$.
\Par

Now we proceed by induction on $i$. As our induction hypothesis,
we'll assume there is a subsequence $\{{\alpha}_{i-1,m}\}$ of
$\{\alpha_n\}$ converging to $\alpha$ and points
$b_{i-1,m}\in{\gamma}_{i-1,m}$ such that $b_{i-1,m}\to b_{i-1}$.

Since ${\cal V}_i\cap {\cal V}_{i+1}$ is open and $\alpha$ is a
limit curve of $\alpha_{i,m}$, there is a subsequence
$\{a_{i,m}\}\subset {\cal V}_i\cap {\cal V}_{i+1}$ converging to $
a_i$ such that $a_{i,m}\in \alpha_{i,m}$. By further taking a
subsequence we may assume $a_{i,m}\in J^+(a_{i-1,m})$ since
otherwise strong causality at $a_{i-1}$ would not hold. Thus since
${\cal V}_i$ is causally convex we have that the entire segment of
$\alpha_{i,m}$ from $a_{i-1,m}$ to $a_{i,m}$ is contained in
${\cal V}_i$, hence in ${\cal U}_i$.\Par

Since $b_{i-1,m}\to b_{i-1}$ then $\hat{\cal U}_i$ will contain
$b_{i-1,m}$ for all $m$ large. Since $\pi\vert_{\hat{\cal U}_i}$
is a homeomorphism and $\gamma_{i,m}$ is the unique lift of
$\alpha_{i,m}$ with base point $x_{i,m}$ we conclude
$b_{i,m}:=\pi^{-1}(a_{i,m})\cap\gamma_{i,m}\in\hat{\cal U}_i$ and
$b_{i,m}\to b_i$.\Par

We can now repeat this process on each $a_i$ and at the end,
causality and the unique lifting property will guarantee
$b_{k,m}=y_{k,m}$ and $b_k=b$ as claimed. $\Box$\Par

\begin{remark}
As can be seen from the proof, this result holds for open
spacetimes as well.
\end{remark}

\begin{remark}
 Notice that if $\hat{M}$ happens to be a spacetime in which
$\pi$ is a local isometry, we have that the curve $\gamma$ joining
$x$ to $y$ is causal, thus $y\in J^+(x)$.
\end{remark}

\begin{prop}\label{cover}
Let $M$ be a globally hyperbolic spacetime-with-timelike-boundary
and  let $\pi\colon\hat{M}\to M$ be a (topological) covering map.
Then there is a smooth structure on $\hat{M}$ in which $\hat{M}$
is a globally hyperbolic spacetime-with-timelike-boundary and
$\pi$ is a local isometry.
\end{prop}

\noindent \textit{Proof:} Let $p\in M$ and consider
$\hat{p}\in\pi^{-1}(p)$. Let $\hat{\cal U}$ be a neighborhood of
$\hat{p}$ such that $\pi{\vert}_{\hat{\cal U}}$ is a homeomorphism
and consider a coordinate chart $({\cal V},\phi )$ around $p$.
Then it is clear that the collection $\{ (\hat{\cal U}\cap
\pi^{-1}({\cal V}),\phi\circ\pi\}$ is a smooth atlas for
$\hat{M}$. Notice as well that $\pi^{-1}(\partial M)=\partial
\hat{M}$.\Par

By taking the pullback metric $\pi^*g$ we turn $\hat{M}$ into a
lorentzian manifold in which $\pi$ is a local isometry, hence
$\partial \hat{M}$ is timelike. The orientation and time
orientation of $M$ lift as well to make $\hat{M}$ a
spacetime-with-timelike-boundary. \Par

We now prove $\hat{M}$ is strongly causal. Thus, let $x\in
\hat{M}$ and $\hat{\cal U}$ a neighborhood containing $x$. Let
$p=\pi (x)$ and choose  a neighborhood ${\cal V}$ of $p$ such that
${\cal V}\subset \pi(\hat{\cal U})$ and it is evenly covered by
$\pi$. By further reducing ${\cal V}$ if necessary, we can assume
the component $\hat{\cal V}$ of $\pi^{-1}({\cal V})$ containing
$x$ is a subset of $\hat{\cal U}$. Since $M$ is strongly causal,
there is a causally convex neighborhood ${\cal W}$ of $p$
contained in ${\cal V}$, thus ${\pi}^{-1}({\cal W})\cap{\cal V}$
is a causally convex neighborhood of $x$.\Par

Finally, let us show the sets $J^+(\hat{p},\hat{M})\cap
J^-(\hat{q},\hat{M})$ are compact. Thus let us consider a sequence
$\{{\hat{a}}_n\}$ in $J^+(\hat{p},\hat{M})\cap
J^-(\hat{q},\hat{M})$ and let $\hat{\alpha_n}$, $\hat{{\beta}_n}$
be causal curves joining $\hat{p}$ to ${\hat{a}}_n$ and
$\hat{a_n}$ to $\hat{q}$, respectively. Now denote by non hatted
symbols the corresponding objects on the base spacetime $M$. Thus
the sequence $\{ a_n\}$ is contained in the compact set
$J^+(p)\cap J^-(q)$, as a consequence, there is a subsequence
$\{a_m\}$ converging to $x\in J^+(p)\cap J^-(q)$.\Par

Now we are in a situation in which we can apply lemma
$\ref{monodromy}$. We first apply it to the sequence of curves
$\{\hat{\alpha}_m\}$ to obtain a subsequence, also denoted by
$\{\hat{a}_m\}$ for brevity, converging to some
$\hat{x}\in\pi^{-1}(x)$ such that $\hat{x}\in
J^+(\hat{p},\hat{M})$. Then an application of the time dual
version to the sequence of reverse curves
$\{{{\hat{\beta}_m}}^-\}$ will yield a subsequence
$\{{\hat{b}}_k\}$ of $\{{\hat{a}}_m\}$ converging to $\hat{y}\in
J^-(\hat{q},\hat{M})$. But since $\{a_m\}$ is already a convergent
sequence we must have $\hat{x}=\hat{y}$. The proof is complete.
$\Box$\Par

\begin{prop}\label{cover2}
Let $M$ be a globally hyperbolic spacetime-with-timelike-boundary
with $\partial M$ connected. Then there exists a covering map
$\pi\colon\hat{M}\to M$ with $\hat{M}$ globally hyperbolic such
that

\noindent 1.- $\hat{M}$ contains a copy of $\partial M$, i.e.
there is a component $S$ of $\pi^{-1}(\partial M)$ such that
$\pi{\vert}_S$ is an isometry onto $\partial M$.

\noindent 2.- The induced map ${\imath}_*\colon \Pi_1(S,x)\to
\Pi_1(\hat{M},x)$ is surjective.

\end{prop}

\noindent\textit{Proof:} For any $p\in M$ and a curve $\gamma$
from $\partial M$ to $p$, let us denote by $[\gamma ]$ the
homotopy class of $\gamma$ modulo $\partial M$ and $p$. Define
$\hat{M}$ as the set of all pairs $(p,[\gamma ])$ and given a
simply connected neighborhood $V$ of $p$, let $(p,[\gamma
],V)\subset \hat{M}$ denote the set of all pairs of the form
$(q,[\gamma *\sigma])$ where $q\in V$ and $\sigma$ is a curve in
$V$ joining $p$ to $q$. Endow $\hat{M}$ with the topology
generated by the collection $(q,[\gamma ],V)$.\Par

In this context $\cite{Massey}$ shows that $\hat{M}$ is a
topological covering space of $M$ such that ${\imath}_*\colon
\Pi_1(S,x)\to \Pi_1(\hat{M},x)$ is surjective. Hence by
proposition $\ref{cover}$ above we conclude $\hat{M}$ is a
globally hyperbolic spacetime-with-timelike boundary as well.

Moreover, it can be shown (see $\cite{mess}$) that given
$p\in\partial M$, the component $S$ of $\pi^{-1}(\partial M)$
containing $(p,[1_p])$ is homeomorphic to $\partial M$ via $\pi$.
Further, since $\pi{\vert}_S$ is both a homeomorphism and a local
isometry, it is an isometry from $S$ onto $\partial M$. $\Box$\Par

\begin{remark}
The spacetime-with-timelike-boundary $\hat{M}$ can be
characterized as the largest spacetime-with-timelike-boundary
covering $M$ containing an isometric copy of $\partial M$ (see
proposition $2.4$ in $\cite{mess})$.
\end{remark}

\begin{remark}
Notice also that any curve in $\hat {M}$ with endpoints in $S$ is
fixed endpoint homotopic to a curve in $S$. This property is
closely related to the concept of Topological Censorship that will
be explored in the next section.\Par
\end{remark}


\subsection{Smooth null hypersurfaces}

We finish our discussion on spacetimes-with-timelike-boundary by
proving a couple of technical lemmas on smooth mull surfaces.

\begin{lemma}
Let $M$ be a strongly causal spacetime-with-timelike-boundary and
let $\gamma\colon [0,a)\to M$ be a future null geodesic orthogonal
to a spacelike surface $S\subset\partial M$ such that
$\gamma\cap\partial M=\{\gamma (0)\}$.  Let $0<T<a$, then there
exists a neighborhood $U$ of $p:=\gamma (0)$ in S and a future and
normal null vector field $X$ along $U$ such that
$X_p=\gamma^{\prime}(0)$ and all the null geodesics with initial
velocity $X_q$ remain within a neighborhood of $\gamma\vert_0^T$
and intersect $\partial M$ only at the base point $q\in U$.
\end{lemma}

\noindent \textit{Proof:} Let us consider an extension
$\overline{M}$ of $M$ and let us denote by $\exp^\perp$ the normal
exponential map to $S$ in $\overline{M}$. Then PDE theory
guarantees the existence of a neighborhood $V$ of $p$ in $S$ and a
null future vector field $Y$ along $V$ extending
$\gamma^{\prime}(0)$ such that $\gamma_q(t):=\exp^\perp (tX_q)$ is
defined  up to $t=T$ for all $q\in V$. Moreover, since geodesics
depend smoothly on initial conditions, the geodesic segments
$\gamma_q\vert_0^T$ remain within an open neighborhood of
$\gamma\vert_0^T$. It remains to show that we can choose $V$ in
such a way that all the geodesics $\gamma_q$ stay in $M$ and touch
$\partial M$ only at $q$. \Par

Assume the contrary, then there would be a sequence $\{p_n\}$
converging to $p$ such that the geodesics $\gamma_n:=\gamma_{p_n}$
meet $\partial M$ at a time $t_n\in (0,T]$. By compactness of
$[0,T]$ there is a subsequence $\{t_m\}$ converging to $T_0\in
[0,T]$. Since $\{p_m\}$ converges to $p$  we  have
$x_m:=\gamma_m(t_m)\to \gamma (T_0)$. Notice  that $\gamma
(T_0)\in
\partial M$ since $\partial M$ is closed, thus $T_0=0$ since
by hypothesis $\gamma$ intersects $\partial M$ only at $p$.\Par

Let ${{\cal N}}_0$ be a normal neighborhood of $S$ in
$\overline{M}$. We proceed to show that $p$ has no causally convex
neighborhood contained in ${{\cal N}}_0\cap M$, and hence $M$ can
not be strongly causal at $p$. Thus, let ${\cal U}\subset{{\cal
N}}_0\cap M$ be a $M$-neighborhood of $p$ and let us consider $U$,
a $\partial M$-neighborhood of $p$  such that $D(U,\partial M )$
is globally hyperbolic with Cauchy surface $S\cap U$, and a
$\overline{M}$-neighborhood ${{\cal U}}_0$, ${{\cal U}}_0\cap
M\subset{\cal U}$ with ${{\cal U}}_0\cap\partial M\subset U$ and
such that $S\cap{{\cal U}}_0$ is acausal in ${\cal U}$ (for a
construction of such neighborhoods $U$ and ${{\cal U}}_0$, refer
to the proof of proposition $\ref{normal}$).\Par

Since both sequences $\{x_m\}$ and $\{p_m\}$ converge to $p$, we
have that ${{\cal U}}_0\cap M$ contains $x_m$ and $p_m$ for all
$m\ge N$. Notice that $\gamma_N$ can not intersect $\partial M$ to
the past of $S\cap U$ because of the acausality of $S\cap {{\cal
U}}_0$ in ${{\cal U}}_0$, therefore we must have $x_N\in I^+(S\cap
U,U)$. It follows $\tau_{\overline{N}} (S,x_N)>0$, thus $\gamma_N$
fails to realize Lorentzian distance in ${{\cal N}}_0$ between
$x_N$ and $S$. However, by properties of normal neighborhoods with
respect to $\exp^{\perp}$, the initial segment of
$\gamma_N\cap{{\cal N}}_0$ is distance realizing in
${\cal\overline{N}}$. Therefore
$\gamma_N\vert_0^{x_N}\not\subset{{\cal N}}_0$, hence
$\gamma_N\vert_0^{x_N}$ intersects ${\cal U}$ in a disconnected
set. The proof is complete. $\Box$\Par

\begin{lemma}\label{nullsurf}
Let $M$ be a globally hyperbolic spacetime-with-timelike-boundary
and $\gamma\colon [0,a)\to M$ be a null geodesic that meets
$\partial M$ only at $p:=\gamma (0)$. Further, let $\partial_0$ be
the component of $\partial M$ containing $p$ and let $S$ be a
Cauchy surface of $\partial_0$ passing through $p$; and assume
$\gamma (t)\in J^+(S)-I^+(S)$ $\forall t\in [0,a)$. Then for every
$T\in (0,a)$ there exists a smooth null hypersurface $N\subset M$
for which $S\cap N$ is a spacelike cut and $\gamma\vert_0^T$ is
one of its null generators.
\end{lemma}\Par

\noindent\textit{Proof:} First notice that  by propositions
$\ref{geo1}$ and $\ref{normal}$, $\gamma$ is a null geodesic that
meets $S$ orthogonally. Furthermore, by proposition $\ref{focal}$
we know the segment $\gamma\vert_0^T$ is focal point free.\Par

Next, let us consider a neighborhood $U\subset S$ of $p$ and a
null normal vector field $X$ along $U$ as established by the
previous lemma and consider a neighborhood $V$ of $p$ with compact
closure such that $\overline{V}\subset U$. Let us define  $\Phi
\colon U\times [0,a)\to M$ by $\Phi (q,t)=\exp^{\perp}(tX_q)$,
then $t\mapsto \Phi (q,t)$ is just the null geodesic normal to $S$
with initial velocity $X_q$.\Par

Let $t\in [0,T]$. Since  $\gamma (t)$ is not a focal point of $S$
along $\gamma$,  $\exp^{\perp}$ is not singular at
$t\gamma^{\prime}(0)$ thus in virtue of theorem \ref{focalnullpts}
and the inverse function theorem, there exists a neighborhood
${\cal U}_t$ of $t\gamma^{\prime}(0)$ in the normal null cone
$\mathfrak{N}(S)$ such that $\exp^{\perp}\vert_{{\cal U}_t}$ is a
diffeomorphism, hence injective.

 Now we show there is a neighborhood $U_t\subset \Phi (\overline{V}\times [0,T])$ of
$\gamma (t)$,  such that $\Phi (q_1,t_1)=\Phi (q_2,t_2)$ implies
$(q_1,t_1)=(q_2,t_2)$. The existence of such a neighborhood for
$t=0$ is assured by the existence of normal neighborhoods (see
lemma 7.26 on \cite{ON}), so let us consider now the case $t>0$.
Assume no such $U_t$ exists, then there would be a sequence of
points $\{x_n\}$ converging to $\gamma (t)$ and two sequences
$\{(p_n,t_n)\}$, $\{(q_n,s_n)\}$ such that $(p_n,t_n)\neq
(q_n,s_n)$ and $\Phi (p_n,t_n)=x_n=\Phi (q_n,s_n)$. By compactness
of $\overline{V}$ there are subsequences $\{p_m\}$, $\{q_m\}$
converging to $p^{\prime}$ and $q^{\prime}$ respectively. Then,
since $x_n\in J^+(p_n)$ and $p^{\prime}\neq \gamma (t)$, by
proposition $\ref{curvelemma}$ there is a limit causal curve
$\alpha$ joining $p^{\prime}$ to $\gamma (t)$. Moreover, since
$\alpha\subset \Phi (U\times [0,T])$ we know $\alpha$ only meets
$\partial M$ at $\alpha (0)$. Hence, since $\gamma (t)\in
J^+(S)-I^+(S)$ propositions $\ref{geo1}$ and $\ref{normal}$
guarantee that $\alpha$ is a null geodesic normal to $S$. If
$\alpha\neq \gamma$ then $\gamma (t+\varepsilon )\in I^+(S)$ by
$\ref{2null}$. Hence $\alpha =\gamma$ and therefore $p_m\to p$.
Thus by further taking a subsequence we can assume $X_{p_m}\to
\gamma^{\prime}(0)$. A similar argument applied to the sequence
$\{q_m\}$ shows $q_m\to q$ and $X_{q_m}\to \gamma^{\prime}(0)$.
This in turn implies that $\exp^{\perp}$ is not injective in any
neighborhood of $t\gamma^{\prime}(0)$ in $\mathfrak{N}(S)$. A
contradiction.\Par

Choose now $U_t$ so that $U_t\subset \exp^{\perp}({{\cal U}_t})$
and consider neighborhoods $V_t$ of $\gamma (t)$ such that
$\overline{V_t}\subset U_t$. Since $\overline{V}\times [0,T]$ is
compact and $\Phi$ is continuous, there are finitely many $U_i$
such that $\Phi (\overline{V}\times [0,T])\subset \cup_iV_i$. Let
$W^{\prime}=\bigcup_i\Phi^{-1}(U_i)$ and
$W=\bigcup_i\overline{V_i}$, then $\Phi\vert_{W^{\prime}\times
[0,T]}$ is a one to one local diffeomorphism, hence a
diffeomorphism onto its image. In fact, by the compactness of
$\overline{W}$ we conclude $\Phi\vert_{W\times [0,T]}$ is an
embedding. Therefore $N=\Phi (W\times [0,T])$ is the desired null
hypersurface. $\Box$\Par

\begin{remark}\label{rem01}
 Notice that the affine parameter on the null generators of $N$
can be rescaled so that the slices $\Phi (W\times\{s\})$ are all
diffeomorphic to $W\subset S$. Notice as well that such $N$ always
exists locally. Indeed, because of properties of the null normal
exponential map,  given $v\in T_pM\cap\mathfrak{N}(S)$ we can
always find a $S$-neighborhood $U$ of $p$, a null normal vector
field $V$ along $U$ extending $v$ and $T_0>0$ small enough such
that $\exp^{\perp}(tV)$, $t\in [0,T_0]$ defines a smooth null
surface.
\end{remark}

Let $S\subset \partial M$ be spacelike hypersurface of $\partial
M$. Then there are, up to  positive scaling, two future directed
null normal vector fields along $S$, one of which is inward
pointing while the the other is outward pointing. Let $K_1$ and
$K_2$ be these vector fields, with $K_1$ inward pointing. For
$i=1,2$ define the \textit{null Weingarten map of $S$ relative to
$K_i$} by $b_i\colon T_pSt\to T_pS$, $b_i(X):=
\text{tan}\nabla_XK_i$, where $\text{tan}\colon T_pM\to T_pS$ is
the standard projection. The corresponding \textit{null second
fundamental forms}  $B_i\colon T_pS\times T_pS\to \R$ are defined
by $B_i(X,Y)=g(b_i(X),Y)$. Finally the \textit{null expansions of
$S$},  are given by $\theta_i=\text{Tr}(b_i)$.

\begin{defi}
We say $S$ is null convex if $B_1$ is negative definite and $B_2$
is positive definite. We say $S$ is weakly null convex if
$\theta_1<0$ and $\theta_2\ge 0$.
\end{defi}

\begin{remark}
Let $N\subset M$ is a null surface that meets $\partial M$
transversally at $p\in\partial M$. Then $S:=N\cap\partial M$ is
spacelike near $p$. Let $K$ be the null vector field $K$
associated to $N$. If $K\vert_S$ agrees with $K_i$ then by
equation $(\ref{nullmeanc})$ we readily see that the null
expansion $\theta_K$ of $K$ agrees with $\theta_i$ at $p$.
\end{remark}



\section{The Principle of Topological Censorship}

The singularity theorems of Gannon $\cite{gannon}$ and Lee
$\cite{lee}$ establish that non trivial spatial topology on a
spacetime leads to the formation of singularities. More precisely,
if a spacetime $M$ satisfying the null energy condition admits an
asymptotically flat Cauchy surface $S$ with $\pi (S)\neq 0$ then
$M$ is null geodesically incomplete.

On the other hand, one expects that our universe does not present
``naked" singularities. In other words, the physical processes
responsible of  the formation of singularities (e.g. gravitational
collapse) would also induce the formation of an event horizon
hiding the singularity from view.

\eject

Thus, in view of the aforementioned singularity theorems we should
expect that the region ${\cal D}$ of spacetime outside all black
holes and white holes to be topologically trivial. This is the
essence of the Principle of Topological Censorship. Following
$\cite{FSW}$, we state the Principle of Topological Censorship in
the following way:

\par\vspace{.3cm}

\noindent\textbf{Principle of Topological Censorship (PTC):} Let
$(\tilde{M},\tilde{g})$ be a spacetime admitting a conformal
completion $({M},{g})$. Then every causal curve on ${M}$ whose
initial and final endpoints lie on $\scri$ is fixed endpoint
homotopic to a curve on $\scri$.

\par\vspace{.3cm}

Topological censorship has proved to be an important tool to
relate the topology at infinity with the topology of black hole
horizon. Actually, under the assumption of topological censorship
the famous result by S. Hawking on the spherical topology of
stationary black holes $\cite{HBH}$ has been generalized in
different directions $\cite{CW,JV,galbh}$.


\section{PTC on spacetimes with timelike boundary}

In the seminal work of J. Friedman, K. Schleich and D. Witt
$\cite{FSW}$, the PTC is showed to hold for asymptotically flat
spacetimes satisfying an energy condition. Later developments
extended the field of application of the PTC to globally
hyperbolic spacetimes with a \textit{timelike} conformal structure
$\cite{GSWW,BG}$.

It is of special interest the  quasilocal version of the PTC
presented in $\cite{galfi}$. In this context, the conformal
boundary $\scri$  is replaced by a timelike boundary $\partial M
\approx S^{n-2}\times \R$ with null convex Cauchy surfaces. PTC
then follows as a consequence of the following theorem:

\begin{teo}\label{finiteinf}
Let $(M,g)$ be a globally hyperbolic
spacetime-with-timelike-boundary which obeys the null energy
condition, such that each component $T_{\alpha}$ of $\partial M$
has a spacelike Cauchy surface $S_{\alpha}\approx
S^{n-2}$\symbolfootnote[2]{ Notice  that by proposition
$\ref{partialgh}$ each component of $(\partial M,\imath^*g)$ is a
globally hyperbolic spacetime in its own right. Hence each
component of $\partial M$ has a Cauchy surface and a smooth time
function.}. Further, assume each of these Cauchy surfaces is null
convex and acausal in $M$. Then for all $\alpha\neq\beta$
$J^+(T_{\alpha})\cap J^-(T_{\beta})=\emptyset$.
\end{teo}

In physical terms, this result implies the non traversability of
wormholes whose ``mouths" are weakly null convex spheres in
globally hyperbolic spacetimes satisfying the null energy
condition. In fact, if $S$ is a Cauchy surface of spacetime and
$N\subset S$ is such a wormhole, then $M=N\times\R$ is a
spacetime-with-timelike-boundary satisfying all the hypotheses of
theorem $\ref{finiteinf}$.

In this section we will improve the main result on $\cite{galfi}$
in two different ways. More precisely, we will show PTC holds in
the finite infinity setting when the spatial sections of $\partial
M$ are \textit{compact} and \textit{weakly null convex.} In
particular, no assumption on its homotopy type is made. Moreover,
we fill in a technical gap in the proof of theorem
$\ref{finiteinf}$.

\subsection{Fastest causal curves}

\enlargethispage{.5cm}

A fundamental problem related to the PTC consists in proving the
existence of ``fastest" curves communicating two different
boundary components of $M$. In other words, if we assume
${\partial}_{\alpha}$ and $\partial_{\beta}$ are two components of
$\partial M$ with $J^+(S)\cap
\partial_{\beta}\neq\emptyset$, where $S$ is a Cauchy surface of
$\partial_{\alpha}$; then we want to find a  null geodesic
$\eta\subset
\partial I^+(S)$ connecting $S$ to ${\partial}_{\beta}$ at the
earliest time possible.\Par

To be more precise, let us denote by ${\Sigma}_t$ the level sets
of the time function $t$ of ${\partial}_{\beta}$ and let $B=\{
t\in\R\mid J^+(S)\cap {\Sigma}_t\neq\emptyset\}$, hence
$B\neq\emptyset$. Then if $t_0:=\inf B$ exists, we would like to
be able to construct a null geodesic $\eta\subset \partial I^+(S)$
joining $S$ to $\Sigma_{t_0}$. Such a geodesic will be called
\textit{a fastest geodesic from $S$ to $\partial_{\beta}$.}\Par

Before proceeding any further, let us prove some useful
lemmas.\Par

\begin{lemma}\label{nullgeo}
Suppose $t_0\neq -\infty$ and $\gamma$ is a causal curve joining
$S$ to $\Sigma_{t_0}$ that intersects ${\partial}M$ only at its
endpoints. Then $\gamma$ is a null geodesic contained in $\partial
I^+(S)$ that meets both $\Sigma_{t_0}$ and $S$ orthogonally.
\end{lemma}

\noindent \textit{Proof:} Let $q\in \Sigma_{t_0}$ be the future
endpoint of $\gamma$. First notice that $q\not\in I^+(S)$. In
fact, if this was not the case by proposition $\ref{open}$ we
would have $I^+(S)\cap{\Sigma}_t\neq\emptyset$ for $t$ slightly
less than $t_0$. Moreover, this fact and proposition $\ref{crono}$
guarantee that no pair of points of $\gamma$ can be
chronologically related. Therefore $\gamma\subset
J^+(S)-I^+(S)$.\Par

Further, since $J^+(S)$ is closed in virtue of proposition
$\ref{causimple}$, we have $J^+(S)\supset\overline{I^+(S)}$. On
the other hand $J^+(S)\subset\overline{I^+(S)}$ by proposition
$\ref{closure}$, then $J^+(S)=\overline{I^+(S)}$ so
\begin{equation}
\partial I^+(S)=\overline{I^+(S)}-I^+(S)=J^+(S)-I^+(S),
\end{equation}
thus $\gamma\subset\partial I^+(S)$ as claimed.\Par

Finally, since $\gamma$ only meets $\partial M$ at endpoints we
can apply propositions $\ref{geo1}$ and $\ref{normal}$ above to
conclude $\gamma$ is a null geodesic meeting $\Sigma_{t_0}$ and
$S$ orthogonally. $\Box$\Par

\begin{lemma}\label{raychad}
Let $M$ be a globally hyperbolic spacetime-with-timelike-boundary
satisfying the null energy condition and ${\partial}_{\alpha}$,
${\partial}_{\beta}$ two different connected components of
$\partial M$. Further, let $S_{\alpha}\subset
{\partial}_{\alpha}$, $S_{\beta}\subset {\partial}_{\beta}$ be
weakly null convex Cauchy surfaces with $J^+(S_{\alpha})\cap
S_{\beta}\neq \emptyset$.  If $I^+(S_{\alpha})\cap
S_{\beta}=\emptyset$ then every future causal curve joining
$S_{\alpha}$ and $S_{\beta}$ must meet $\partial M$ at a point
other than its endpoints.
\end{lemma}

\noindent \textit{Proof:} We proceed by contradiction. Assume
there is a future causal curve $\gamma\colon [0,1]\to M$ from
$S_{\alpha}$ to $S_{\beta}$ that meets $\partial M$ only at
endpoints. By propositions $\ref{geo1}$ and $\ref{focal}$ $\gamma$
must be a null geodesic, orthogonal to both $S_{\alpha}$ and
$S_{\beta}$ and with no focal points to $S_{\alpha}$ on $[0,1)$.
\Par

First notice that by remark $\ref{rem01}$ above, given
$-\gamma^{\prime}(1)\perp S_{\beta}$ there is a
$S_{\beta}$-neighborhood $U_{\beta}$ of $q$, $\varepsilon >0$ and
a smooth surface $N_{\beta}$ of the form $N_{\beta}=\Psi
(U_{\beta}\times (1-2\varepsilon ,1])$, $\Psi
(y,s)=\exp^{\perp}(sV_y)$ such that
$\gamma\vert_{1-2\varepsilon}^1$ is one of its generators. We can
further assume that the slices $U_{\beta ,s}:=\Psi
(U_{\beta}\times\{s\})$ are all diffeomorphic to $U_{\beta}$.\Par

On the other hand, by proposition $\ref{nullsurf}$ there is a
surface $N_{\alpha}$ emanating from a neighborhood
$U_{\alpha}\subset S_{\alpha}$ of the form $N_{\alpha}=\Phi
(U_{\alpha}\times [0,1-\varepsilon /2))$, $\Phi (x,t)=\exp^\perp
(tZ_x)$ having $\gamma\vert_0^{1-\alpha /2}$ as one of its
generators. Moreover, we can assume $U_{\alpha}\approx U_{\alpha
,t}$, where $U_{\alpha ,t}:=\Phi (U_{\alpha}\times\{t\})$. Observe
$N_\alpha$ and $N_{\beta}$ meet on $r=\gamma (1-\varepsilon
)$.\Par

Furthermore, since $S_{\beta}$ is null convex and
$\gamma^{\prime}(1)$ is future outward pointing we have that
$\theta_{\beta}(1)\ge 0$, where $\theta_{\beta}$ is the null
expansion of $U_{\beta ,s}$ along $\gamma$. Hence, by the
Raychaudhuri equation and the null energy condition we have
\begin{equation}
 \frac{d\theta_{\beta}}{dt}\le -\text{Ric}(\alpha^{\prime},\alpha^{\prime})\le 0
\end{equation}
thus $\theta_{\beta}(s)$ is non increasing, therefore
$\theta_{\beta}(r)=\theta_{\beta}(1-\varepsilon) \ge 0$. In a
similar fashion, since $S_{\alpha}$ is null convex and
$\gamma^{\prime}(0)$ is future inward pointing we have
$\theta_{\alpha}(0 )<0$, thus
$\theta_{\alpha}(r)=\theta_{\alpha}(1-\varepsilon )<0$.\Par

Finally, let us notice $N_{\alpha}$ is to the future of
$N_{\beta}$ near $r$. To this end, let us a consider a
neighborhood ${\cal V}$ of $r$ such that $N_{\beta}$ is closed and
achronal in ${\cal V}$, hence $r\not\in\text{edge}(S)$, thus there
exists a neighborhood ${\cal W}\subset{\cal V}$ of $r$ such that
every timelike curve from $I^-(r,{\cal W})$ to $I^+(r,{\cal W})$
intersects $N_{\beta}$. Let $a\in I^+(r,{\cal W})$ and $b\in
I^-(r,{\cal W})$ and let ${\cal U}:=I^-(a,{\cal W})\cap
I^+(b,{\cal W})$. We will show $N_{\alpha}\cap{\cal U}\subset
J^+(N_{\beta}\cap{\cal U},{\cal U})$.\Par

Let $x\in N_{\alpha}\cap{\cal U}$, then there is a future timelike
curve $\sigma$ from $b\in I^-(r,{\cal W})$ to $a\in I^+(r,{\cal
W})$ passing through $x$. Thus, by the way ${\cal W}$ was chosen,
$\sigma$ has to intersect $N_{\beta}$ at some point, say $y$. If
$y\in I^+(x,{\cal W})$ then by concatenating $\sigma\vert_x^y$ to
the generator of $N_{\alpha}$ through $y$ and the corresponding
generator through $x$ we get a causal curve from $S_{\alpha}$ to
$S_{\beta}$ that is not a null geodesic and meets $\partial M$
only at endpoints. As a consequence, $I^+(S_{\alpha})\cap
S_{\beta}\neq\emptyset$, contradicting one of our hypothesis. Thus
$y$ comes before $x$ along $\sigma$, hence $x\in
J^+(N_{\beta}\cap{\cal U},{\cal U})$ as desired.\Par

Therefore, by the maximum principle for smooth null hypersurfaces
$\ref{MP}$ we have that $\theta_{\alpha}(r)=0=\theta_{\beta}(r)$.
This contradicts $\theta_{\alpha}(r)<0$. $\Box$

Returning to our original problem; we analyze the simplest case of
all, that is, when $\partial M$ consists of two components
$\partial_1$ and $\partial_2$, both acausal in $M$.

\begin{teo}\label{fastestcurve}
Let $M$ be a globally hyperbolic spacetime-with-timelike-boundary
and assume $\partial M$ has only two components ${\partial}_1$ and
${\partial}_2$. Further assume the Cauchy surfaces of both
${\partial}_1$ and ${\partial}_2$ are compact and acausal in $M$.
Let $S$ be a Cauchy surface of ${\partial}_1$ with $J^+(S)\cap
{\partial}_2\neq \emptyset$. Then there exists a fastest null
geodesic $\eta\subset\partial I^+(S)$ from $S$ to $\partial_2$.
Moreover, $\eta$ is normal to both $S$ and $\Sigma_{t_0}$.
\end{teo}

\eject

In order to be able to use lemma $\ref{nullgeo}$, we must first
prove that $\inf B$ exists under the hypotheses of the
theorem.\Par

\begin{lemma}\label{inf}
Let $B=\{ t\mid J^+(S)\cap {\Sigma}_t\neq\emptyset\}$. Then $B$ is
bounded below and $J^+(S)\cap {\Sigma}_{t_0}\neq \emptyset$, where
$t_0=\inf B$.
\end{lemma}
\Par

\noindent \textit{Proof:} We proceed by contradiction. If $B$ is
unbounded from below, then there exist sequences $t_n\searrow
-\infty$ and $q_n\in {\Sigma}_{t_n}$ with $q_n\in J^+(S)$. Fix
$T\ge t_1$ and let $p_n\in{\Sigma}_T$ be the projection of $q_n$
under the time function $t$ of ${\partial}_2$. Since ${\Sigma}_T$
is compact, $\{p_n\}$ has a convergent subsequence $p_m\to p$. Let
${\gamma}_m\colon [0,T-t_m]\to{\partial}_2$ be the integral line
segments of $\nabla t$ from $p_m$ to $q_m$. Then by the limit
curve lemma there is a limit curve $\gamma\colon
[0,a]\to{\partial}_2$ of $\{{\gamma}_m\}$. Since $t_m\to -\infty$
we have that $a=\infty$, thus $\gamma$ is inextendible.\Par

Let $s\in[0,\infty )$ and consider $N$ such that ${\gamma}_m$ is
defined at $s$ for all $m\ge N$. Since ${\gamma}_m(s)\in J^+(S)$
for all $m\ge N$ and $J^+(S)$ is closed we have $\gamma
(s)={\lim}_m{\gamma}_m(s)\in J^+(S)$. Thus $\gamma\subset J^+(S)$.
On the other hand $\gamma (0)=p$ so $\gamma\subset J^-(p)$.
Therefore $\gamma\subset J^-(p)\cap J^+(S)\cap\partial_2$.\Par

Now note $J^-(p)\cap J^+(S)$ is a compact set by proposition
$\ref{causimple1}$ and $\gamma$ is a future inextendible causal
curve contained in it. This contradicts strong causality on
$\partial_2$.

Finally, let us consider a sequence $\{t_n\}\subset B$ converging
to $t_0=\sup B$ and $y_n\in J^+(S)\cap {\Sigma}_{t_n}$. Let $x_n$
be the projection of $y_n$ to ${\Sigma}_{t_0}$ by the time
function $t$. Since ${\Sigma}_{t_0}$ is compact there is a
subsequence $\{x_m\}$ converging to $x\in{\Sigma}_{t_0}$, hence
$y_m\to x$ as well. Thus causal simplicity implies that $x\in
J^+(S)$. $\Box$\Par

\noindent\textit{Proof:} (of theorem $\ref{fastestcurve}$) Since
$J^+(S)\cap{\Sigma}_{t_0}\neq\emptyset$, let $\gamma$ be a causal
curve from $p\in S$ to $q\in{\Sigma}_{t_0}$. By  lemmas
$\ref{nullgeo}$ and $\ref{inf}$, it suffices to show $\gamma$
meets $\partial M$ only at $p$ and $q$.\Par

Let $p^{\prime}\in\partial M$ be the last point in which $\gamma$
intersects ${\partial}_1$. If $p^{\prime}\neq p$ then since $S$ is
acausal in $M$ we must have $p^{\prime}\in J^+(p)-S$. Even
further, since $\partial_1$ is globally hyperbolic, we can use the
time function of $\partial_1$ to show $p^{\prime}\in
I^+(S,\partial_1 )$. Hence by proposition $\ref{crono}$ we have
$q\in I^+(S)$, which will lead to a contradiction to the
minimality of $t_0$. Therefore $p=p^{\prime}$, i.e. $\gamma$ only
touches ${\partial}_1$ at its initial point.\Par

On the other hand, let $q^{\prime}\in \Sigma_T$ the first point in
which $\gamma$ meets $\partial_2$. Clearly $T\ge t_0$ is by the
way $t_0$ is defined. Finally, if $q^{\prime}\neq q$ then the
acausality of $\Sigma_T$ is violated. This completes the proof.
$\Box$\Par

As an application of theorem $\ref{fastestcurve}$ we prove the non
traversability of wormholes in the two boundary component case.

\begin{cor}
Let $M$ be a globally hyperbolic spacetime-with-timelike-boundary
and assume $\partial M$ has only two components ${\partial}_1$ and
${\partial}_2$. Further assume the Cauchy surfaces of both
${\partial}_1$ and ${\partial}_2$ are compact, weakly null convex
and acausal in $M$. Then $J^+(\partial_1)\cap
J^-(\partial_2)=\emptyset$.
\end{cor}

\noindent\textit{Proof:} Assume $J^+(\partial_1)\cap
J^-(\partial_2)\neq\emptyset$, then there is a causal curve from
$p\in\partial_1$ to $q\in\partial_2$. Let $S$ be a Cauchy surface
of $\partial_1$ containing $p$. Then by theorem
$\ref{fastestcurve}$ there is a future null geodesic $\eta$
joining $S$ to $\Sigma_{t_0}$. Moreover, the proof of theorem
$\ref{fastestcurve}$ also shows that $\eta$ meets $\partial M$
only at its endpoints. Hence by the contrapositive of lemma
$\ref{raychad}$ we have $I^+(S)\cap\Sigma_{t_0}\neq\emptyset$.
Since $I^+(S)$ is open, there would be a point $x\in\partial_2\cap
I^+(S)$ slightly to the past of $\Sigma_{t_0}$, thus contradicting
the choice of $t_0$ as the infimum of the set $\{ t\mid J^+(S)\cap
{\Sigma}_t\neq\emptyset\}$. $\Box$



\subsection{Strong form of the PTC}

We prove now the main result of this chapter, namely that a
stronger version of the PTC holds on the domain of outer
communication relative to a boundary component.

\begin{defi}
Let $(M,g)$ be a spacetime-with-timelike-boundary and
$A\subset\partial M$. The domain of outer communication relative
to $A$ is given by ${\cal D}(A):= I^+(A)\cap I^-(A)$.
\end{defi}

Thus, let $(M,g)$ be a spacetime with a connected timelike
boundary. ${\cal I}:=\partial M$. First notice that since ${\cal
I}$ is timelike we have ${\cal I}\subset {\cal D}({\cal I})$,
hence ${\cal D}({\cal I})$ is a spacetime-with-timelike-boundary
and $\partial{\cal D}({\cal I})={\cal I}$.\Par

\noindent\textbf{Strong form of PTC} \textit{Any} curve in ${\cal
D}({\cal I})$ with endpoints in ${\cal I}$ is fixed endpoints
homotopic to a curve in ${\cal I}$.

As pointed out in $\cite{GSWW}$, the strong form of the PTC is a
consequence of the PTC in its standard form and the following
topological result (see lemma $3.2$ in $\cite{GSWW}$):

\begin{prop}\label{last}
Let $M$ and $S$ be topological manifolds, $\imath\colon
S\hookrightarrow M$ an embedding and $\pi\colon M^*\to M$ the
universal cover of $M$. If $\pi^{-1}(S)$ is connected then the
induced group homomorphism $\imath_*\colon \pi_1(S)\to \pi_1(M)$
is surjective.
\end{prop}

We thus begin by showing that the standard version of PTC holds on
${\cal D}({\cal I})$.

\begin{teo}\label{qlptc}
Let $M$ be a spacetime-with-timelike-boundary with ${\cal
I}=\partial M$ connected and assume ${\cal D}:= {\cal D}({\cal
I})$ is globally hyperbolic. Further assume the Cauchy surfaces of
${\cal I}$ are compact, weakly null convex and acausal in ${\cal
D}$. Then the Principle of Topological Censorship holds on ${\cal
D}$.
\end{teo}

\noindent\textit{Proof:} Let $\gamma\colon [0,1]\to {\cal D}$ be a
causal curve from $p\in{\cal I}$ to $q\in{\cal I}$, we want to
show that $\gamma$ is fixed endpoint homotopic to a curve
$\gamma_0\colon [0,1]\to {\cal I}$.\Par

First of all, by proposition $\ref{cover2}$ there is a globally
hyperbolic covering spacetime-with-timelike-boundary
$\pi\colon\hat{\cal D}\to {\cal D}$ having a copy $\hat{\cal I}$
of ${\cal I}$. Now let $\hat{p}\in\hat{\cal I}$ such that
$\pi(\hat{p})=p$ and lift $\gamma$ to a curve $\hat{\gamma}\colon
[0,1]\to\overline{\cal D}$ starting at $\hat{p}$, then
$\hat{q}:=\hat{\gamma}(1)\in
\partial\hat{M}$.\Par

 If $\hat{\gamma}(1)\in\hat{\cal I}$,  by remark $\ref{cover2}$
we know $\hat{\gamma}$ is fixed endpoints homotopic to a curve
$\hat{\gamma}_0\colon [0,1]\to \hat{\cal I}$, in which case we are
done since $\gamma_0=\pi\circ\hat{\gamma}_0$ would be the desired
curve.\Par

Now, let us consider the case $\hat{\gamma}(1)\in \partial
\hat{M}-\hat{\cal I}$.\Par

Let $T$ be a time function on ${\cal I}$ such that the Cauchy
surface $S_0:=T^{-1}(0)$ contains $p$, and let us define
$S_t:=T^{-1}(t)$. Consider now
$\hat{S_0}:=\pi^{-1}(S_0)\cap\hat{\cal I}$ and
$B:=J^+(\hat{S_0},\hat{\cal D})\cap(\partial\hat{M}-\hat{\cal
I})$, hence $\hat{q}\in B$. \Par

Since $B\neq\emptyset$ then $(T\circ\pi )(B)\neq\emptyset$.
Further, $(T\circ\pi )(B)$ is bounded from below by $0$, since the
acausality of $S_0$ in $M$ would be violated otherwise. Hence
$t_0=\inf \{(T\circ\pi )(B)\}$ exists.

By definition, consider a decreasing sequence $\{t_n\}\subset
(T\circ\pi )(B)$ converging to $t_0$. Thus, let $\hat{{\gamma}}_n$
be a future causal curve in $\hat{D}$ joining $\hat{p}_n\in
\hat{S_0}$ to $\hat{q}_n\in
\partial\hat{\cal D}-\hat{\cal I}$, $t_n=(T\circ \pi )(\hat{q}_n)$ and let us denote by unhatted
symbols the corresponding objects in ${\cal D}$.\Par

Since $S_0$ is compact and $\pi\vert_{\hat{\cal I}}$ is a
homeomorphism we know $\hat{S_0}$ is compact as well, hence there
is a subsequence $\{\hat{p}_m\}$ converging to $\hat{p}_0\in
\hat{S_0}$.\Par

On the other hand, let $q^{\prime}_m$ be the projection of $q_m$
into $S_{t_0}$ by the time function $T$. By compactness of
$S_{t_0}$, after further taking a subsequence, we may assume
$q^{\prime}_m$ converges to $q_0\in S_{t_0}$. Hence $q_m\to q_0$
as well.\Par

 Further notice that $q_0\neq p_0$. Assume the contrary,
and then consider a neighborhood ${\cal U}$ of $p_0$ evenly
covered by $\pi$ and let ${\cal V}\subset{\cal U}$ be a causally
convex neighborhood of $p_0$. Since both sequences $\{p_m\}$ and
$\{q_m\}$ converge to $p_0$ we have that $p_N,q_N\in{\cal V}$ for
some $N$ sufficiently large. Thus $\gamma_N\subset {\cal V}$,
hence $\hat{\gamma}_N\subset\hat{\cal U}$, where $\hat{\cal U}$ is
the unique component of $\pi^{-1}({\cal U})$ containing
$\hat{p_0}$. It follows $\hat{q}_N\in\hat{\cal I}$ which is a
contradiction.\Par

Therefore, all the hypothesis of the monodromy lemma (proposition
$\ref{monodromy}$) are met, hence we can assert the existence of a
future causal curve $\hat{\gamma_0}$ from $\hat{p_0}$ to
$\hat{q_0}$ and thus $T\circ\pi (\hat{q_0})=t_0$. Hence $t_o\in
(T\circ\pi )(B)$.\Par

Next, let us denote by $\partial_0$ the component of
$\partial\hat{M}$ containing $\hat{q_0}$ and let $\Sigma_{0}$ be a
Cauchy surface of $\partial_0$ through $\hat{q_0}$. Notice that
$\hat{\gamma_0}$ touches $\partial\hat{\cal D}-\hat{\cal I}$ only
at $\hat{q_0}\in\partial_0$. In fact, if
$\hat{x}\in\hat{\gamma_0}\cap \partial\hat{\cal D}-\hat{\cal I}$,
$\hat{x}\neq\hat{q_0}$, then $t_1:=T\circ\pi (\hat{x})\ge t_0$ and
thus the segment of $\gamma_0$ from $x$ to $q_0$ would be a future
causal curve connecting $S_{t_1}$ to $S_{t_0}$, violating the
acausality of $S_{t_0}$.\Par

Finally, note that since $\pi$ is a local isometry, then both
$\hat{\cal S}_0$ and $\Sigma_0$ are null convex. Hence by lemma
$\ref{raychad}$ above we conclude
$I^+(\hat{S_0})\cap\Sigma_0\neq\emptyset$, but then since
$I^+(\hat{S_0})$ is an open set we would have $I^+(\hat{S_0})\cap
I^-(\Sigma_0,\partial_0)\neq\emptyset$; which in turns yield a
contradiction to the fact that $t_0=\inf T\circ\pi (B)$.\Par

Hence we just ruled out the case $\hat{\gamma}(1)\in \partial
\hat{M}-\hat{\cal I}$, thus $\hat{\gamma}(1)\in \hat{\cal I}$ and
the proof is complete. $\Box$

\begin{teo}
Let ${\cal D}$ be as in theorem $\ref{qlptc}$, then the strong
form of the principle of topological censorship holds on ${\cal
D}$.
\end{teo}

\noindent\textit{Proof:} Let $\pi\colon{\cal D}^*\to {\cal D}$ be
the universal cover of ${\cal D}$. Then ${\cal D}^*$ is a
spacetime with timelike boundary ${\cal I}^*:=\pi^{-1}({\cal I})$.
By proposition $\ref{last}$ it suffices to show ${\cal I}^*$ is
connected.

Thus, let $\{{\cal I}^*_{\alpha}\}$, $\alpha\in A$, be the
collection of connected components of ${\cal I}^*$ and let us
define ${\cal D}^*_{\alpha}:= I^+({\cal I}^*_{\alpha},{\cal
D}^*)\cap I^-({\cal I}^*_{\alpha},{\cal D}^*)$.

Let $x\in{\cal D}^*$. Since $\pi (x)\in{\cal D}$, there is
 a future causal curve $\gamma^*\colon I\to{\cal D}^*$  through
$x$ with endpoints on ${\cal I}^*$. Hence $\gamma
:=\pi\circ\gamma^*$ is a future causal curve in ${\cal D}$ with
endpoints in ${\cal I}$, then by theorem $\ref{qlptc}$ $\gamma$ is
fixed endpoint homotopic to a causal curve $\beta\colon I \to{\cal
I}$. By lifting this homotopy to ${\cal D}^*$, we find that
$\gamma^*$ is homotopic to a curve in ${\cal I}^*$. Thus the
endpoints of $\gamma^*$ lie on the same component ${\cal
I}^*_{\alpha}$, hence $x\in {\cal D}^*_{\alpha}$. It follows
$\{{\cal D}^*_{\alpha}\}$ is an open cover of ${\cal D}^*$.

Moreover, we have also shown that $I^+({\cal I}^*_{\alpha},{\cal
D}^*)\cap I^-({\cal I}^*_{\beta},{\cal D}^*)=\emptyset$ if
$\alpha\neq\beta$. As a consequence the sets ${\cal D}^*_{\alpha}$
are disjoint. However ${\cal D}^*$ is connected, thus we must have
$\vert A\vert =1$, i.e. ${\cal I}^*$ is connected.

Hence by proposition $\ref{last}$ we have that any loop in ${\cal
D}$ based at ${\cal I}$ is homotopic to a loop in ${\cal I}$ based
at the same point. Let now $\gamma$ be any curve in ${\cal D}$
from $p\in{\cal I}$ to $q\in{\cal I}$. Since ${\cal I}$ is
connected, there is a curve in $\sigma$ in ${\cal I}$ from $q$ to
$p$, hence the concatenation $\gamma\ast \sigma$ is a loop in
${\cal D}$ based at $p\in{\cal I}$. Therefore there is a loop $c$
in ${\cal I}$ so that $\gamma\ast\sigma$ is fixed endpoint
homotopic to $c$. Thus $\gamma$ is homotopic to $c\ast\sigma^-$,
completing the proof. $\Box$ \Par

\section{Future research}

We end this dissertation by making some comments on possible
directions for future research related to the new results
established here.\par\vspace{.4cm}

\noindent \textbf{I. Rigidity in asymptotically flat spacetimes}

A spacetime $(M,g)$ admitting a conformal completion is
\textit{asymptotically flat} if the conformal boundary $\scri$ is
null. Some of the classical models in general relativity, such as
Schwarzchild, are asymptotically flat.  The study of such
spacetimes is an important branch in mathematical relativity, that
dates back to the origins of the field. Moreover, some notions
like topological censorship  were first developed in the
asymptotically flat setting.

Recall that in section $\ref{ghdss}$ we extended the rigidity
result presented in the form of theorem $\ref{rigds}$ from the
asymptotically simple and de Sitter to the more relaxed globally
hyperbolic and asymptotically de Sitter case. However, theorem
$\ref{rigds}$ is also valid for asymptotically flat and simple
spacetimes. Thus is natural to expect to get a rigidity result
analogous to theorem $\ref{teo1}$ in the asymptotically flat case.

Notice however, that the proof of theorem $\ref{teo1}$ relies
heavily on the construction of a globally hyperbolic extension and
on the rather simple causal structure of this new spacetime near
the endpoints of a null line $\eta$ connecting $\scri^+$ and
$\scri^-$. A more careful analysis is needed in the asymptotically
flat setting, since $\partial I^+(\eta)$ intersects $\scri$ in a
null geodesic in this case.

\eject

\noindent\textbf{II. Other models}

In section $\ref{mattersec}$ we explored asymptotically de Sitter
spacetimes which were solution to the Einstein equations with an
energy-momentum tensor satisfying some properties. We believed
similar results may be obtained in the asymptotically flat case as
well, but furthermore, by suitable changing the restrictions on
$T$ we can obtain rigidity results for other choices of $T$, such
as electromagnetic fields.\par\vspace{.4cm}

\noindent \textbf{III. Asymptotically Anti de Sitter black
strings}

Intuitively, a black string is formed when the result of
gravitational collapse is not a point (i.e.  a black hole) but
rather a one dimensional object. Lately, the conjectured AdS/CFT
correspondence has sparked much interest in the study of black
strings that approach Anti de Sitter space outside  a compact set.

More formally, an asymptotically de Sitter black string is a
spacetime $(M,g)$ asymptotic to $AdS^n\times N$, where
$(AdS^n,g_0)$ is the standard Anti de Sitter spacetime and $(N,h)$
is a compact Riemannian manifold, whose metric $g$ approaches
$g_0+ h$ near $\scri$.

In this setting, we feel a quasi-local form of topological
censorship can be established as a consequence of theorem
$\ref{qlptc}$, once appropriate decay conditions on the metric are
imposed. Such result would settle a question on the traversability
of worm holes posed by J. Maldacena $\cite{mal}$.

\linespread{2} \small\normalsize

\appendix

\chapter{Extensions of spacetimes with boundary}

A good part of the constructions and results in Lorentzian
geometry rely on ``open conditions" such as variational principles
or the existence of open normal neighborhoods.

In the process of extending the main results of Lorentzian
geometry to the spacetime-with-boundary setting, we are faced with
the technical difficulty that such open conditions in principle
may not hold in the boundary. Hence a much more delicate analysis
has to be performed.

Another way to deal with this issue consists in embedding our
spacetime-with-boundary in an open spacetime of the same
dimension. This latter approach enables us to apply the standard
results and methods in the extended spacetime and then interpret
the results so obtained back in to our original
spacetime-with-boundary setting.

As the next result shows, any spacetime-with-boundary can be
extended to an open spacetime.

\begin{teo}\label{openex}
Every spacetime-with-boundary $(M,g)$ admits an extension to a
spacetime $(\overline{M},\overline{g})$.
\end{teo}
\par

\noindent\textit{Proof:} First extend $M$ to a smooth manifold
$M^{\prime}$ by means of attaching collars to all the components
of $\partial M$. Since $M$ is time orientable, there exists a
timelike vector field $V\in{\cal X}(M)$. Let us extend $V$ to all
of $M^{\prime}$ and let $W=\{p\in M^{\prime}\mid V_p\neq 0\}$.
Clearly $W$ is an open subset of $M^{\prime}$ containing all of
$M$, so without loss of generality we can assume
$M^{\prime}=W$.\par

Let $p\in\partial M$ and choose a $M^{\prime}$-chart ${\cal U}_p$
around it. Now let $g=g_{ij}dx^idx^j$ be the coordinate expression
of $g$ in the $M$-chart $M\cap{\cal U}_p$. Since the $g_{ij}$'s
are smooth functions on $M\cap{\cal U}_p$, they can be smoothly
extended to an $M^{\prime}$-neighborhood ${\cal
U}^{\prime}_p\subset{\cal U}_p$ with $M\cap{\cal
U}^{\prime}_p=M\cap{\cal U}_p$. Let us denote by $g^{\prime}_{ij}$
such extensions. It is important to notice that ${\cal
U}^{\prime}_p$ can be chosen in such a way that
$g^{\prime}=g^{\prime}_{ij}{dy}^i{dy}^j$ is a Lorentz metric
 on ${\cal U}^{\prime}_p$ with $g^{\prime}(V,V)<0$.\par

Now, let $e_0$ be the unit vector field (with respect to
$g^{\prime}$) in the direction of $V$ and let $B=\{e_0, e_1, e_2,
e_3,\}$ be an orthonormal basis of $T_pM^{\prime}$. Choose an
$M^{\prime}$-neighborhood ${\cal V}_p\subset {\cal U}^{\prime}_p$
in which a local frame field relative to $g^{\prime}$ and adapted
to $B$ is defined. Thus
\begin{equation} g^{\prime}=-e_0^*\otimes
e_0^*+\sum_ie_i^*\otimes e_i^*\ \textrm{on}\ {\cal
V}_p,\end{equation} where ${e_i}^*$ denotes the covector
$g^{\prime}$-related to $e_i$. Choose a cover $\{{\cal
V}_{\alpha}\}$ of $\partial M$ by such open sets and  let
$h_{\alpha}=2e_0^*\otimes e_0^*+g^{\prime}_{\alpha}$. Further
consider a smooth partition of unity $f_{\alpha}$ subordinated to
$\{{\cal V}_{\alpha}\}$, thus
$h=\sum_{\alpha}f_{\alpha}h_{\alpha}$ is a Riemannian metric on
${\cal V}=\cup_{\alpha}{\cal V}_{\alpha}$.\par

Finally,  let $X$ be the unit vector field (with respect to $h$)
in the direction of $V$, let $\omega$ be the covector $h$-related
to $X$ and let $g^{\prime\prime}=h-2\omega\otimes\omega$. It is a
straightforward computation to check that $g^{\prime\prime}$ is a
Lorentz metric on ${\cal V}$ that agrees with $g$ on the overlap
${\cal V}\cap M$. Thus by gluing $g^{\prime\prime}$ and $g$
together we obtain a Lorentz metric $\overline{g}$ on
$\overline{M}={\cal V}\cup M$. Notice $\overline{g}$ is smooth
since ${\cal V}$ is open. $\Box$\par

\chapter{Focal points along null geodesics}

Here we specialize the classical result relating focal points,
normal variations and degeneracy of the normal exponential map to
the case in which the base geodesic is null.\par

In what follows,  $\sigma$ is a geodesic normal to $P$ at
$p=\sigma (0)$ and $P\subset M$ is a spacelike submanifold of
codimension at least 2.\par

Let us first recall some definitions and standard results
concerning Jacobi fields. For a reference, consult chapters $8$
and $10$ on $\cite{ON}$ or chapter $10$ on $\cite{BE}$.

\begin{lemma}\label{jaco1}
Let $p\in M$ and  $x\in T_pM$. For $v_x\in T_x(T_pM)$ we have
\begin{equation}
({\rm exp}_p)_*(v_x)=V(1)
\end{equation}
where $V$ is the unique Jacobi field on the geodesic ${\gamma}_x$
such that $V(0)=0$ and $V^{\prime}(0)=v\in T_pM$.
\end{lemma}
\par

\begin{lemma}\label{jaco2}
Let $Y$ be a Jacobi field on the geodesic $\gamma$. Then
$Y\perp\gamma$ if and only if there exist $a\neq b$ such that
$Y(a)\perp \gamma$ and $Y(b)\perp\gamma$.
\end{lemma}
\par

\begin{defi} The
tensor $\widetilde{{\rm II}}\colon {\mathcal X}(P)\times {\mathcal
X}(P)^\perp\to {\mathcal X}(P)$ is defined by
\begin{equation}
\widetilde{{\rm II}}(X,Y)={\rm tan}\nabla_XY
\end{equation}
where $\nabla$ is the Levi-Civita connection on $M$.
\end{defi}

\begin{defi}
A $P$-Jacobi field on a geodesic $\sigma$ is a Jacobi field that
satisfies the following initial conditions:
\begin{enumerate}
\item $V(0)$ is tangent to $P$. \item ${\rm tan}\,
V^{\prime}(0)=\widetilde{{\rm
II}}\left(V(0),{\sigma}^{\prime}(0)\right)$
\end{enumerate}
\end{defi}
\par

\begin{defi}
The point $q=\sigma (r)$, $r\neq 0$ is a focal point of $P$ along
$\sigma$ provided there exists a nonzero $P$-Jacobi field $J$ on
$\sigma$ with $J(r)=0$.
\end{defi}
\par

\begin{defi}
The normal null cone to $P$ is the manifold
\begin{equation}
\mathfrak{N}( P)=\{v\in NP\mid g(v,v)=0\, ,\, v\neq 0\}
\end{equation}
\end{defi}
\par

\begin{teo}\label{focalnullpts}
Let $\sigma$ be a null geodesic normal to $P$. The following are
equivalent:
\begin{enumerate}
\item $\sigma (1)$ is a focal point of $P$ along $\sigma$. \item
There exists a nontrivial variation $\mathbf{x}$ of $\sigma$
through $P$-normal null geodesics such that $V(1)=0$. \item The
normal exponential map ${\rm exp}^{\perp}$ restricted to
$\mathfrak{N}( P)$ is singular. (i.e. there is $x\neq 0$ tangent
to ${\mathfrak{N}( P)}$ such that ${\rm exp}^{\perp}_*(x)= 0$).
\end{enumerate}
\end{teo}
\par

\noindent\textit{Proof:} We divide the proof in four parts.

 \noindent{\bf I.} $(1)$ implies $(2)$:\par

Let $Y$ be a $P$-Jacobi vector field on $\sigma$ such that
$Y(1)=0$. Then $Y(0)\perp\sigma$ and $Y(1)\perp\sigma$ so by lemma
\ref{jaco2} above $Y\perp\sigma$. Thus
\begin{equation}
0=\frac{d}{dt}g(Y(t),{\sigma}^{\prime}(t))=g(Y^{\prime},{\sigma}^{\prime})+
g(Y,{\sigma}^{\prime\prime})=g(Y^{\prime},{\sigma}^{\prime}).
\end{equation}
In particular, since ${\sigma}^{\prime}(0)\perp P$ we have
\begin{equation}\label{normaleq1}
0=g(Y^{\prime}(0),{\sigma}^{\prime}(0))=g({\rm nor}\,
Y^{\prime}(0),{\sigma}^{\prime}(0)).
\end{equation}
\par

Now let $A=T_p(P)^{\perp}$ and $W=\,{\rm
span}_A\,\{{\sigma}^{\prime}(0) \}$. It is clear that $W^\perp
:=\{x\in A\mid g(x,w)=0\, \forall w\in W\}$ is a degenerate
subspace of $A$ hence by remark $\ref{degiso}$ translation by
${\sigma}^{\prime}(0)$ gives rise to an isomorphism
\begin{equation}\label{iso}
W^{\perp}\approx T_{{\sigma}^{\prime}(0)}\left(
{{{\Lambda}}_p(A)}\right). \end{equation}\par

As we have noted before ${\rm nor}\, Y^{\prime}(0)\in W^{\perp}$,
thus by the isomorphism just mentioned there exists a curve
$\lambda$ in ${{{\Lambda}}_p}(A)$ with $\lambda
(0)={\sigma}^{\prime}(0)$ and
${\displaystyle\frac{d{\lambda}}{ds}}(0)={\rm nor}\,
Y^{\prime}(0)$.\par

Let $\{e_i\}$ be a basis of $T_p(P)^{\perp}$, then since $\lambda
(s)\in T_p(P)^{\perp}$ we have $\lambda (s)={\lambda}^i(s)e_i$.
Consider now a curve $\alpha$ in $P$ such that $\alpha (0)=p$ and
${\alpha}^{\prime}(0)=Y(0)$ and let $\{E_i\}$ be a normal parallel
frame along $\alpha$ adapted to $\{e_i\}$ and define $Z\in
\mathcal{X}({\alpha})$ by $Z(t)={\lambda}^i(t)E_i$. Hence ${\rm
nor}\, Z^{\prime}(t)={\displaystyle
\frac{d{\lambda}^i}{dt}}E_i(t)$, then
\begin{equation}
{\rm nor}\, Z^{\prime}(0)={\lambda}^{\prime}(0)={\rm nor}\,
Y^{\prime}(0).
\end{equation}
Notice also
\begin{equation}
{\rm tan}\, Z^{\prime}(0)=\widetilde{{\rm II}} \left(
{\alpha}^{\prime}(0),Z(0)\right)=\widetilde{{\rm II}}\left(
Y(0),{\sigma}^{\prime}(0)\right)={\rm tan}\, Y^{\prime}(0)
\end{equation}
Thus $Z$ is a null vector field on $\alpha$ normal to $P$ such
that
\begin{equation}
Z(0)={\sigma}^{\prime}(0)\quad\quad {\rm and}\quad\quad
Z^{\prime}(0)=Y^{\prime}(0). \end{equation}\par

Let $\mathbf{x}(u,v)={\rm exp}^{\perp}(uZ(v))$. Note
$\mathbf{x}_u(0,v)=Z(v)$ therefore ${\sigma}_v(u):
=\mathbf{x}(u,v)$ is a null geodesic normal to $P$ with initial
velocity $Z(v)$. In particular ${\sigma}_0(u)={\rm
exp}^{\perp}(u{\sigma}^{\prime}(0))=\sigma (u)$ so $\mathbf{x}$ is
a variation of $\sigma$.\par

Finally, let $V$ be the variation vector field of $\mathbf{x}$. By
standard results, $V$ is a Jacobi vector field on $\sigma$. Notice
$\mathbf{x}(0,v)={\rm exp}^{\perp}\left(0_{\pi\circ
Z(v)}\right)=\alpha (v)$ thus
\begin{equation}
V(0)=\mathbf{x}_v(0,v)= {\alpha}^{\prime}(0)=Y(0).
\end{equation}
Moreover,
\begin{equation}
V^{\prime}(0)=\mathbf{x}_{vu}(0,0)=\mathbf{x}_{uv}(0,0)=Z^{\prime}(0)=Y^{\prime}(0)
\end{equation}
Thus, $V(u)=Y(u)$ by uniqueness of Jacobi fields. Hence
$V(1)=Y(1)=0$ so $\mathbf{x}$ is a variation of $\sigma$ with the
desired properties.\par

\noindent{\bf II.} $(2)$ implies $(1)$:\par

Let $\mathbf{x}(u,v)$ be such a variation. Let $\alpha$ be the
curve in $P$ defined by $\alpha (v)=\mathbf{x}(0,v)$ and define
$Z\in\mathcal{X}(\alpha )$ by $Z(v)=\mathbf{x}_u(0,v)$.\par

Consider now the variation vector field $V$ of $\sigma$. Then
\begin{equation}
V(0)=\mathbf{x}_v(0,0)={\alpha}^{\prime}(0)\in T_p(P).
\end{equation}
 Moreover,
\begin{equation}
V^{\prime}(0)=\mathbf{x}_{vu}(0,0)=\mathbf{x}_{uv}(0,0)=Z^{\prime}(0)
\end{equation}
thus, since $Z(0)=\mathbf{x}_u(0,0)={\sigma}^{\prime}(0)$ we have
\begin{equation}
{\rm tan}\, V^{\prime}(0)= {\rm tan}\,
Z^{\prime}(0)=\widetilde{{\rm II}}\left(
{\alpha}^{\prime}(0),Z(0)\right)= \widetilde{{\rm II}}\left(
V(0),{\sigma}^{\prime}(0)\right)
\end{equation}
so $V(u)$ is a $P$-Jacobi field that vanishes at $u=1$.\par

\noindent{\bf III.} $(3)$ implies $(2)$:\par

Let $x\in T_{{\sigma}^{\prime}(0)}\mathfrak{N}( P)$, $x\neq 0$,
with ${\rm exp}^{\perp}_*(x)=0$. Let $\pi\colon NP\to P$ be the
standard projection. Since $\pi$ is a submersion, we have that
$\pi_* (x)=0$ if and only if $x$ is tangent to the fiber
$T_p(P)^{\perp}$.\par

Let us first consider the case $\pi_* (x)=0$. Then $x\in
T_{{\sigma}^{\prime}(0)}(T_p(P)^{\perp})$, so let $x_0\in
T_p(P)^{\perp}$ be the  vector that corresponds canonically to
$x$. By lemma \ref{jaco1}
\begin{equation}
0={\rm exp}^{\perp}_*(x)=({\rm exp}_p)_*(x)=J(1)
\end{equation}
where $J$ is the unique Jacobi field on $\sigma$ with the initial
conditions $J(0)=0$ and $J^{\prime}(0)=x_0$. \par

Since $x_0\in T_p(P)^{\perp}$, then
\begin{equation}
{\rm tan}\, J^{\prime}(0)=0=\widetilde{{\rm
II}}(J(0),{\sigma}^{\prime}(0)).
\end{equation}
Thus $J$ is a $P$-Jacobi field on $\sigma$ vanishing at $J(1)$. It
is clear that $J\not\equiv 0$ since $J^{\prime}(0)=x_0\neq 0$ and
the result follows from ({\bf I}) above.\par

We now consider the case $\pi_* (x)\neq 0$. Let
$Z:I\to\mathfrak{N}( P)$ be a curve such that
$Z(0)={\sigma}^{\prime}(0)$ and $Z^{\prime}(0)=x$. Let
$\mathbf{x}(u,v)={\rm exp}^{\perp}(uZ(v))$ and note
$\mathbf{x}(u,0)={\rm exp}^{\perp}(u{\sigma}^{\prime}(0))=\sigma
(u)$, so $\mathbf{x}$ is a variation of $\sigma$ through null
geodesics normal to $P$.\par

Further, $\mathbf{x}(0,v)={\rm exp}^{\perp}(0_{\pi\circ
Z(v)})=(\pi\circ Z)(v)$, then
\begin{equation}
V(0)=\mathbf{x}_v(0,0)=(\pi\circ Z)^{\prime}(0)=\pi_* (x)\neq 0
\end{equation}
and, as a consequence, $V\not\equiv 0$ .\par

Finally, observe that $\mathbf{x}(1,v)={\rm exp}^{\perp}(Z(v))$
hence
\begin{equation}
V(1)={\mathbf{x}}_v(1,0)={\rm exp}^{\perp}_*(Z^{\prime}(0))={\rm
exp}^{\perp}_*(x)=0
\end{equation}\par

\noindent {\bf IV.} $(2)$ implies $(3)$:\par

Let $\mathbf{x}$ be a variation with the given properties and
denote by $\alpha$ its initial curve (i.e.
${\alpha}(v)=\mathbf{x}(0,v)$). Let us define a variation on
$\mathfrak{N}(P)$ by $\tilde{\mathbf{x}}(u,v)=u\mathbf{x}_u(0,v)$
and notice $\mathbf{x}={\rm exp}^\perp\circ\tilde{\mathbf{x}}$.
Thus
\begin{equation}
\mathbf{x}_v=\mathbf{x}_*(\partial_v)= ( {\rm
exp}^\perp\circ\tilde{\mathbf{x}})_*(\partial_v)= {\rm
exp}^\perp_*(\tilde{\mathbf{x}}_*(\partial_v))= {\rm
exp}^\perp_*(\tilde{\mathbf{x}}_v)
\end{equation}
hence
\begin{equation}
{\rm exp}^\perp_*(\tilde{\mathbf{x}}_v(0,1)=\mathbf{x}_v(0,1)=0.
\end{equation}
Now let $\beta\colon I\to \mathfrak{N}(P)$ be the final
transversal curve of $\tilde{\mathbf{x}}$, that is, $\beta
(v)=\tilde{\mathbf{x}}(1,v)$. Then we have
$\beta^{\prime}(0)=\tilde{\mathbf{x}}_v(1,0)$. Moreover, by
definition $\tilde{\mathbf{x}}(1,v)=\mathbf{x}_u(0,v)$ hence
$\alpha =\pi\circ\beta$. Therefore
\begin{equation}
\mathbf{x}_v(0,0)=\alpha^{\prime}(0)=\pi_*(\beta^{\prime}(0))=\pi_*(\tilde{\mathbf{x}}_v(1,0)).
\end{equation}
In virtue of the last two equations, we have that $x_v(0,0)\neq 0$
implies $\tilde{\mathbf{x}}_v(1,0)\neq 0$ and we are done.

On the other hand assume $\mathbf{x}_v(0,0)=0$ and consider the
variation vector field $V=\mathbf{x}_v(u,0)$, hence $V(0)=0$. By
({\bf II}) above $V$ is a $P$-Jacobi field, so in particular
\begin{equation}
{\rm tan}\, V^{\prime}(0)=\widetilde{{\rm
II}}\left(V(0),{\sigma}^{\prime}(0)\right)=0
\end{equation}
thus $V^{\prime}(0)$ is normal to $P$ at $p$. Since $V$ is
perpendicular to $\sigma$ at both $u=0$ and $u=1$ the relation
$g(V^{\prime}(0),\sigma^{\prime}(0))=0$  follows easily from lemma
\ref{jaco2} (see equation (\ref{normaleq1})). Further notice
$V^{\prime}(0)\neq 0$, since otherwise $V\equiv 0$. Hence
$V^{\prime}(0)$ is the desired vector due to the isomorphism
(\ref{iso}). $\Box$

\chapter{Conformally related metrics}

Here we carry out some computations that will enable us see how
the mathematical objects present in the Einstein equations change
under conformal transformations of the metric. Thus, let us
consider
\begin{equation}
g={\Omega}^2\tilde{g} \end{equation} and let us denote by a tilde
$\tilde{\ }$ the quantities related to the physical metric
$\tilde{g}$.

First, we write down the formula for conformally related curvature
tensors. For a proof, refer to $\cite{handbook}$.

\begin{prop}\label{confcurva}
 Let $g$ and $g_0$ be two conformally related Lorentz metrics
in $M^{n}$, i.e. $g={\Omega}^2 \tilde{g}$ for some positive
$\Omega \in {\cal C}^{\infty}(M)$, then
\begin{equation}
\widetilde{{\text {Ric}}}=\text{Ric}+\frac{n-2}{\Omega}{\text
Hess}_{\ \Omega}+\left[
\frac{\Delta\Omega}{\Omega}-\frac{n-1}{{\Omega}^2}g(\nabla\Omega
,\nabla\Omega )\right]g
\end{equation}
\end{prop}

Now, we turn our attention to the energy-momentum tensor.

\begin{prop}\label{confdiv}
Let $T$ be any symmetric $(0,2)$ tensor on $M$. Then
\begin{equation}
\mbox{\text {div}}\, T(X)=\frac{1}{{\Omega}^2}{\widetilde{\text {
div}}}\, T(X) +\frac{(n-2)}{\Omega}T(\nabla\Omega
,X)-\frac{X(\Omega )}{\Omega^3}\widetilde{\mbox{ Tr}}\, T
\end{equation}
\end{prop}

\noindent\textit{Proof:} Let $\{E_i\}$ be a $g$-orthonormal frame.
Then $\{\tilde{E_i}\}$ is a $\tilde{g}$-orthonormal frame, where
$\tilde{E_i}=\Omega E_i$ and notice also that
$\tilde{\nabla}\Omega =\Omega^2\nabla\Omega$. Consider now the
formula for the divergence of $T$ in an orthonormal frame
$\cite{ON}$:
\begin{eqnarray}\label{div001}
\textrm{div}T(X)&=&\sum_{i=1}^n\epsilon_i\nabla_{E_i}T(E_i,X)\\
\nonumber
&=&\sum_{i=1}^n\epsilon_i[E_i(T(E_i,X))-T(\nabla_{E_i}E_i,X)-T(E_i,\nabla_{E_i}X)].
\end{eqnarray}

We proceed to compute each one of the summands in the right hand
side of (\ref{div001}). The first term is easy to deal with:

\begin{eqnarray}\label{div017}
E_i(T(E_i,X))&=& \frac{1}{\Omega}\tilde{E_i}\left(
\frac{1}{\Omega}T(\tilde{E_i},X\right)  \nonumber\\ &=&
\frac{1}{\Omega^2}\tilde{E_i}(T(\tilde{E_i},X))+\frac{1}{\Omega}\tilde{E_i}\left(
\frac{1}{\Omega}\right)
T(\tilde{E_i},X)\\
&=& \frac{1}{\Omega^2}\tilde{E_i}(T(\tilde{E_i},X))
-\frac{1}{\Omega^3}\tilde{E_i}(\Omega )T(\tilde{E_i},X)\nonumber
\end{eqnarray}

To find a formula for the third term, let us recall first that the
connections $\nabla$ and $\tilde{\nabla}$ are related by (see
1.159 in \cite{besse}):
\begin{equation}\nabla_XY=\tilde{\nabla}_XY+\frac{X(\Omega
)}{\Omega}Y+\frac{Y(\Omega
)}{\Omega}X-\frac{\tilde{g}(X,Y)}{\Omega}\tilde{\nabla}(\Omega
),\end{equation} hence
\begin{equation}
\nabla_{\tilde{E_i}}X=\tilde{\nabla}_{\tilde{E_i}}X+\frac{\tilde{E_i}(\Omega
)}{\Omega}X+\frac{X(\Omega
)}{\Omega}\tilde{E_i}-\tilde{g}(\tilde{E_i},X)\frac{\tilde{\nabla}\Omega}{\Omega}.
\end{equation}
Thus
\begin{equation}
T(E_i,\nabla_{E_i}X)=T\left(
\frac{\tilde{E_i}}{\Omega}\nabla_{\tilde{E_i}/\Omega}X
\right)=\frac{1}{\Omega^2}T(\tilde{E_i},\nabla_{\tilde{E_i}}X)
\end{equation}
and then
\begin{eqnarray}\label{div010}
 T(E_i,\nabla_{E_i}X)&=&\frac{1}{\Omega^2}T(\tilde{E_i},\tilde{\nabla}_{\tilde{E_i}}X)+\frac{\tilde{E_i}(\Omega
 )}{\Omega^3}T(\tilde{E_i},X)\\&+&\frac{X(\Omega
 )}{\Omega^3}T(\tilde{E_i},\tilde{E_i})-\frac{\tilde{g}(\tilde{E_i},X)}{\Omega^3}T(\tilde{E_i},\tilde{\nabla}\Omega
 ).\nonumber
\end{eqnarray}

Finally, let us tackle the second term. First notice that
\begin{equation}
T(\nabla_{E_i}E_i,X)=T (
\tilde{\nabla}_{\tilde{E_i}/\Omega}\tilde{E_i}/\Omega+\frac{2\tilde{E_i}(\Omega)}{\Omega^3}
\tilde{E_i}-\tilde{g} (
{\tilde{E_i}}/{\Omega},{\tilde{E_i}}/{\Omega})\frac{\tilde{\nabla}\Omega}{\Omega},X
) ,
\end{equation}
and also
\begin{eqnarray}\label{div002}
T(\tilde{\nabla}_{\tilde{E_i}/\Omega}\tilde{E_i}/\Omega ,X) &=&
\frac{1}{\Omega}T\left(
\frac{1}{\Omega}\tilde{\nabla}_{\tilde{E_i}}\tilde{E_i}+\tilde{E_i}\left(\frac{1}{\Omega}\right)\tilde{E_i},X
\right)\nonumber \\ &=&
\frac{1}{\Omega^2}T(\tilde{\nabla}_{\tilde{E_i}}\tilde{E_i},X)+\frac{1}{\Omega}T(\tilde{E_i}\left(\frac{1}{\Omega}\right)
\tilde{E_i},X)\\
&=&\frac{1}{\Omega^2}T(\tilde{\nabla}_{\tilde{E_i}}\tilde{E_i},X)-\frac{1}{\Omega^3}\tilde{E_i}(\Omega
)T(\tilde{E_i},X),\nonumber
\end{eqnarray}

\begin{equation}\label{div003}
T\left( \frac{2}{\Omega\tilde{E_i}(\Omega)}{\Omega^3}
\tilde{E_i},X\right) = \frac{2}{\Omega^3}\tilde{E_i}(\Omega
)T(\tilde{E_i},X),
\end{equation}

\begin{equation}\label{div004}
T(\tilde{g}(
{\tilde{E_i}}/{\Omega},{\tilde{E_i}}/{\Omega})\frac{\tilde{\nabla}\Omega}{\Omega},X)
=\frac{\tilde{g}(\tilde{E_i},\tilde{E_i})}{\Omega}T\left(\frac{\tilde{\nabla}\Omega}{\Omega^2}
,X\right)=\frac{\epsilon_i}{\Omega}T({\nabla}\Omega ,X).
\end{equation}

\noindent Hence equations (\ref{div002}), (\ref{div003}) and
(\ref{div004}) imply
\begin{equation}\label{div014}
T(\nabla_{E_i}E_i,X)=\frac{1}{\Omega^2}T(\tilde{\nabla}_{\tilde{E_i}}\tilde{E_i},X)+\frac{1}{\Omega^3}\tilde{E_i}(\Omega
)T(\tilde{E_i},X)-\frac{\epsilon_i}{\Omega}T(\nabla\Omega ,X)
\end{equation}

Now we group all the above terms. First notice
\begin{eqnarray}\label{div005}
\sum_{i=1}^n\epsilon_i\left(\frac{1}{\Omega^3}\tilde{E_i}(\Omega
)T(\tilde{E_i},X)\right)&=&\frac{1}{\Omega^3}T\left(\sum_{i=1}^n\epsilon_i\tilde{E_i}(\Omega
)\tilde{E_i},X\right)\\
&=&\frac{1}{\Omega^3}T(\tilde{\nabla}\Omega,X)=\frac{1}{\Omega}T(\nabla\Omega
,X)\nonumber
\end{eqnarray}
and similarly
\begin{eqnarray}\label{div006}
\sum_{i=1}^n\epsilon_i\left(
\frac{\tilde{g}(\tilde{E_i},X)}{\Omega^3}T(\tilde{E_i},\tilde{\nabla}\Omega
 )\right) &=&\frac{1}{\Omega^3}T\left( \sum_{i=1}^n\epsilon_i\tilde{g}(\tilde{E_i},X)\tilde{E_i},\tilde{\nabla}\Omega)
 \right)\\
&=& \frac{1}{\Omega^3}T(X,\tilde{\nabla}\Omega
)=\frac{1}{\Omega}T(X,\nabla\Omega ).\nonumber
\end{eqnarray}
On the other hand
\begin{equation}\label{div008}
\sum_{i=1}^n\epsilon_i\left( \frac{X(\Omega
)}{\Omega^3}T(\tilde{E_i},\tilde{E_i})\right) =\frac{X(\Omega
)}{\Omega^3}\sum_{i=1}^n\epsilon_iT(
{\tilde{E_i}},{\tilde{E_i}})=\frac{X(\Omega
)}{\Omega^3}\widetilde{\textrm{Tr}}\, T,
\end{equation}
hence by putting (\ref{div010}), (\ref{div005}), (\ref{div006})
and (\ref{div007}) together we get
\begin{equation}\label{div011}
\sum_{i=1}^n\epsilon_iT(E_i,\nabla_{E_i}X)=\frac{X(\Omega
)}{\Omega^3}\widetilde{\textrm{Tr}}+T\frac{1}{\Omega^2}\sum_{i=1}^nT(\tilde{E_i},{\tilde{\nabla}}_{\tilde{E_i}}X).
\end{equation}

Moreover, since $\epsilon_i^2=1$ we have
\begin{equation}\label{div007}
\sum_{i=1}^n\epsilon_i[\frac{\epsilon_i}{\Omega}T(\nabla\Omega
,X)]=\frac{n}{\Omega}T(\nabla\Omega ,X),
\end{equation}
thus by (\ref{div014}), (\ref{div005}) and (\ref{div007}) we have
\begin{equation}\label{div012}
\sum_{i=1}^n\epsilon_iT(\nabla_{E_i}E_i,X)=\frac{1-n}{\Omega}T(\nabla\Omega
,X)+\frac{1}{\Omega^2}\sum_{i=1}^n\epsilon_iT(\tilde{\nabla}_{\tilde{E_i}},X).
\end{equation}

Further combining (\ref{div017}) and (\ref{div005}) yields
\begin{equation}\label{div016}
\sum_{i=1}^n\epsilon_iE_i(T(E_i,X))=-\frac{1}{\Omega}T(\nabla\Omega
,X)+\frac{1}{\Omega^2}\sum_{i=1}^n\epsilon_i\tilde{E_i}(T(\tilde{E_i},X))
\end{equation}

Finally,
\begin{equation}\label{div009}
\frac{1}{\Omega^2}\sum_{i=1}^n\epsilon_i\left(
\tilde{E_i}(T(\tilde{E_i},X))-T(\tilde{\nabla}_{\tilde{E_i}}\tilde{E_i},X)-
T(\tilde{E_i},\tilde{\nabla}_{\tilde{E_i}}X) \right) =
\frac{1}{\Omega^2}\tilde{\textrm{div}}T(X).
\end{equation}

Hence, by substituting (\ref{div011}), (\ref{div012}),
(\ref{div016}) and (\ref{div009}) into (\ref{div001}) we get
\begin{equation}
\mbox{\text {div}}\, T(X)=\frac{1}{{\Omega}^2}{\widetilde{\text {
div}}}\, T(X) +\frac{(n-2)}{\Omega}T(\nabla\Omega
,X)-\frac{X(\Omega )}{\Omega^3}\widetilde{\mbox{ Tr}}\, T
\end{equation}
as desired. $\Box$

\end{document}